\RequirePackage{silence}
\documentclass[twocolumn, times]{aastex631}

\usepackage{amsmath}
\usepackage{xspace}
\AtBeginDocument{\hfuzz=100pt\vfuzz=20pt\hbadness=10000}

\shorttitle{Radiative Transfer Modeling of IRAS~15398$-$3359}
\shortauthors{Ishizuka et al.}

\begin{document}
\title{Radiative Transfer Modeling of Multi-Shell Molecular Outflows in the Class 0 Protostar IRAS~15398$-$3359}

\author[0009-0009-7258-4985]{Yushi Ishizuka}
\correspondingauthor{Yushi Ishizuka}
\email{yushi.ishizuka@gmail.com}
\affiliation{Department of Physics and Astronomy, Graduate School of Science and Engineering, Kagoshima University, 1-21-35 Korimoto, Kagoshima, Kagoshima 890-0065, Japan}

\author[0000-0003-0845-128X]{Shigehisa Takakuwa}
\affiliation{Department of Physics and Astronomy, Graduate School of Science and Engineering, Kagoshima University, 1-21-35 Korimoto, Kagoshima, Kagoshima 890-0065, Japan}
\affiliation{Academia Sinica Institute of Astronomy \& Astrophysics,
11F of Astronomy-Mathematics Building, AS/NTU, No. 1, Sec. 4, Roosevelt Rd, Taipei 10617, Taiwan, R.O.C.}

\author[0000-0003-1549-6435]{Kazuya Saigo}
\affiliation{Academia Sinica Institute of Astronomy \& Astrophysics,
11F of Astronomy-Mathematics Building, AS/NTU, No. 1, Sec. 4, Roosevelt Rd, Taipei 10617, Taiwan, R.O.C.}

\author[0000-0003-4361-5577]{Jinshi Sai}
\affiliation{Department of Physics and Astronomy, Graduate School of Science and Engineering, Kagoshima University, 1-21-35 Korimoto, Kagoshima, Kagoshima 890-0065, Japan}

\author[0009-0001-1889-9216]{Shu Ishibashi}
\affiliation{Department of Physics and Astronomy, Graduate School of Science and Engineering, Kagoshima University, 1-21-35 Korimoto, Kagoshima, Kagoshima 890-0065, Japan}

\begin{abstract}

Recent high-angular-resolution observations of protostellar outflows have revealed complex spatial and
velocity structures that are difficult to describe with a single shell.
We have investigated the primary outflow of the Class 0 protostar IRAS~15398$-$3359 in the H$_2$CO and CS emission observed with ALMA and identified four shell-like components, which trace the cavity walls of the more extended CO outflows.
Non-LTE radiative-transfer calculations of H$_2$CO and synthetic ALMA
observations of the three-dimensional model show that a common ellipsoidal geometrical form and
Hubble-law-like velocity field can broadly reproduce the channel-map and
position--velocity structures of the identified shells
when the geometry, spatial configuration, and
kinematics of each shell are allowed to vary independently.
The kinematic timescales and inclination angles of the inner and outer shells are $\sim$1000 yr ($19^\circ$ and $15^\circ$) and $\sim$750 yr ($5^\circ$ and $8^\circ$), respectively.
The momentum injection rate
of the outer bipolar pair is approximately 2.2 times that of the inner pair.
The inferred $\dot{M}_{\rm wind}/\dot{M}_{\rm acc}$ ratios are $0.7$--$3\%$ and $2$--$7\%$ for the inner and outer shells.
The observed shell-to-shell differences suggest episodicity of the mass ejection events, where the strength
and direction of the outflow activity may vary over short timescales of
order $10^2$--$10^3$~yr.

\end{abstract}

\section{Introduction}
\label{sec:introduction}
\begin{table*}
\centering

\caption{Observed molecular transitions and imaging parameters.}
\label{table:IRAS15398-3359_molecular_line}

\begin{tabular*}{\textwidth}{@{\extracolsep{\fill}}cccccccc@{}}
\hline\hline

Molecule
&
Transition
&
Frequency
&
$E_u^{a,b}$
&
$A^{b,c}$
&
$n_{\rm crit}^{b,d}$
&
Baseline length range
&
rms$^{e}$
\\

&
&
(GHz)
&
(K)
&
(s$^{-1}$)
&
(cm$^{-3}$)
&
(m)
&
(mJy beam$^{-1}$)
\\

\hline

$^{12}$CO
&
2--1
&
230.53800
&
16.60
&
$6.910 \times 10^{-7}$
&
$1.1 \times 10^{4}$
&
14.3--8202.7
&
10.5
\\

$^{13}$CO
&
2--1
&
220.39868
&
15.87
&
$6.038 \times 10^{-7}$
&
$9.4 \times 10^{3}$
&
14.3--8202.7
&
6.84
\\

H$_2$CO
&
$3_{0,3}$--$2_{0,2}$
&
218.22220
&
21.0
&
$2.818 \times 10^{-4}$
&
$2.6 \times 10^{6}$
&
7.4--1310.7
&
1.16
\\

CH$_3$OH
&
$4_{2,2}$--$3_{1,2}$
&
218.44005
&
45.5
&
$4.686 \times 10^{-5}$
&
$2.0 \times 10^{7}$
&
7.4--1310.7
&
1.14
\\

CS
&
5--4
&
244.93556
&
35.3
&
$2.981 \times 10^{-4}$
&
$3.8 \times 10^{6}$
&
8.4--2507.1
&
1.60
\\

\hline
\end{tabular*}

\vspace{1.5mm}

\begin{minipage}{\textwidth}
\footnotesize
\textit{Notes.}
$^a$ Upper-state energy of the transition.
$^b$ Values taken from the LAMDA database \citep{2005Schoier}.
$^c$ Einstein $A$ coefficient.
$^d$ Critical density calculated at a kinetic temperature of 20~K
using the Einstein $A$ coefficients and collisional rate coefficients
provided in the LAMDA database.
$^e$ rms noise adopted for the channel-map analysis
at a spatial resolution of $0.48'' \times 0.39''$ and velocity spacing of $0.4$ km s$^{-1}$.
\end{minipage}

\end{table*}

During early stages of star formation, protostars accrete material
through disks while driving jets and outflows from the vicinity of the
protostar--disk system.
These ejections remove angular momentum from the accretion flow and inject
mass, momentum, and energy into the surrounding environment, making their
formation and launching mechanisms central to understanding protostellar
mass growth and interactions with the surrounding material
\citep[e.g.,][]{2019Machida,2020Machida,2023Tsukamoto}.
Several scenarios have been proposed for the origin of molecular outflows, including
wind-driven shells, jet-driven bow shocks, and slow disk winds
\citep[e.g.,][]{2000Lee,1993Raga,2018Louvet},
while recent ALMA observations have revealed morphologies and velocity
structures that are difficult to explain with a single mechanism
\citep[e.g.,][]{2026FeeneyJohansson}.
Multiple shell-like structures can provide important clues to protostellar
ejection histories.
They have been associated with episodic ejection or episodic entrainment
\citep[e.g.,][]{2019Zhang,2021Vazzano,
2024LopezVazquez},
but can also arise in continuous flows
\citep[e.g.,][]{2025Liu,2025Hanawa}.
Thus, multiple-shell morphology alone cannot establish episodic ejection;
individual shells must be separated spatially and kinematically, and their
three-dimensional structures and kinematics must be constrained simultaneously.
A key observational question is therefore whether the apparent
multiple-shell morphology can be decomposed into spatially and
kinematically distinct three-dimensional shell components, and what
differences among these components imply for the outflow history.

We investigated this problem toward the nearby low-mass Class~0 protostar
IRAS 15398$-$3359 in the Lupus I molecular cloud \citep{1989HeyerGraham,2000Shirley,2007Chapman},
at a distance of 155.5~pc \citep{2023Thieme}.
The protostellar mass is only $\sim0.022\,M_\odot$
\citep{2023Thieme}, and the bolometric temperature and luminosity are $T_{\rm bol}=50$ K and $L_{\rm bol}=1.4\,L_\odot$, respectively \citep{2023Ohashi},
indicating that the source is at a very early stage
of protostellar evolution.
Multiple shell-like structures have been reported in its NE--SW primary outflow
from the $^{12}$CO and JWST/MIRI observations
\citep{2021Vazzano, 2022Yang, 2025Okoda}.
While \citet{2021Vazzano} interpreted them as signatures of episodic
ejection, \citet{2025Hanawa} showed that an overpressured jet with a continuous gas supply can
produce similar structures through repeated expansion and compression.
Testing these scenarios requires tracers that can isolate individual
shell-like structures.
Although $^{12}$CO is useful for tracing the overall outflow morphology and
kinematics, it can sample multiple physical components
\citep[e.g.,][]{2025Sharma}.
In contrast, H$_2$CO and CS clearly trace cavity-wall-like and shell-like
structures within the primary outflow
\citep{2014Oya,2020Okoda,2021Okoda,2025Sharma}.
H$^{13}$CO$^+$ and CH$_3$OH observations further suggest that IRAS~15398$-$3359 has experienced
an accretion burst within
the past $\sim100$--1000~yr \citep{2013Jorgensen}.
These properties make IRAS~15398$-$3359 well suited for testing the competing
formation scenarios and probing the three-dimensional ejection history
at a very early stage of protostellar evolution.

Previous multiple-shell studies have mainly constrained shell morphology and
kinematics using channel maps and position--velocity (PV) diagrams
\citep[e.g.,][]{2019Zhang,2020FernandezLopez,2021Vazzano,
2026FeeneyJohansson}.
However, fewer studies have tested three-dimensional
models of multiple shells directly against molecular-line emission including radiative-transfer
and interferometric effects.
Here, we constrain the three-dimensional morphology and kinematics of the
multiple-shell structure in IRAS~15398$-$3359 with our modeling incorporating the radiative-transfer calculations and the ALMA observing simulations.
We have identified four shell-like components from their continuity across
velocity channels and PV structures, and constructed a multiple-shell model
in which the geometry, spatial configuration, and kinematics of each shell
are specified independently.
Non-LTE radiative-transfer calculations of H$_2$CO and synthetic ALMA
observations are adopted to reproduce the observed channel maps and PV diagrams.
We then discuss the ejection history and origin of the
multiple-shell structure from the resulting differences among the shells.

This paper is organized as follows.
Sect.~\ref{sec:observations} describes the ALMA and JWST data.
Sect.~\ref{sec:shell_identification} presents the molecular outflow structure and identification of the
shell-like components.
Sect.~\ref{sec:model} describes the three-dimensional multiple-shell model and
radiative-transfer calculations, and Sect.~\ref{sec:results} presents the comparison
between the observations and the model.
Sect.~\ref{sec:discussion} discusses the origin of the multiple-shell structure, implications
for protostellar outflow models, and molecular tracers.
Sect.~\ref{sec:conclusions} summarizes our conclusions.

\begin{figure*}[!t]
\centering
\includegraphics[width=0.98\textwidth]{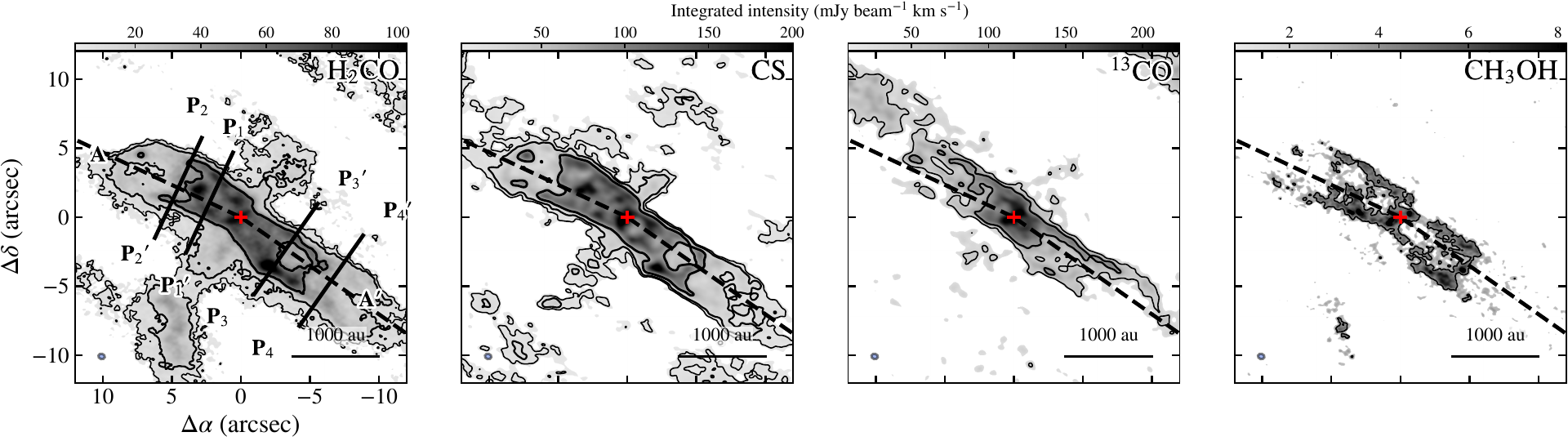}
\caption{
Integrated-intensity maps of H$_2$CO $3_{0,3}$--$2_{0,2}$, CS $J=5$--4, $^{13}$CO $J=2$--1, and CH$_3$OH $4_{2,2}$--$3_{1,2}$
toward IRAS~15398$-$3359.
A common synthesized beam of $0.48'' \times 0.39''$
(P.A. = $74.3^\circ$) is adopted for all the images,
which is shown with the blue ellipses to the bottom-left corners.
The integrated velocity ranges are
$V_{\rm LSR} = 2.4$--$7.2~{\rm km\,s^{-1}}$ for H$_2$CO and CS,
3.2--$7.2~{\rm km\,s^{-1}}$ for $^{13}$CO,
and 4.8--$5.6~{\rm km\,s^{-1}}$ for CH$_3$OH.
The grayscale starts from the $3\sigma$ level,
while the contours are at 5, 10, and $30\sigma$, where the rms noise levels of the
integrated-intensity maps are 1.64, 2.30, 8.99, and
0.77~mJy~beam$^{-1}$~km~s$^{-1}$ for H$_2$CO, CS,
$^{13}$CO, and CH$_3$OH, respectively.
The red cross marks the protostellar position
(R.A. = 15:43:02.232, Dec. = $-34$:09:06.96; \citealt{2023Thieme}).
Dashed lines show the outflow axis, and the other lines shown
in the H$_2$CO panel indicate the PV cut lines shown in Fig.~\ref{fig:h2co_cs_pv_diagrams}.
}
\label{fig:mom0}
\end{figure*}

\section{Observational Data}
\label{sec:observations}

The ALMA Band~6 archival data obtained as part of the FAUST project
\citep{2021Codella}
(Project ID: 2018.1.01205.L)
and the eDisk project
\citep{2023Ohashi}
(Project IDs: 2019.1.00261.L and 2019.A.00034.S)
are adopted for our study.
The data were reduced with the standard ALMA pipeline calibration with
CASA version 6.2.1.7 for the FAUST data, and versions 6.2.1 and
6.1.1-15 for the eDisk data.
The FAUST observations used the 12 m and ACA 7 m arrays, with baseline
lengths ranging from 7.4 to 2507.1 m,
while the eDisk observations used the 12 m array with baselines ranging
from 14.3 to 8202.7 m. The angular scales corresponding to the shortest
baselines, estimated as $\sim \lambda/B_{\min}$, are $\gtrsim 30''$
for the FAUST data and $\sim 20''$ for the eDisk data.
We analyzed H$_2$CO $3_{0,3}$--$2_{0,2}$,
CS $J=5$--4, and CH$_3$OH $4_{2,2}$--$3_{1,2}$ from the FAUST data,
and
$^{12}$CO $J=2$--1 and
$^{13}$CO $J=2$--1 from the eDisk data.
H$_2$CO is adopted as a primary tracer of the shell morphology and kinematics,
with CS for comparison,
while $^{13}$CO and CH$_3$OH are used to examine the shell-gas density
and chemical properties, respectively.
$^{12}$CO is adopted to compare the entire CO outflow components
to the shell features traced by the other molecular emission.

Imaging was performed with CASA \citep{2007McMullin} version 6.5.5.21.
After the continuum subtraction with \texttt{uvcontsub}, the molecular-line cubes
covering $V_{\rm LSR} = -4.9$--$14.9~{\rm km~s^{-1}}$ were produced with \texttt{tclean}
using natural weighting and multiscale deconvolution with scales of
[0, 10, 50] pixels.
The image and cell sizes were $1350\times1350$ pixels and
$0.03''$, respectively, and \texttt{auto-multithresh} masking and a common
restoring beam were adopted.
The cubes were produced at a velocity spacing of
0.2~km~s$^{-1}$.
The systemic velocity of
$V_{\rm sys}=5.2$~km~s$^{-1}$ is adopted \citep{2017Yen}.
These cubes were used for the PV analysis.
For the channel maps and comparisons among different molecular species,
two adjacent channels were averaged with \texttt{imrebin} to produce cubes
with a velocity spacing of 0.4~km~s$^{-1}$, improving the
signal-to-noise ratio for the shell-like emission.
For direct comparison, the CS, $^{12}$CO
and $^{13}$CO cubes were convolved with
the H$_2$CO synthesized beam of
$0.48'' \times 0.39''$ using \texttt{imsmooth}.
The molecular transitions and resulting image properties are summarized
in Table~\ref{table:IRAS15398-3359_molecular_line}.

To examine the infrared morphology of the outflow,
we also used the calibrated Stage~3 I2D F1000W image obtained with the
JWST Mid-Infrared Instrument
(MIRI; \citealt{2015Bouchet,2015Rieke,2023Wright})
as part of the CORINOS program
(PI: Y.-L. Yang)
\citep{2022Yang}.
The image was produced by the JWST calibration pipeline
(version~1.20.2; CRDS context \texttt{jwst\_1464.pmap}).

\begin{figure*}[p]
\includegraphics[width=1.00\textwidth]{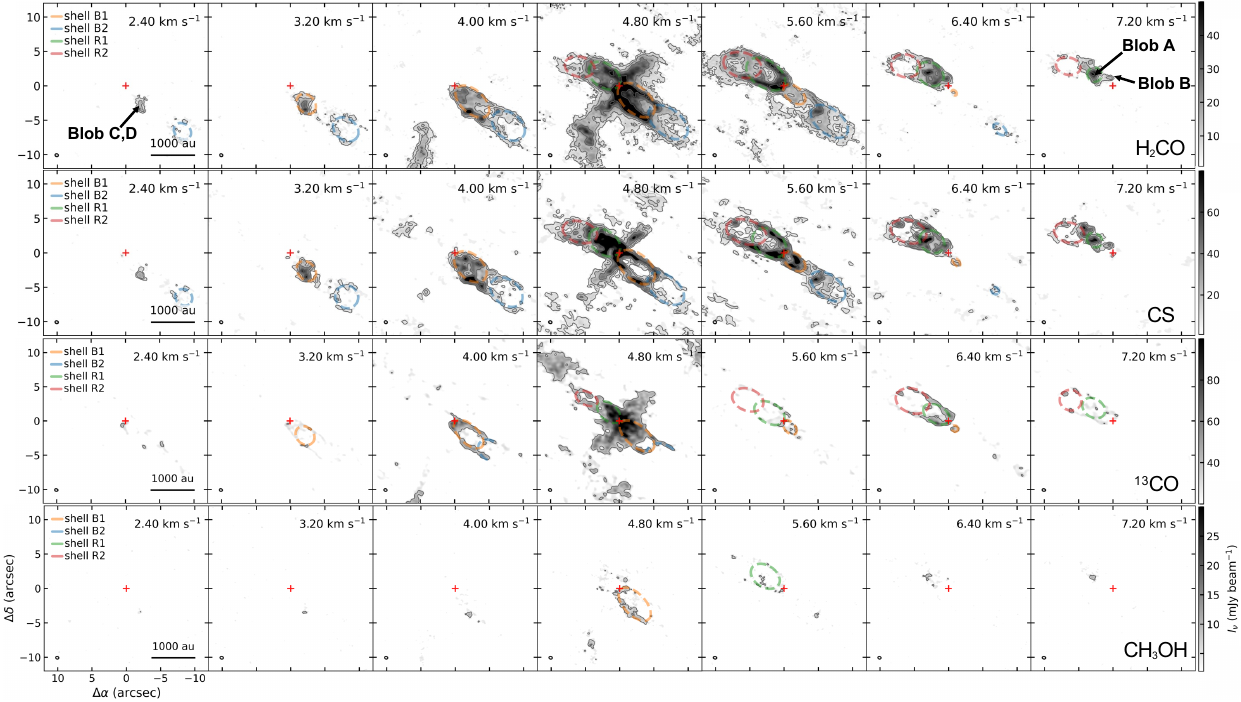}
\caption{
Velocity channel maps of H$_2$CO $3_{0,3}$--$2_{0,2}$, CS $J=5$--4, $^{13}$CO $J=2$--1, and CH$_3$OH $4_{2,2}$--$3_{1,2}$
toward IRAS~15398$-$3359.
The velocity spacing adopted is
$\Delta v = 0.4~{\rm km\,s^{-1}}$, and the spatial resolution is unified
with a common synthesized beam of $0.48'' \times 0.39''$.
The grayscale starts from the $3\sigma$ level,
while the contours are at 5, 10, 20, and $40\sigma$.
The dashed colored ellipses are visual guides to the four shell-like
structures traced by H$_2$CO and CS across successive velocity channels.
Blobs A--D are from \citet{2020Okoda}.
}
\label{fig:h2co_cs_channelmap}
\end{figure*}

\begin{figure*}[p]
\centering
\includegraphics[width=0.85\textwidth]{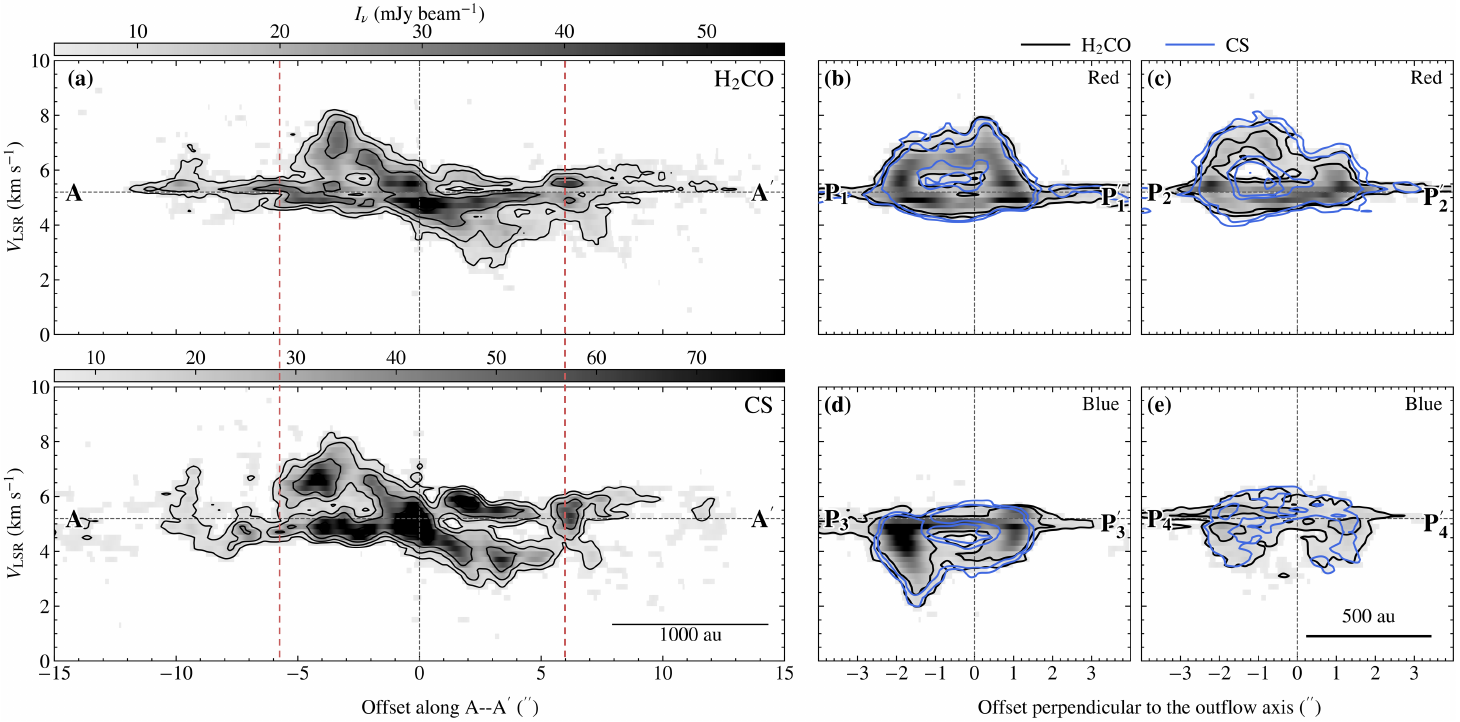}
\caption{
PV diagrams of H$_2$CO $3_{0,3}$--$2_{0,2}$ and CS $J=5$--4 toward IRAS~15398$-$3359.
The velocity spacing is
$\Delta v = 0.2~{\rm km\,s^{-1}}$,
and
the grayscale starts from the $3\sigma$ level.
The rms noise levels are
1.40 and
1.98~mJy\,beam$^{-1}$
for H$_2$CO and CS, respectively.
Panel~(a) shows H$_2$CO and CS PV diagrams along the
outflow axis A--A$'$.
Contours are drawn at
5, 10, 20, and $40\sigma$.
The horizontal dashed line indicates the systemic velocity
(\(V_{\rm sys}=5.2~{\rm km\,s^{-1}}\)),
and the red vertical dashed lines indicate the approximate spatial boundaries between the shell components as traced by dashed lines in Fig.~\ref{fig:h2co_cs_channelmap}.
Panels~(b)--(e) show PV diagrams along cuts perpendicular to the
outflow axis at P1--P4.
The grayscale shows the H$_2$CO emission, while black and blue contours show the
H$_2$CO and CS emission, respectively, at
5 and $10\sigma$.
The horizontal dashed line indicates the systemic velocity.
}
\label{fig:h2co_cs_pv_diagrams}
\end{figure*}

\section{Outflow Structure and Shells}
\label{sec:shell_identification}

Figure~\ref{fig:mom0} shows the integrated-intensity (moment~0) maps of the four molecular lines.
As a whole, all the molecular lines show elongated features along the northeast to southwest direction.
As shown by dashed lines in Figure~\ref{fig:mom0}, the axes of the elongated features appear to bend between the northeast and southwest direction, where the PAs are $64.9^\circ$ \citep{2021Vazzano} and $234.9^\circ$ \citep{2025Okoda}, respectively.
These elongated components correspond to the primary outflow driven from IRAS~15398$-$3359 \citep[e.g.,][]{2014Oya}.
The H$_2$CO and CS emission also exhibit another elongated component along the northwest to the southeast direction; the secondary outflow \citep{2021Okoda}.

Figure~\ref{fig:h2co_cs_channelmap} compares velocity channel maps of H$_2$CO, CS,
$^{13}$CO, and CH$_3$OH. Among them, H$_2$CO and CS most clearly trace
the velocity-dependent shell-like structures. Visually, the H$_2$CO
and CS emission suggests two principal elliptical structures in both the blueshifted (SW) and redshifted (NE) lobes, as indicated by the colored ellipses in Figure~\ref{fig:h2co_cs_channelmap}.
Emission components originating from the blueshifted and redshifted shells
are detected over $V_{\rm LSR}=2.4$--$6.4~{\rm km\,s^{-1}}$ and
$4.8$--$7.2~{\rm km\,s^{-1}}$, respectively.
Their detected velocity ranges overlap around the systemic velocity,
suggesting the three-dimensional configurations of those shells.
The elliptical emission can be traced across multiple velocity channels, which are more spatially
extended near the systemic velocity and progressively smaller with
increasing velocity offset.
At larger velocity offsets from the systemic velocity, localized
emission enhancements remain and coincide with Blobs A--D identified
by \citet{2020Okoda}, which are interpreted as locally shocked
regions produced by the interaction of the outflow with the ambient gas.
Because the present analysis focuses on the global shell-like structure
of the primary outflow, these localized shocked features and the
secondary-outflow component are not included in our modeling.

The individual shell-like structures in the primary outflow as seen in the $^{13}$CO emission are, however, not as clear as those seen in the H$_2$CO and CS emission. This is partly due to the higher noise level of the eDisk observations, as well as a possible effect of the spatial filtering.
The CH$_3$OH emission is weak and is seen only toward the inner shells.
We therefore use H$_2$CO and CS as the primary tracers for identifying
the individual shell components and characterizing their kinematics.
The $^{13}$CO emission is instead used as an additional constraint on
the shell-gas density in the modeling described in
Sect.~\ref{sec:model_description}.

Figure~\ref{fig:h2co_cs_pv_diagrams} shows the H$_2$CO and CS
position--velocity (PV) diagrams along and perpendicular to the outflow axis.
In the PV diagram along the outflow axis (Fig.~\ref{fig:h2co_cs_pv_diagrams}(a)), within the spatial offset of
$\sim\pm6''$ (dashed red lines), there appears a tilted, X-shaped morphology.
One of the strokes shows a velocity gradient from redshifted to blueshifted
from $-6''$ to $+6''$, while the other stroke shows a moderate velocity
gradient from blueshifted to redshifted.
At the $\pm6''$ edges, these two components appear to merge.
These structures comprise the inner shells as seen in Fig.~\ref{fig:h2co_cs_channelmap}.
Beyond $\pm6''$, distinct velocity components are seen in the PV diagram,
tracing the outer shells.

In the PV diagrams perpendicular to the outflow axis
(Figs.~\ref{fig:h2co_cs_pv_diagrams}(b)--(e)),
both the H$_2$CO and CS lines exhibit elliptical or partially elliptical structures,
consistent with the velocity structure expected from
expanding shells
\citep[e.g.,][]{2000Lee}.
Outflow rotation can produce a transverse velocity gradient
\citep[e.g.,][]{2017Hirota,2017Tabone},
but no clear systematic velocity gradient is detected perpendicular to the outflow axis.

Based on the channel-map and the PV structures, we define a shell-like component
using the following observational criteria:
(i) it can be traced across multiple velocity channels as a continuous
arc-like or elliptical structure, and (ii) it can be distinguished as a
separate structure in the PV diagrams.
Applying these criteria to structures detected in both the H$_2$CO and CS
emission, four shell-like components, two in each lobe, are identified.
The inner and outer blueshifted components are referred to as
B1 and B2, and their redshifted counterparts as R1 and R2, respectively.
B1 and R1 lie at projected distances of approximately $0''$--$6''$
from the protostar, whereas B2 and R2 lie at approximately
$6''$--$11''$.

\begin{figure}
\includegraphics[width=0.45\textwidth]{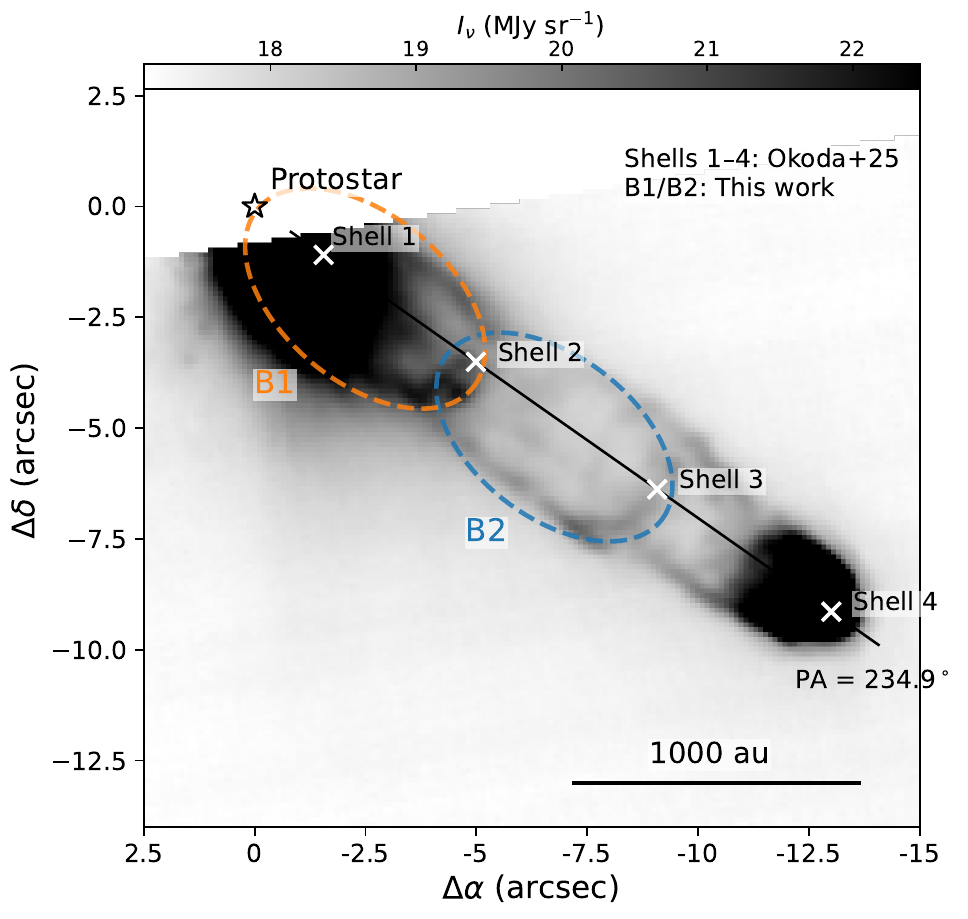}
\caption{
Comparison of the shell-like structures in the southwestern outflow
seen with JWST/MIRI and ALMA.
The grayscale shows the JWST/MIRI F1000W emission.
Shells 1--4 are from \citet{2025Okoda}, and the white crosses mark
the tips of the corresponding shells identified in their study.
The orange and blue ellipses indicate the largest fitted ellipses of B1 and B2, respectively.
}
\label{fig:jwst_shell}
\end{figure}

Figure~\ref{fig:jwst_shell} highlights the JWST/MIRI F1000W image in
comparison with B1 and B2 identified in the present H$_2$CO and CS study.
Dashed lines in Fig.~\ref{fig:jwst_shell} trace the largest ellipses of
B1 and B2 at the relevant velocities (see Fig.~\ref{fig:h2co_cs_channelmap}).
From the MIRI image, \citet{2025Okoda} identified four arc-like structures
at the tips of the shells (Shells~1, 2, 3, and 4).
The B1 shell as seen in the molecular lines appears to span from the
protostellar position to Shell~2, while B2 extends from around Shell~2
to Shell~3.
No clear molecular counterparts are seen toward Shells~1 and 4.

\begin{figure*}
\centering
\makebox[\textwidth][c]{%
  \hspace*{1.6em}%
  \includegraphics[width=0.95\textwidth]{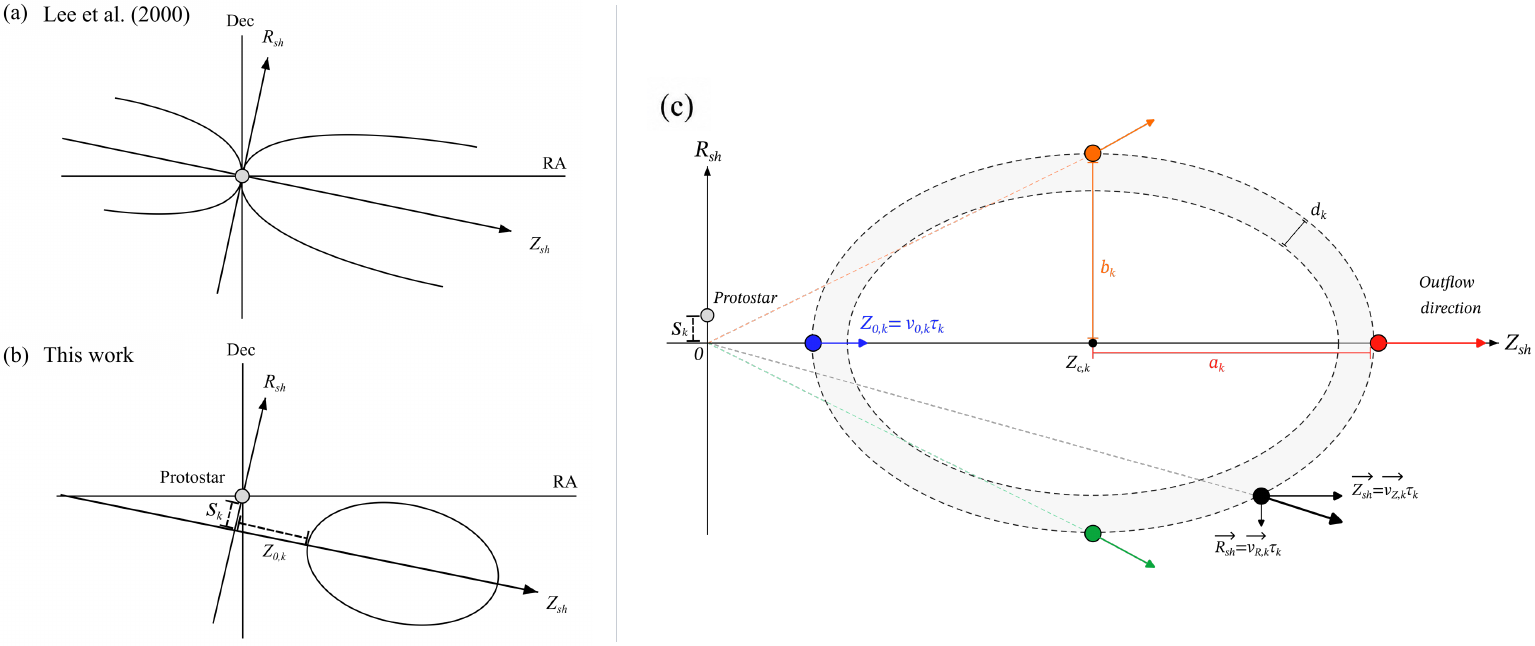}%
}
\caption{
Schematic illustration of the shell model.
Panel (a) shows the parabolic wind-driven shell model of
\citet{2000Lee}, and panel (b) shows the finite axisymmetric
ellipsoidal geometry adopted in this study while retaining the same
basic Hubble-law-like velocity prescription.
Panel (c) illustrates the geometrical and kinematic parameters of an
individual shell in the $Z_{\rm sh}$--$R_{\rm sh}$ cross section.
The velocity field is defined with respect to the origin of this
coordinate system and is given by
$v_{R,k}=R_{{\rm sh},k}/\tau_k$ and
$v_{Z,k}=Z_{{\rm sh},k}/\tau_k$.
}
\label{fig:shell_parameters}
\end{figure*}

\section{Model}
\label{sec:model}

\subsection{Model Description and Parameters}
\label{sec:model_description}

In this section, we describe the three-dimensional structure and kinematics
of the four shell-like components identified in
Sect.~\ref{sec:shell_identification} within a simple physical framework
consistent with existing wind-driven shell (WDS) models.
The surrounding envelope and material inside the cavities are not included
in the model.
The WDS model of \citet{2000Lee} describes the outflow morphology and
kinematics using a parabolic shell geometry,
\(Z=CR^2\), and a Hubble-law-like velocity field in which the velocity is
proportional to position,
\(V_R=R/\tau\) and \(V_Z=Z/\tau\).
In this study, we retain this basic velocity prescription but extend
the shell geometry to a finite axisymmetric ellipsoid to describe
the observed elliptical morphology
(Figs.~\ref{fig:shell_parameters}(a) and (b)).
A finite ellipsoidal geometry is also consistent with the physical framework
of WDSs demonstrated by \citet{2001Lee} and therefore does not depart from
the basic picture of conventional WDS models.

For each shell \(k\), we define a shell-specific Cartesian coordinate system
\((X_{{\rm sh},k},Y_{{\rm sh},k},Z_{{\rm sh},k})\),
where \(Z_{{\rm sh},k}\) is the symmetry axis of the shell and
\(R_{{\rm sh},k}=(X_{{\rm sh},k}^{2}+Y_{{\rm sh},k}^{2})^{1/2}\)
is the distance from this axis.
Following the Lee-type WDS model, we adopt the velocity field
\begin{equation}
v_{R,k}
=
\frac{R_{{\rm sh},k}}{\tau_k},
\qquad
v_{Z,k}
=
\frac{Z_{{\rm sh},k}}{\tau_k}.
\end{equation}
Here, \(\tau_k\) is the inverse velocity gradient and represents a
kinematic timescale defined by the present-day position--velocity relation
of each shell.
Because the velocity is proportional to position, this velocity field
corresponds to self-similar expansion where the shell preserves its
shape as it expands.
We describe the present-day shape of a shell with this velocity field
as an axisymmetric ellipsoid
(Fig.~\ref{fig:shell_parameters}(c)),
\begin{equation}
\frac{(Z_{{\rm sh},k}-Z_{c,k})^2}{a_k^2}
+
\frac{R_{{\rm sh},k}^2}{b_k^2}
=
1 ,
\label{eq:elliptical_shell_geometry}
\end{equation}
where \(a_k\) and \(b_k\) are the axial and transverse semiaxes,
respectively.
The inner boundary is defined by a concentric ellipsoid with semiaxes
\(a_k-d_k\) and \(b_k-d_k\), where \(d_k\) represents the shell-wall
thickness.
Defining \(Z_{0,k}\) as the axial position of the protostellar-side endpoint
of the shell gives
\begin{equation}
Z_{c,k}=Z_{0,k}+a_k .
\end{equation}
The parameter \(Z_{0,k}\) specifies the root of each shell,
allowing outer shells such as B2 and R2 to be described using the same geometrical
form as the inner shells.
Applying the Hubble-law-like velocity field defined above to this root gives
\begin{equation}
v_{Z,k}(Z_{0,k})
=
\frac{Z_{0,k}}{\tau_k}.
\end{equation}
Since the coordinate system is defined for each shell and the
\(Z_{{\rm sh},k}\) axis corresponds to that shell axis, the
\(Z_{{\rm sh},k}\) axis does not necessarily intercept the protostellar
position.
An additional parameter, \(s_k\), is introduced, which describes the
distance between the protostellar position and the shell axis.
The origin of the Hubble-law-like velocity field is taken to be at the
position on the \(Z_{{\rm sh},k}\) axis nearest to the protostar.
No additional bulk motion of the entire shell is incorporated.
The driving position of each shell must be the protostellar position,
and the displacement between the protostellar position and the shell axis
should reflect temporal variation of the shell morphology.
The three-dimensional orientation of each shell is specified by its
position angle \({\rm PA}_k\) and inclination \(i_k\) of the shell axis.
The position and velocity vectors in the shell-specific coordinate system
are then transformed into the observer's coordinate system.

To reflect the large-scale density gradient along the shell axis,
the H$_2$ number density within each shell wall, i.e., within the thickness \(d_k\), is assumed to follow
a power law that depends only on \(Z_{{\rm sh},k}\):
\begin{equation}
n_{{\rm H}_2,k}
=
n_{0,k}
\left(
\frac{|Z_{{\rm sh},k}|}{1~{\rm au}}
\right)^{-p},
\label{eq:shell_h2_density}
\end{equation}
where \(n_{0,k}\) is the density at
\(|Z_{{\rm sh},k}|=1~{\rm au}\), and \(p\) is the density power-law
index.
For simplicity, no additional density gradient is assumed along
\(R_{{\rm sh},k}\), so that the density is uniform at a given
\(Z_{{\rm sh},k}\) within the shell wall.
To avoid introducing unnecessary shell-to-shell degrees of freedom,
\(p\) is taken to be common to all shells.
The number density of molecular species \(m\) is given by
\(n_{m,k}=X_{m,k}n_{{\rm H}_2,k}\), with a constant molecular
abundance \(X_{m,k}\) within each shell.
The gas temperature in each shell is obtained as
$T_{\mathrm{gas}} = f_{T,k} \times T_{\mathrm{dust}}$,
where $T_{\mathrm{dust}}$ is the dust temperature calculated
from the thermal Monte Carlo simulation with the central
protostar as a heating source, and $f_{T,k}$ is the coefficient
to account for gas temperature enhancement with respect to
the dust temperature in thermal equilibrium.

\begin{table*}[t]
\centering
\caption{
Geometrical, kinematic, density, and H$_2$CO parameters
adopted in the Fiducial and Temperature Models.
}
\label{tab:model_parameters}

\small
\renewcommand{\arraystretch}{1.15}
\setlength{\tabcolsep}{4pt}

\begin{tabular*}{\textwidth}{@{\extracolsep{\fill}}cccccccccccc@{}}
\hline\hline

&
\multicolumn{9}{c}{Common Parameters}
&
\multicolumn{1}{c}{Fiducial Model}
&
\multicolumn{1}{c}{Temperature Model}
\\

\cline{2-12}

shell
& \(a_k\)
& \(b_k\)
& \(d_k\)
& \(s_k^{a}\)
& \({\rm PA}_k^{b}\)
& \(i_k^{c}\)
& \(\tau_k\)
& \(Z_{0,k}^{d}\)
& \(n_{0,k}\)
& \(X_k({\rm H_2CO})^{e}\)
& \(f_{T,k}^{f}\)
\\

&
(arcsec)
&
(arcsec)
&
(arcsec)
&
(arcsec)
&
(deg)
&
(deg)
&
(yr)
&
(arcsec)
&
(cm\(^{-3}\))
&
&
\\

\hline

B1
& \(3.51\)
& \(1.80\)
& \(0.80\)
& \(0.00\)
& \(231\)
& \(19\)
& \(1000\)
& \(0.10\)
& \(2.3\times10^{6}\)
& \(3.5\times10^{-8}\)
& \(2.0\)
\\

B2
& \(3.70\)
& \(2.00\)
& \(0.80\)
& \(\llap{-}0.53\)
& \(237\)
& \(5\)
& \(750\)
& \(5.20\)
& \(1.5\times10^{6}\)
& \(3.5\times10^{-8}\)
& \(1.4\)
\\

R1
& \(3.15\)
& \(1.80\)
& \(0.80\)
& \(0.00\)
& \(61\)
& \(15\)
& \(1000\)
& \(\llap{-}0.20\)
& \(2.3\times10^{6}\)
& \(2.5\times10^{-8}\)
& \(1.8\)
\\

R2
& \(3.00\)
& \(2.00\)
& \(0.80\)
& \(1.38\)
& \(77\)
& \(8\)
& \(750\)
& \(\llap{-}3.80\)
& \(1.5\times10^{6}\)
& \(1.5\times10^{-8}\)
& \(1.0\)
\\

\hline
\end{tabular*}

\vspace{1.5mm}

\noindent
\begin{minipage}{\textwidth}
  \footnotesize
  \textit{Notes.}
  \(^{a}\) Positive \(s_k\) is toward the northwest.
  \(^{b}\) The position angle \({\rm PA}_k\) is measured east of north.
  \(^{c}\) The inclination \(i_k\) is measured from the plane of the sky.
  \(^{d}\) Positive \(Z_{0,k}\) is toward the southwest.
  \(^{e}\) The listed H$_2$CO abundances refer to para-H$_2$CO relative to H$_2$
  and are adopted in the Fiducial Model.
  In the Temperature Model, a common para-H$_2$CO
  abundance of $1.5\times10^{-8}$ is adopted
  for all the shells.
  \(^{f}\) The listed temperature scaling factors
  $f_{T,k}$ are adopted in the Temperature Model,
  whereas $f_{T,k}=1.0$ is assumed for all the shells
  in the Fiducial Model.
  \end{minipage}%

\end{table*}

Thus, each shell is described by
eleven
parameters:
seven parameters for its geometry and spatial configuration,
\((a_k,b_k,d_k,Z_{0,k},s_k,{\rm PA}_k,i_k)\);
\(\tau_k\) for the velocity field; and
\(n_{0,k}\) ,
\(X_{m,k}\)
, and $f_{T,k}$
for the physical structure (see Table~\ref{tab:model_parameters}).
In addition, the density power-law index \(p\) is common to all shells.

\subsection{Radiative Transfer and Synthetic ALMA Observations}
\label{sec:radiative_transfer}

For each trial set of model parameters, we constructed the corresponding
three-dimensional density, velocity, and molecular
distributions and performed radiative-transfer calculations of the H$_2$CO and $^{13}$CO emission with
RADMC-3D
\citep{2012Dullemond}.
The basic procedures for the radiative-transfer calculations are the same as those in
\citet{2024Takakuwa}.
The fixed parameters adopted for the radiative-transfer
calculations are summarized in
Table~\ref{tab:radiative_transfer_parameters}.

We first calculated the dust-temperature distribution with
\texttt{mctherm}, assuming heating by the central protostar.
Molecular data for para-H$_2$CO and $^{13}$CO are taken from the
LAMDA database
\citep{2005Schoier}.
Non-LTE level populations were calculated using the Large Velocity
Gradient (LVG) approximation in RADMC-3D, with the H$_2$
ortho-to-para ratio fixed at 3.
Ray tracing was then performed to generate the molecular-line model cubes.
The systemic velocity,
\(V_{\rm sys}=5.2~{\rm km\,s^{-1}}\)
\citep{2017Yen},
was added to the model velocities to convert them to the observed velocity
frame.
For direct comparison with the observations, we generated synthetic ALMA
data from the RADMC-3D model cubes using CASA.
The model brightness distribution was converted into model visibilities
at the same \(uv\) sampling as the observations using \texttt{ft}.
Thermal noise was then added to match the observed rms, and the synthetic
visibilities were imaged with \texttt{tclean} using the same settings as
for the observational data.
The resulting synthetic ALMA cubes therefore include the effects of
\(uv\) sampling, thermal noise, the synthesized beam,
and imaging.
All subsequent model--observation comparisons were performed using these
synthetic cubes.
For each trial parameter set, the resulting synthetic cubes were compared
with the observations, and the model parameters were adjusted manually
and iteratively.

In principle, the H$_2$ gas density, gas temperature,
and molecular abundances cannot be constrained simultaneously
with single molecular-line transitions available in the present study.
To constrain the H$_2$ density distribution,
including the density power-law index $p$
and the density normalization $n_{0,k}$,
the $^{13}$CO data is adopted on the assumption of
$T_{\rm gas}=T_{\rm dust}$ and
a constant $^{13}$CO abundance
$X(^{13}{\rm CO})=1.4\times10^{-6}$
\citep{1982Frerking,2005Milam}.
We tested several values of \(p\) and adopted a common value of
\(p=0.35\), which broadly reproduced the major $^{13}$CO intensity
distributions across all shells.
A common density normalization is adopted for B1 and R1,
and likewise for B2 and R2, because the $^{13}$CO emission
of each pair is broadly reproduced by the same value.
The resulting synthetic $^{13}$CO channel maps are shown in
Appendix~\ref{app:additional_channel_maps}.
After constraining the H$_2$ density distribution, 
we tested two models with different H$_2$CO abundance
and gas-temperature assumptions, while keeping
the geometrical, kinematic, and density parameters fixed.
In the first model, referred to as the Fiducial Model,
the para-H$_2$CO abundance, $X_k({\rm H_2CO})$,
was adjusted independently for each shell to broadly
reproduce the observed H$_2$CO line intensities,
assuming $T_{\rm gas}=T_{\rm dust}$.
In the second model, referred to as the Temperature Model,
we adopted a common para-H$_2$CO abundance of
$1.5\times10^{-8}$ for all the four shells,
corresponding to the lowest abundance
in the Fiducial Model.
The gas temperature of each shell
was instead modified as
\begin{equation}
T_{{\rm gas},k}=f_{T,k}T_{\rm dust},
\label{eq:temperature_scaling}
\end{equation}
where $T_{\rm dust}$ is the dust temperature calculated
from the thermal Monte Carlo simulation by RADMC-3D, and $f_{T,k}$ is the temperature scaling
factor for each shell $k$.
The values of $f_{T,k}$ were adjusted independently
for each shell to broadly reproduce the observed
H$_2$CO line intensities.
We also checked that the raised temperatures do not affect the $^{13}$CO intensities due to its lower rotational energy level, and thus the adopted gas densities derived from the Fiducial Model are still applicable.
The adopted parameters for both models
are summarized in
Table~\ref{tab:model_parameters}.
Note that these two different intensity scaling techniques
do not affect our discussion on the structure and kinematics
of the observed shells.

\begin{figure*}
\centering
\includegraphics[width=0.95\textwidth]{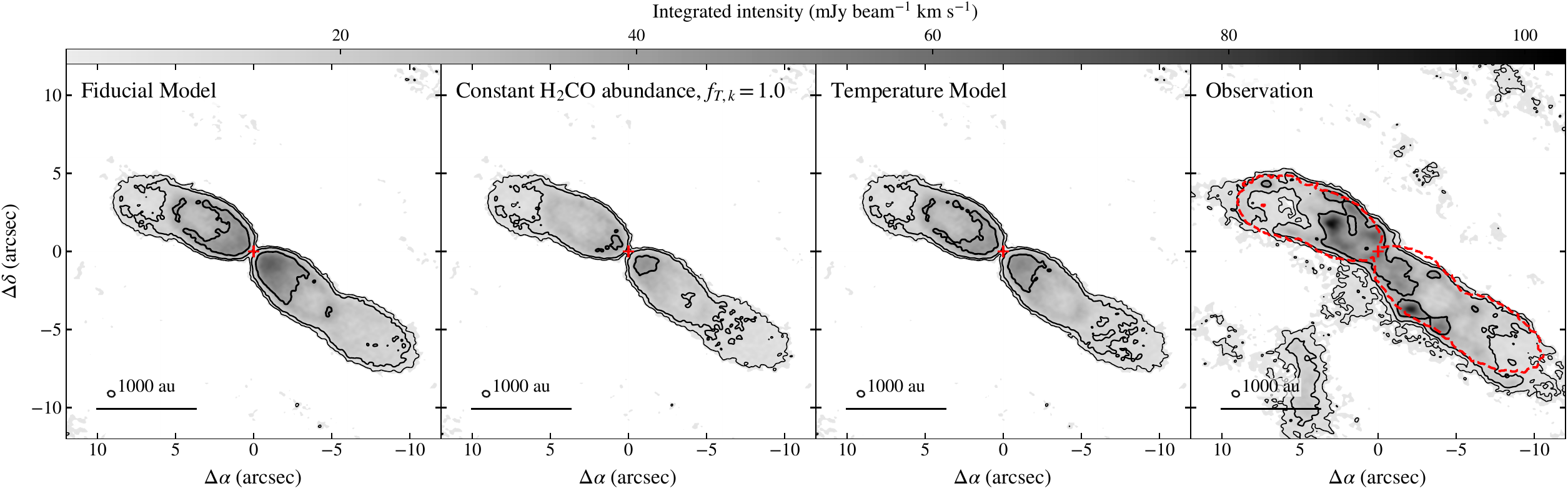}
\caption{
Integrated-intensity maps of H$_2$CO $3_{0,3}$--$2_{0,2}$
toward IRAS~15398$-$3359
for the Fiducial Model, the $f_{T,k}=1.0$ model,
the Temperature Model, and the observations.
The integrated velocity range is
$V_{\rm LSR} = 2.4$--$7.2~{\rm km\,s^{-1}}$,
excluding the channels at 4.8 and $5.2~{\rm km\,s^{-1}}$,
which are strongly contaminated by extended emission
components not included in our model.
The grayscale starts from the $3\sigma$ level,
while the contours are at 5, 10, and $25\sigma$,
where the observational rms noise level of the integrated-intensity maps
is $1.47~{\rm mJy\,beam^{-1}\,km\,s^{-1}}$.
The red dashed contours in the observed map
indicate the $5\sigma$ contour of the Fiducial Model.
}
\label{fig:mom0_temp}
\end{figure*}

\begin{figure*}[t]
\centering
\hspace*{-2.5em}
\includegraphics[
  height=0.99\textheight,
  keepaspectratio
]{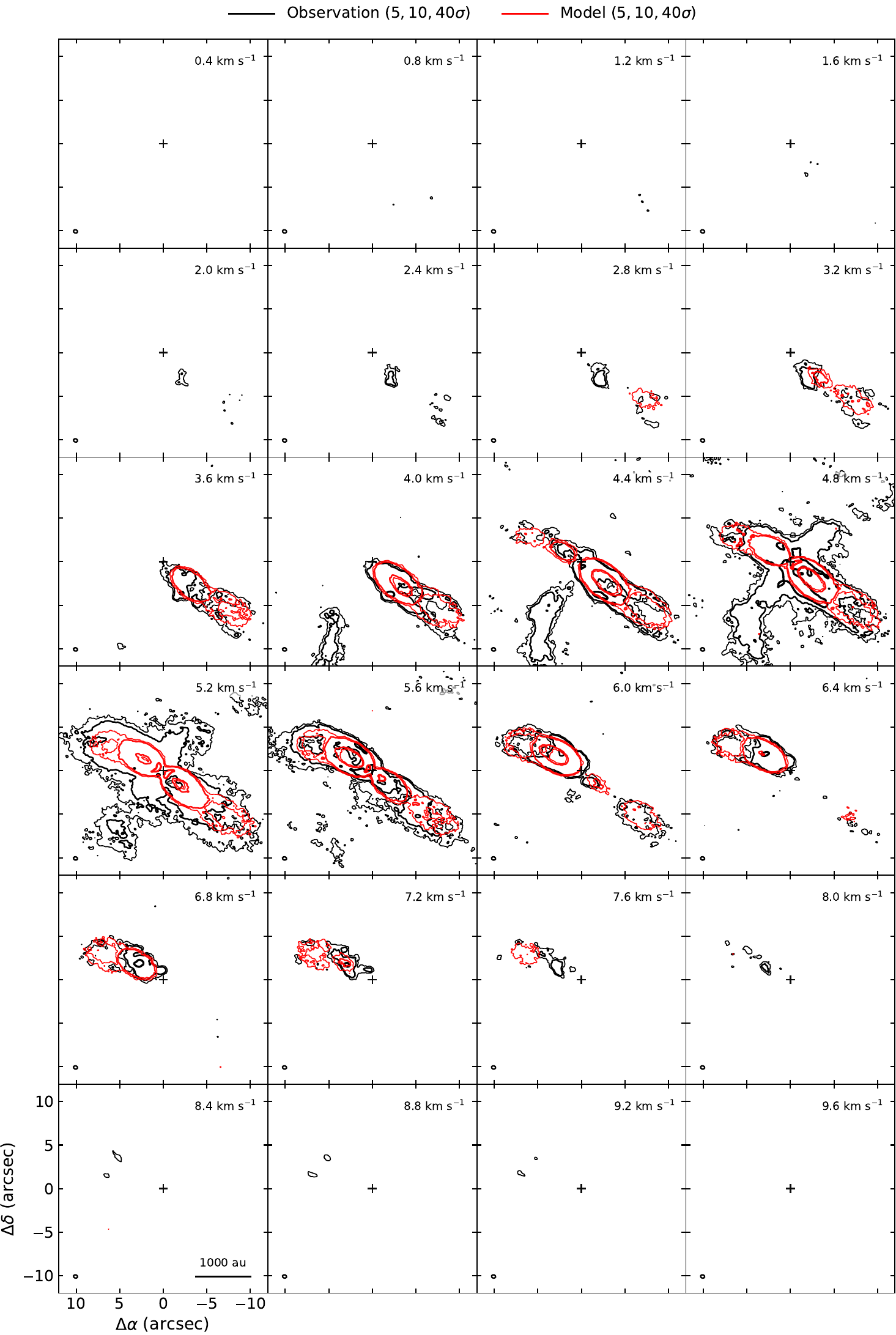}
\caption{
Comparison of the observed and modeled H$_2$CO $3_{0,3}$--$2_{0,2}$ channel maps.
Black and red contours show the ALMA observations and Temperature Model, respectively. Contours are drawn at 5, 10, and \(40\sigma\), using the observational rms.
}
\label{fig:obs_model_channelmap_temp}
\end{figure*}

\begin{figure*}[pt]
\centering
\hspace*{-0.6em}
\includegraphics[width=0.92\textwidth]{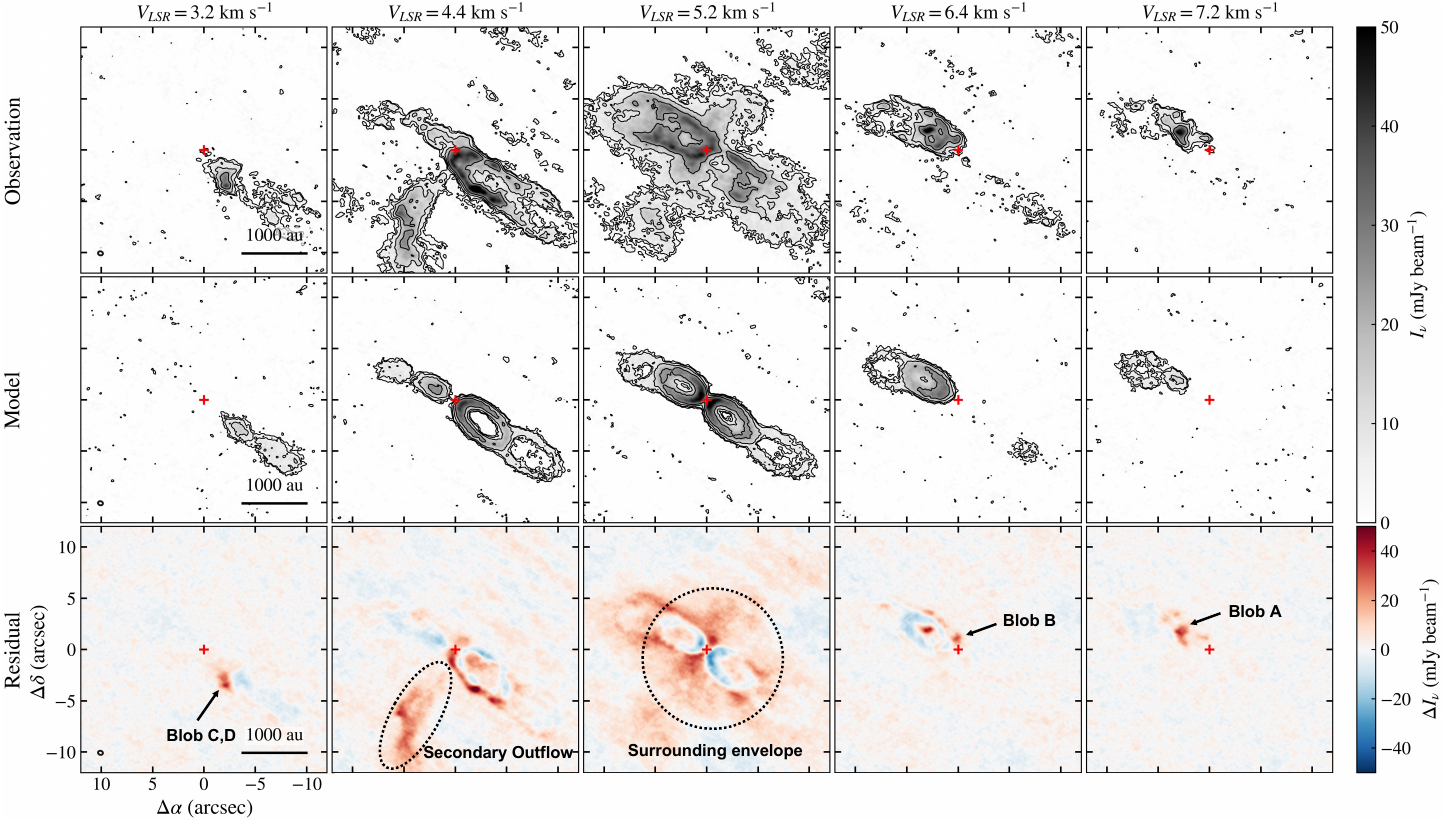}
\caption{
Velocity channel maps of H$_2$CO $3_{0,3}$--$2_{0,2}$
for the observations (top row), the Temperature Model (middle row),
and residuals (observations $-$ model; bottom row).
For the observations and model, grayscale images are shown with contours at 3, 5, 10, and $20\sigma$
($1\sigma = 1.16$ mJy\,beam$^{-1}$).
}
\label{fig:residual_temp}
\end{figure*}

\begin{figure*}
\centering
\hspace*{-1.1em}
\includegraphics[width=0.93\textwidth]{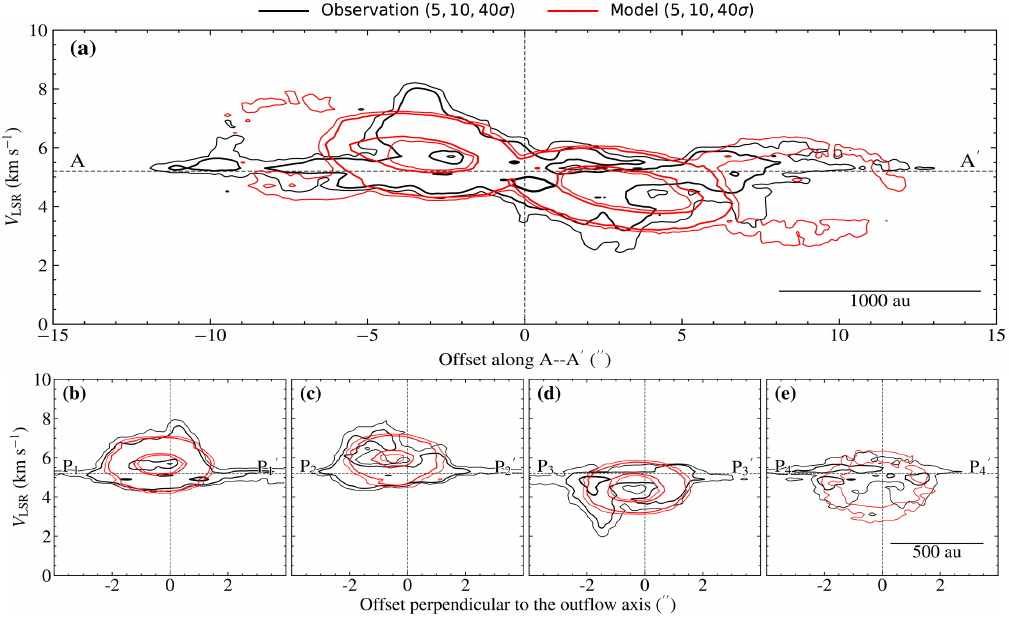}
\caption{
Comparison of the observed and modeled H$_2$CO $3_{0,3}$--$2_{0,2}$ PV diagrams.
Black and red contours show the ALMA observations and Temperature Model, respectively.
Panel (a) shows the cut along the outflow axis, and panels (b)--(e) show cuts perpendicular to it.
The cut locations are shown in Fig.~\ref{fig:mom0}.
Contours are drawn at 5, 10, and \(40\sigma\), using the observational rms.
}
\label{fig:obs_model_PV_temp}
\end{figure*}

\begin{figure*}
\centering
\includegraphics[width=0.95\textwidth]{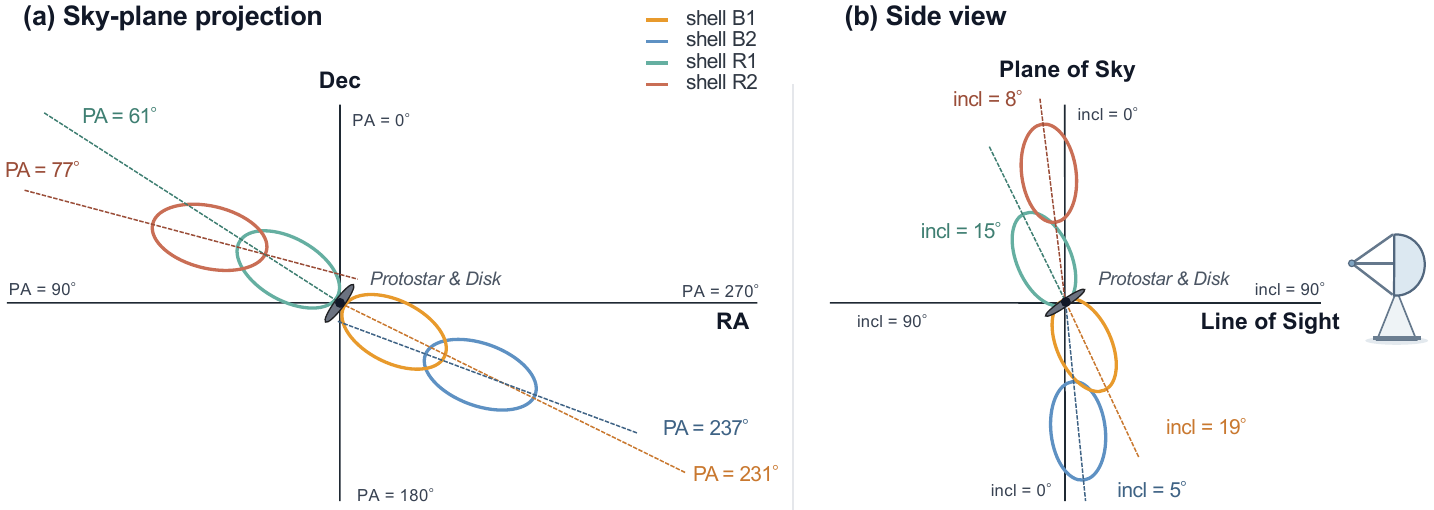}
\caption{
Sky-plane and side-view projections of the four modeled outflow shells.
}
\label{fig:shematic_overview}
\end{figure*}

The multiple-shell model presented here is a phenomenological model
designed to reproduce the observed morphology and kinematics in three
dimensions, rather than to calculate the formation and temporal evolution
of the shells hydrodynamically.
It assumes smooth, axisymmetric shells and does not account for
small-scale or non-axisymmetric structures such as clumps, knots, or
inhomogeneities in the surrounding medium.
Because each trial parameter set requires non-LTE radiative-transfer
calculations and synthetic ALMA observations, a formal statistical
optimization over the full parameter space is computationally
impractical.
The adopted parameter sets should therefore
not be regarded as unique best-fit solutions.
Dependence of the inclination angles on the radiative-transfer models is further discussed in Appendix~\ref{app:inclination_test}.

\begin{table}[htbp]
\centering
\caption{
Parameters adopted for the radiative-transfer calculations.
}
\label{tab:radiative_transfer_parameters}

\scriptsize
\setlength{\tabcolsep}{4pt}

\begin{tabular*}{\columnwidth}{@{\extracolsep{\fill}}lll}
\hline
Parameter & Adopted value & Reference \\
\hline

Distance
& \(155.5~{\rm pc}\)
& \citet{2023Thieme}
\\

Protostellar mass
& \(0.022~M_\odot\)
& \citet{2023Thieme}
\\

Protostellar luminosity
& \(1.4~L_\odot\)
& \citet{2023Ohashi}
\\

Protostellar radius
& \(1~R_\odot\)
& This work
\\

Gas-to-dust mass ratio
& \(100\)
& \citet{1990Beckwith}
\\

Mean molecular weight per H$_2$
& \(2.8\)
& \citet{2008Kauffmann}
\\

Dust opacity model
& DSHARP
& \citet{2018Birnstiel}
\\

Number of photon packets
& \(1.5\times10^{7}\)
& This work
\\

Number of grids ($r$, $\theta$, $\phi$)
& $(512, 512, 512)$
& This work
\\

Inner grid radius
& \(15~{\rm au}\)
& This work
\\

Outer grid radius
& \(5000~{\rm au}\)
& This work
\\

\hline
\end{tabular*}
\end{table}

\section{Results}
\label{sec:results}

\subsection{Reproduction of the Observed Shell Structures}
\label{sec:results_shell_reproduction}

Figure~\ref{fig:mom0_temp} compares
the H$_2$CO moment-0 maps of the Fiducial Model,
the model with a common H$_2$CO abundance of
$1.5\times10^{-8}$ and unchanged gas temperatures
($f_{T,k}=1.0$),
the model with the same abundance but modified
gas temperatures for each shell (the Temperature Model),
and the observations.
In the Temperature Model,
$f_{T,\mathrm{B1}}=2.0$,
$f_{T,\mathrm{B2}}=1.4$,
$f_{T,\mathrm{R1}}=1.8$, and
$f_{T,\mathrm{R2}}=1.0$ (Table~\ref{tab:model_parameters}) are adopted.
This model broadly reproduces the overall intensity
distribution of the observations, similarly to
the Fiducial Model.
This distribution could not be reproduced by adopting
a common H$_2$CO abundance alone ($f_{T,k}=1.0$).
These results suggest that the differences in the
H$_2$CO abundance adopted in the Fiducial Model
may also be explained by differences in gas temperature.
Both models adopt the same geometrical, kinematic,
and density parameters (Table~\ref{tab:model_parameters}).
Thus, in the following, we compare the spatial and velocity
structures of the observed shells with those of
the Temperature Model.
The corresponding comparisons for the Fiducial Model
are presented in Appendix~\ref{app:fiducial_model}.

Figure~\ref{fig:obs_model_channelmap_temp} compares the observed H$_2$CO
channel maps with the synthetic ALMA maps from the
Temperature Model.
The model broadly reproduces the projected locations and major elliptical
or partially elliptical morphologies of B1, B2, R1, and R2.
It also captures the systematic changes in the apparent positions and
morphologies of the shells across the velocity channels.

Figure~\ref{fig:residual_temp} shows the residuals between
the observed H$_2$CO channel maps and
those of the Temperature Model.
As already shown in Fig.~\ref{fig:obs_model_channelmap_temp},
the
Temperature Model
broadly reproduces
the spatial distribution of the shell-like emission
and the overall intensity trends across velocity channels.
However, there are significant residuals near the systemic
velocities, which arise from the secondary outflow and the
envelope component surrounding the modeled cavity walls.
There are other residuals at higher velocities, which correspond
to Blobs A--D identified by \citet{2020Okoda}.
Our model is not intended to reproduce all of these observed
components, but rather to reproduce the spatial and velocity
structures of the characteristic shell walls associated with
the outflow as discussed in Sect.~\ref{sec:shell_identification}.

Figure~\ref{fig:obs_model_PV_temp} compares the observed and modeled PV
diagrams along and perpendicular to the outflow axis.
In the PV diagram along the outflow axis
(Fig.~\ref{fig:obs_model_PV_temp}(a)), the observed, tilted X-shaped morphology is reproduced with two elliptical features of the model. The observed outer components beyond approximately $\pm6''$
are shown as additional elliptical features in the model.
In the PV diagrams perpendicular to the shell axes
(Figs.~\ref{fig:obs_model_PV_temp}(b)--(e)),
the elliptical emission distributions and velocity ranges associated with
the individual shells are also broadly reproduced.

Figure~\ref{fig:shematic_overview} shows the three-dimensional
orientations of the four shells
inferred from our modeling.
The inner shells B1 and R1 have inclinations of $19^\circ$ and
$15^\circ$, respectively, comparable to the $\sim20^\circ$
inclination previously inferred for the primary outflow as a whole
\citep{2014Oya, 2017Yen}.
In contrast, the outer shells B2 and R2 have smaller inclinations of
$5^\circ$ and $8^\circ$, respectively, and thus lie closer to the
plane of the sky.
Their position angles are also different,
${\rm PA}=231^\circ$, $237^\circ$, $61^\circ$, and $77^\circ$
for B1, B2, R1, and R2, respectively.
These results imply that the individual shells do not align along a
unified axis in the three-dimensional space, which also reflects the
bending between the blueshifted and redshifted outflow axes as seen
in Fig.~\ref{fig:mom0}.
The kinematic timescales are $\tau_k=1000$~yr for the inner shells
and $\tau_k=750$~yr for the outer shells.
Note that these $\tau_k$ values do not necessarily correspond directly
to the true elapsed times since the shells formed
(see Sect.~\ref{sec:discussion_multiple_shells}).
The outer shells also require larger absolute values of the axial configuration parameters
$Z_{0,k}$ and non-zero transverse offsets $s_k$ than the inner shells.
Thus, although all the shells are described with the same ellipsoidal
geometry and Hubble-law-like velocity prescription, their
three-dimensional orientations and
kinematics differ from shell to shell.

\begin{table*}[t]
\centering
\caption{
Derived physical properties of the individual shells.
}
\label{tab:shell_physical_properties}

\renewcommand{\arraystretch}{1.18}

\begin{tabular*}{\textwidth}{@{\extracolsep{\fill}}ccccccc@{}}
\hline\hline

shell
& \(M_k\)
& \(T_{{\rm dust},k}\)
& \(P_k\)
& \(v_{{\rm char},k}\)
& \(\dot{P}_k\)
& \(\dot{M}_{{\rm wind},k}\) \\

&
\((M_\odot)\)
&
\(({\rm K})\)
&
\((M_\odot\,{\rm km\,s^{-1}})\)
&
\(({\rm km\,s^{-1}})\)
&
\((M_\odot\,{\rm km\,s^{-1}\,yr^{-1}})\)
&
\((M_\odot\,{\rm yr^{-1}})\) \\

\hline

B1
& $(2.88 \pm 0.29)$$\times10^{-4}$
& \(24\)
& $(7.47 \pm 1.37)$$\times10^{-4}$
& \(2.59 \pm0.40\)
& $(7.47 \pm 1.37)$$\times10^{-7}$
& $(2.24 \pm 0.41)$$\times10^{-8}$ \\

B2
& $(1.57 \pm 0.16)$$\times10^{-4}$
& \(15\)
& $(1.37 \pm 0.15)$$\times10^{-3}$
& \(8.74 \pm 0.40\)
& $(1.83 \pm 0.20)$$\times10^{-6}$
& $(5.50 \pm 0.60)$$\times10^{-8}$ \\

R1
& $(2.66 \pm 0.27)$$\times10^{-4}$
& \(24\)
& $(6.52 \pm 1.25)$$\times10^{-4}$
& \(2.45 \pm 0.40\)
& $(6.52 \pm 1.25)$$\times10^{-7}$
& $(1.96 \pm 0.38)$$\times10^{-8}$ \\

R2
& $(1.44 \pm 0.14)$$\times10^{-4}$
& \(16\)
& $(9.68 \pm 1.13)$$\times10^{-4}$
& \(6.72 \pm 0.40\)
& $(1.29 \pm 0.15)$$\times10^{-6}$
& $(3.88 \pm 0.45)$$\times10^{-8}$ \\

\hline
\end{tabular*}

\vspace{1.5mm}

\begin{minipage}{\textwidth}
\footnotesize
\textit{Notes.}
\(T_{{\rm dust},k}\) is the dust-mass-weighted mean temperature
of each shell.
The characteristic velocity is
\(v_{{\rm char},k}=P_k/M_k\).
The momentum injection and wind mass-loss rates are
\(\dot{P}_k=P_k/\tau_k\)
and
\(\dot{M}_{{\rm wind},k}=\dot{P}_k/v_{\rm wind}\),
respectively.
The quoted uncertainties are propagated from the adopted 10\% absolute
flux calibration uncertainty and the velocity uncertainty of
0.4~km~s$^{-1}$.
\end{minipage}
\end{table*}

\subsection{Shell Masses and Wind Mass-loss Rates}
\label{sec:shell_physical_properties}

From the three-dimensional density and velocity distributions in the model, the mass and momentum of each shell can be estimated. The shell mass is expressed as;
\begin{equation}
M_k
=
\sum_{j\in k}
\rho_{{\rm gas},j}\Delta V_j ,
\label{eq:shell_mass}
\end{equation}
where \(j\) denotes the three-dimensional model grid cell belonging to
shell \(k\), and $\rho_{{\rm gas},j}$ and $\Delta V_j$ are the density and volume of cell $j$, respectively.
The shell momentum and momentum injection rate are
\begin{equation}
P_k
=
\sum_{j\in k}
m_j |\boldsymbol{v}_j|,
\qquad
\dot{P}_k
=
\frac{P_k}{\tau_k},
\label{eq:shell_momentum}
\end{equation}
respectively, where \(m_j=\rho_{{\rm gas},j}\Delta V_j\) and
\(\boldsymbol{v}_j\) is the model velocity of the shell gas in cell \(j\).
Because the shell mass is not the mass of the driving wind itself but comprises swept-up ambient gas,
\(M_k/\tau_k\) should not be regarded as the wind mass-loss rate. On the assumption of the momentum conservation, the wind mass-loss rate can be approximated as;

\begin{equation}
\dot{M}_{{\rm wind},k}
=
\frac{\dot{P}_k}{v_{\rm wind}}
=
\frac{P_k}{\tau_k v_{\rm wind}},
\label{eq:wind_mass_loss_rate}
\end{equation}
\citep[e.g.,][]{1984Goldsmith,2007DownesCabrit,2016Bjerkeli_a},
where \(v_{\rm wind}\) denotes the characteristic velocity of the wind
driving the molecular shell, distinct from the modeled shell velocity
\(\boldsymbol{v}_j\).
As the lower limit of the wind velocity, the mean of the maximum
line-of-sight velocity offsets of 15.3 and 11.6~km~s$^{-1}$ for the
blueshifted and redshifted $^{12}$CO outflows, corrected for the
outflow inclination of $23.9^\circ$
\citep{2021Vazzano}, \(v_{\rm wind}=33.3~{\rm km\,s^{-1}}\), is adopted.
The lower limit of the wind velocity provides an upper limit to the
wind mass-loss rate.
If the true wind velocity is \(\sim100~{\rm km\,s^{-1}}\)
\citep[e.g.,][]{2012Panoglou}, the wind mass-loss rate is \(\sim3\) times lower.

The derived properties are summarized in Table~\ref{tab:shell_physical_properties}.
Individual shell masses range from $(1.44 \pm 0.14)$$\times10^{-4}$ to
$(2.88 \pm 0.29)$$\times10^{-4}$~$M_\odot$.
The total masses of the inner pair (B1+R1) and outer pair (B2+R2)
are $(5.55 \pm 0.39)$$\times10^{-4}$ and
$(3.01 \pm 0.21)$$\times10^{-4}$~$M_\odot$,
respectively, making the inner pair approximately 1.8 times more
massive. In contrast, their estimated wind mass-loss rates are
$(4.20 \pm 0.56)$$\times10^{-8}$ and
$(9.38 \pm 0.75)$$\times10^{-8}$~$M_\odot~{\rm yr}^{-1}$,
respectively, with the outer-pair rate approximately 2.2 times higher.
This reflects the higher characteristic velocities and shorter
kinematic timescales of the outer pair despite its lower mass. 
The total masses of the blueshifted (B1+B2) and redshifted
(R1+R2) lobes are $(4.46 \pm 0.33)$$\times10^{-4}$ and
$(4.10 \pm 0.30)$$\times10^{-4}$~$M_\odot$, and
their estimated wind mass-loss rates are
$(7.74 \pm 0.73)$$\times10^{-8}$ and
$(5.83 \pm 0.59)$$\times10^{-8}$~$M_\odot\,{\rm yr^{-1}}$, respectively.
These estimates are approximately consistent with those derived from the SMA $^{12}$CO $J=2$--1 observations \citep{2016Bjerkeli_a}.

The dust-mass-weighted mean temperature of each shell is calculated as
\begin{equation}
T_{{\rm dust},k}
=
\frac{
\sum_{j\in k}
\rho_{{\rm dust},j}\Delta V_j T_{{\rm dust},j}
}{
\sum_{j\in k}
\rho_{{\rm dust},j}\Delta V_j
}.
\end{equation}
Here, $\rho_{{\rm dust},j}$ and $T_{{\rm dust},j}$ are the dust
density and dust temperature in cell $j$, respectively.
The resulting dust-mass-weighted temperatures, listed in
Table~\ref{tab:shell_physical_properties}, are 24, 15, 24,
and 16 K for B1, B2, R1, and R2, respectively.
In the Fiducial Model assuming $T_{\mathrm{gas}} = T_{\mathrm{dust}}$,
the observed H$_2$CO emission is reproduced with these relatively
low temperatures.
In the Temperature Model, the corresponding
dust-mass-weighted gas temperatures are approximately
47, 21, 43, and 16 K for B1, B2, R1, and R2, respectively.
The temperatures in both models are within the range 
inferred for molecular outflow cavities in L1157 and VLA1623--2417
\citep{2012Lefloch,2025Mercimek}.
Higher temperatures have also been reported toward the localized
H$_2$CO-enhanced regions in IRAS~15398$-$3359
\citep{2020Okoda,2025Okoda}.

\section{Discussion}
\label{sec:discussion}

\subsection{Multiple-shell Outflows and Their Physical Origin}
\label{sec:discussion_multiple_shells}

In this study, we have succeeded in reproducing the overall spatial and velocity
structures of the 
multiple
H$_2$CO
shells
in IRAS~15398$-$3359 with a series
of ellipsoidal shell models associated with a Hubble-law-like velocity
prescription. 
As shown in Sect.~\ref{sec:results}, the shells differ in orientation, spatial
configuration, and kinematics.
Furthermore, the momentum injection rate of the
outer bipolar pair is approximately 2.2 times that of the inner pair.
These shell-to-shell differences imply temporal variations
in the strength and direction of the outflow activity.

One natural interpretation for these results is time-variable or episodic ejection.
Protostellar mass accretion is expected to vary episodically
\citep[e.g.,][]{2012Dunham}, and multiple outflow shells could also be associated with episodic ejection or entrainment
\citep[e.g.,][]{2019Zhang,2024LopezVazquez}.
In IRAS~15398$-$3359, the ringlike H$^{13}$CO$^+$ emission distribution along with the centrally peaked CH$_3$OH emission suggests that an accretion burst raising the luminosity to about 100 times the present value occurred within the past $\sim100$--$1000$~yr \citep{2013Jorgensen}.
The kinematic timescales derived here for all the four shells,
$\tau_k=750$--$1000$~yr, lie in a comparable range.
Furthermore, a comparison with the current mass accretion rate
estimated by \citet{2023Thieme},
$\dot{M}_{\rm acc}=(1.3\text{--}6.1)\times10^{-6}
~M_\odot\,{\rm yr^{-1}}$,
gives $\dot{M}_{\rm wind}/\dot{M}_{\rm acc}\simeq$ $0.007$--0.03
for the inner pair and $\simeq$ $0.02$-- $0.07$ for the outer pair.
These values are consistent with the characteristic scale of
$\dot{M}_{\rm wind}/\dot{M}_{\rm acc} \sim 0.01$--$0.1$
discussed for accretion--ejection systems in young stars
\citep[e.g.,][]{2000Casse}.

The shell-to-shell differences in orientation provide an additional
indication of time variability.
Previous wind-driven shell modeling of the primary outflow in IRAS~15398$-$3359 suggests a global outflow inclination
of approximately $20^\circ$ from the plane of the sky
\citep{2014Oya, 2017Yen}.
The inclinations of the inner shells, $19^\circ$ and $15^\circ$,
are comparable to this value, whereas those of the outer shells,
$5^\circ$ and $8^\circ$, are substantially smaller.
In addition, secondary and tertiary outflow candidates with orientations
different from that of the primary outflow have been reported toward
this source
\citep{2021Okoda,2024GuzmanCcolque,2024Sai}.
Although these larger-scale components cannot be directly associated
with the individual shells modeled here, they are qualitatively
consistent with the shell-to-shell differences in the outflow orientation that has varied with time.
In particular, \citet{2024GuzmanCcolque} suggested that
the PA variations ($\sim 20^\circ$) among the secondary
and tertiary outflow structures result from precession
of the outflow axis.
The differences in PA and inclination among the shells
identified in this study are also consistent with
this scenario, suggesting that precession could contribute
to the overall outflow structure of IRAS 15398$-$3359.

\begin{figure*}[t]
\centering
\includegraphics[width=0.99\textwidth]{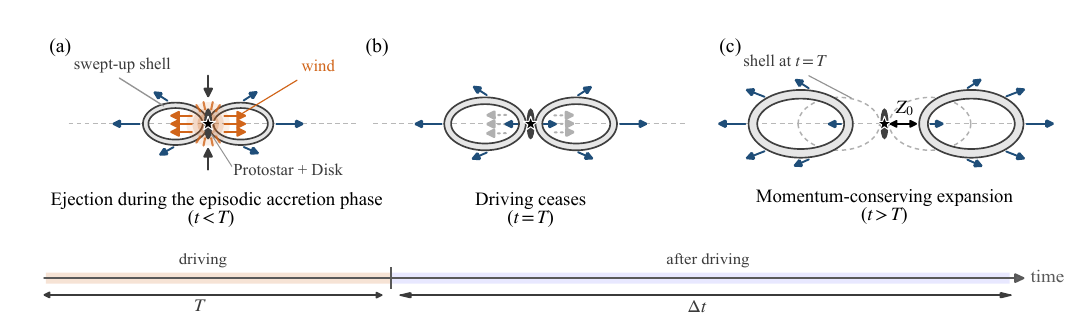}
\caption{
Schematic illustration of the finite-duration driving scenario.
(a) During the driving phase ($t<T$), the wind continuously supplies
momentum to the swept-up shell.
(b) The driving ceases at $t=T$.
(c) The shell may subsequently continue to propagate and expand with
the momentum acquired during the driving phase, producing a detached
configuration described by a non-zero $Z_0$ value.
Here, $T$ is the duration of the driving phase and
$\Delta t=t-T$ is the elapsed time after the driving ceases.
}
\label{fig:finite_duration_schematic}
\end{figure*}

The non-zero $Z_{0,k}$ values required for the outer shells may provide
a further link to time-variable or episodic ejection.
In a steady outflow continuously driven from the protostar,
a swept-up shell is expected to remain connected to the
vicinity of the driving source
\citep[e.g.,][]{1991Shu,2000Lee}.
In contrast, the outer shells in our model have detached configurations
from the protostellar position.
Such a configuration may arise if a shell formed during a
finite-duration ejection continues to propagate and expand with its
acquired momentum after the driving ceases, as schematically
illustrated in Fig.~\ref{fig:finite_duration_schematic}.
In this picture, $Z_{0,k}$ describes the
present location of a shell that has evolved after the initial ejection
episode.
Toward another Class 0 source, IRAS~04166+2706, \citet{2026Tafalla}
have proposed a similar picture to interpret the observed shell geometry.
With their ballistic modeling, they suggest that almost instantaneous,
internal-shock-driven lateral ejection near the protostar produces
shell-like structures that subsequently expand both along and across
the jet axis while sweeping up the surrounding material.

This finite-duration picture also provides a more specific
interpretation of the kinematic timescale $\tau_k$.
We consider an ambient density distribution
\begin{equation}
\rho_{\rm a}\propto r^{-k_\rho}.
\end{equation}
If the driving continues until time $T$ and the shell subsequently
expands while conserving the acquired momentum, the finite-duration solution derived from the
momentum-conserving shell framework of \citet{1999Matzner} gives
\begin{equation}
\tau_k
\equiv
\frac{r}{v}
=
(4-k_\rho)
\left(
t-\frac{T}{2}
\right).
\end{equation}
The elapsed time after the driving ceases,
$\Delta t=t-T$, is therefore
\begin{equation}
\Delta t
=
\frac{\tau_k}{4-k_\rho}
-\frac{T}{2}
<
\frac{\tau_k}{4-k_\rho}.
\end{equation}
For representative protostellar-envelope density profiles with
$k_\rho\simeq1.5$--2.0
\citep[e.g.,][]{1977Shu,2002Jorgensen},
this gives
$\Delta t\lesssim(0.4$--$0.5)\tau_k$.
For the outer shells, $\tau_k=750$~yr, and
this interpretation therefore gives an upper limit on the elapsed
time since the termination of the driving of
$\Delta t\lesssim300$--375 yr.
This upper limit corresponds to the impulsive limit
($T\rightarrow0$), in which the driving duration is negligible.
Thus, $\tau_k$ should be regarded as a kinematic timescale
characterizing the present-day position--velocity relation rather than
as the true elapsed time itself; its relation to the elapsed time depends
on both the ambient density profile and the driving history.

The transverse offset $s_k$ also shows a difference between the inner
and outer shells: whereas $s_k=0$ for the inner shells, non-zero $s_k$
is required for the outer shells, together with non-zero $Z_{0,k}$ values.
This discrepancy of the shell axes from the driving source may imply that those shells propagate through inhomogeneous ambient media.
Another possible interpretation is off-axis outflow driving.
For the Class~I source BHB07-11, \citet{2017Alves} found a bipolar
molecular outflow launched at $\sim$90--130 au from the rotation axis,
outside the $\sim$80 au radius of the disk, and interpreted this
structure as an outflow launched from a narrow region beyond the disk
edge.
MHD simulations have also produced outflows launched from regions
outside the disk edge \citep{2024Basu}.

\citet{2025Hanawa} show that an overpressured jet with a continuous
gas supply can also generate multiple shell-like structures through
repeated lateral expansion and contraction.
The formation timescale of such a shell is, however, on the order of
\(10^4\) yr, much longer than the kinematic timescale of
\(750\)--\(1000\) yr in our shell model.
Considering the shell-to-shell differences in spatial configuration,
the observed shells in IRAS~15398$-$3359 are better explained by
episodic, time-variable outflow activities.

On the other hand, \citet{2023Shang} showed that a steady wind can
produce multiple shell-like structures through interactions
with the ambient gas and magnetic fields in their MHD simulations.
In their model, multiple nested cavities with complex
feather-like structures form within a single large-scale
outflow bubble.
In contrast, the multiple shells identified in IRAS 15398$-$3359
are modeled as distinct, closed shells located along the outflow axis,
each exhibiting a smooth and systematic expansion velocity structure.
In particular, the axial PV diagrams show distinct transitions
in velocity structure between the inner and outer shells
at approximately $\pm 6''$, unlike the complex internal
velocity structures presented by \citet{2023Shang}
(see Sects.~\ref{sec:shell_identification} and
\ref{sec:results_shell_reproduction}).
These results again support the episodic, time-variable outflow
scenario in IRAS 15398$-$3359. Three-dimensional modeling of spatial and velocity structures
of outflow shells is essential to distinguish these different
formation mechanisms.

\begin{figure*}
\centering
\includegraphics[width=0.99\textwidth]{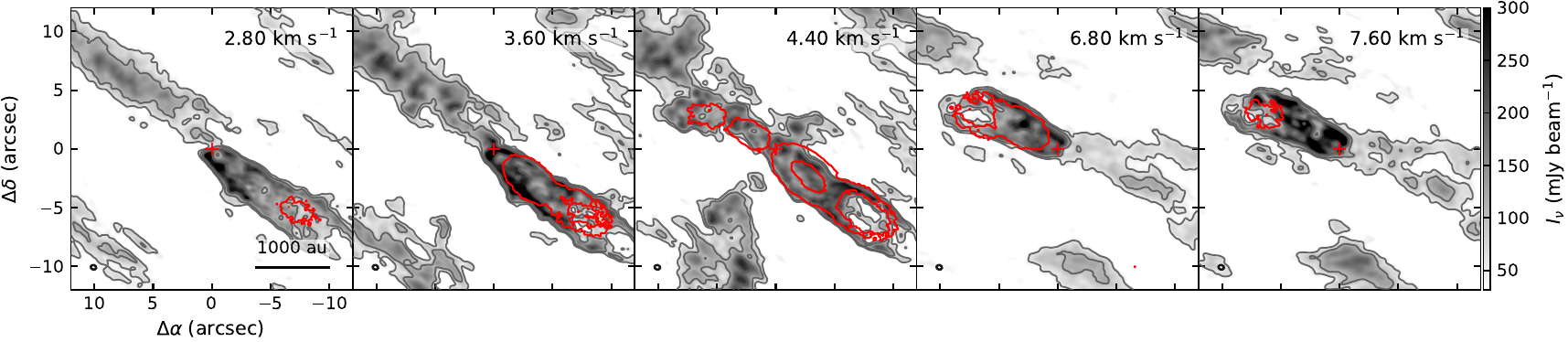}
\caption{
Comparison of the observed $^{12}$CO $J=2$--1
and modeled H$_2$CO $3_{0,3}$--$2_{0,2}$ channel maps from the Temperature Model toward IRAS~15398$-$3359.
The grayscale of $^{12}$CO starts from the $3\sigma$ level,
while the contours are at 5, 10, 20, and $40\sigma$.
The red contours show the modeled H$_2$CO emission
at the $5\sigma$ level.
}
\label{fig:12CO_comparison}
\end{figure*}

The present model of multiple, distinct shells could also alleviate
the discrepancy between the classical single-component, Lee-type shell
model and observed outflow structures.
In HH~46/47, high-velocity emission from some shells cannot be fully
reproduced by the model, and rotation or changes in outflow direction
have been discussed as possible explanations
\citep{2019Zhang}.
In Oph~IRS~63, the velocity of the outer shell is slower than predicted
from the Lee-type model
\citep{2026FeeneyJohansson}.
In DG~Tau~B, even a modified wind-driven shell model fails to reproduce
the observed velocity structures
\citep{2022deValon}.
These results imply that more degrees of freedom are likely required to
fully capture the observed outflow features.
Indeed, the observed outflow features toward HH~30 show multiple
shell-like structures and similar features in the velocity channel maps
and PV diagrams
\citep{2024LopezVazquez},
where our distinct shell model should be applicable.

\subsection{Molecular Tracers of the Shell Walls}
\label{sec:shock_chemistry}

Similar spatial and velocity structures of the CS emission to those of
the H$_2$CO emission (see Figs.~\ref{fig:h2co_cs_channelmap} and
\ref{fig:h2co_cs_pv_diagrams}) show that the H$_2$CO and CS
emission trace the same shell components, and that our multiple-shell
model should be applicable to the CS emission as well.
H$_2$CO can be released into the gas phase through ice-mantle
processing induced by heating or shocks
\citep[e.g.,][]{1994Tielens,2004Watanabe,2008Arce,2016Burkhardt},
whereas the CS abundance depends not only on the physical conditions of
dense gas but also on shock-driven sulfur chemistry
\citep[e.g.,][]{1993PineaudesForets,2005Jimenez-Serra,2014Podio,2025Codella}.
The critical densities of the H$_2$CO $3_{0,3}$--$2_{0,2}$ and
CS $J=5$--4 transitions are comparable,
$2.6\times10^6$ and $3.8\times10^6~{\rm cm^{-3}}$, respectively.
These results imply that the modeled shell features are real physical
structures, rather than specific chemical variations.

On the other hand, the CH$_3$OH emission is weaker than the H$_2$CO emission and is concentrated mainly
in the inner shells, which could be attributed to the higher critical density
of the CH$_3$OH emission (Fig.~\ref{fig:h2co_cs_channelmap}).
Furthermore, H$_2$CO may be released into
the gas phase by weaker shocks than CH$_3$OH \citep{2026Kim}, so this
difference is consistent with variations in shock strength or
ice-mantle processing among the shells.
The spatial distributions of
c-C$_3$H$_2$, SO, and SiO are also different among the shells
\citep{2020Okoda,2025Okoda}.
The H$_2$CO abundances adopted in both models
are of order $10^{-8}$,
within the range inferred for the outflows of young
Class 0 protostars (e.g., \citealt{2010Tafalla}).
In the Temperature Model, the overall intensity distribution
of the observed H$_2$CO emission is broadly reproduced
by adopting a common H$_2$CO abundance
of $1.5 \times 10^{-8}$ for all shells
and varying the gas temperature for each shell.
This result suggests that the shell-to-shell differences
in the H$_2$CO abundance adopted in the Fiducial Model
could also be explained by gas heating associated with
interactions between the outflow and the surrounding envelope.

\citet{2025Sabatini} suggested that
the dust temperature along the outflow shells
of the Class I protostar L1551 IRS 5
is enhanced by a factor of 1.2--1.5
compared to heating by the central protostar alone,
due to additional heating from UV irradiation
and low-velocity shocks.
The gas-temperature scaling factors adopted
in the Temperature Model
($f_{T,k}=1.4$--$2.0$, excluding R2)
are broadly comparable to these values.
The different scaling factors among the shells
($f_{T,\mathrm{B1}}=2.0$,
$f_{T,\mathrm{B2}}=1.4$, and
$f_{T,\mathrm{R1}}=1.8$)
may reflect differences in the interactions
with the surrounding material or local heating conditions.
In particular, the larger scaling factors
for the inner shells B1 and R1 compared to the outer shell B2
are consistent with the possibility that
the inner shells interact with denser surrounding material
and experience additional heating.
These results suggest
that the shells may differ not only in morphology and kinematics, but also
in shock conditions, excitation, and chemical state.

\citet{2026FeeneyJohansson} interpreted the outer high-velocity
$^{12}$CO structure in IRAS~15398$-$3359, which is at a projected
distance of approximately $16.6''$ from the protostar and corresponds
to Shell~4 of \citet{2025Okoda}, as a jet-driven bow shock based on
its correspondence with a JWST/MIRI shock feature.
In contrast, the multiple-shell structures identified here with H$_2$CO
and CS primarily trace molecular outflow components within approximately
$11''$ of the protostar.
$^{12}$CO is sensitive to more extended outflow components
\citep{2025Okoda,2025Sharma},
whereas H$_2$CO and CS can more selectively trace individual shell-like
components and gas under different physical and chemical conditions.
As shown in Fig.~\ref{fig:12CO_comparison},
the $^{12}$CO emission extends not only along the shell walls
identified in this study but also into the shell interiors
and over wider spatial and velocity ranges.
This result further supports the interpretation that
$^{12}$CO traces multiple components, potentially including
the outflow, the shell walls, ambient gas, and the surrounding envelope,
whereas H$_2$CO and CS primarily trace the shell walls.
Understanding the three-dimensional structure and formation of complex
protostellar outflows therefore requires multi-tracer analyses that can
separate distinct physical and chemical components and test whether
multiple formation or driving processes coexist.

\section{Conclusions}
\label{sec:conclusions}

To reproduce the shell-like spatial and velocity structures of the primary outflow as seen in the H$_2$CO and
CS emission in IRAS 15398$-$3359, we constructed a three-dimensional multiple-shell model
and performed the non-LTE radiative-transfer calculations of H$_2$CO and
synthetic ALMA observations.
To separate the density and chemical features in the outflow shells,
the observed $^{13}$CO emission is also modeled with the same framework.
Our main conclusions are summarized below.

\begin{enumerate}

\item
Four shell-like components associated with the primary
outflow are identified in the H$_2$CO and CS emission.
By adopting an ellipsoidal geometrical form and Hubble-law-like
velocity field for all shells while allowing their geometrical parameters,
spatial configurations, and kinematics to vary independently, the
multiple-shell framework broadly reproduces the major channel-map and PV
structures of the
H$_2$CO emission associated with
the shells identified in H$_2$CO and CS.

\item
The four shells differ in kinematic timescale, three-dimensional orientation, and
momentum injection rate.
The inner shells B1 and R1 have inclinations of $19^\circ$ and $15^\circ$,
respectively, whereas the outer shells B2 and R2 have smaller inclinations
of $5^\circ$ and $8^\circ$.
The kinematic timescales are $\tau_k = 1000$~yr for the inner shells and
750~yr for the outer shells, and the momentum injection rate of the outer pair
(B2+R2), $(3.12 \pm 0.25)$$\times10^{-6}$$\ M_\odot\ {\rm km\ s^{-1}\ yr^{-1}}$,
is approximately 2.2 times that of the inner pair
(B1+R1), $(1.40 \pm 0.19)$$\times10^{-6}$$\ M_\odot\ {\rm km\ s^{-1}\ yr^{-1}}$.
Thus, although the four shells can be described using the same basic
geometrical form and velocity prescription, their three-dimensional
configurations and kinematic properties differ.

\item
The similar shell-like spatial and velocity structures traced by H$_2$CO
and CS support the interpretation that both molecules probe the same or
closely adjacent compressed molecular structures associated with the
mass ejection.
The $^{13}$CO $J=2$--1 emission provides an additional constraint
on the shell-gas density scale, with typical densities near the
shell bases of $\sim10^6$ and $\sim10^5~{\rm cm^{-3}}$ for the inner
and outer shell pairs, respectively.
In contrast, the CH$_3$OH emission is weaker and concentrated mainly toward
the inner shells, suggesting shell-to-shell differences in shock
conditions, excitation, or the chemical condition.
The H$_2$CO shell intensities can be broadly reproduced
by both the Fiducial Model, with different H$_2$CO abundances
among the shells, and the Temperature Model,
with a common abundance and different gas temperatures.
This suggests that the abundance differences
may also be explained by additional gas heating.
These results demonstrate the importance of multi-tracer analyses for
separating the physical and chemical components of complex molecular
outflows.

\item
Taken together, the shell-to-shell differences of the physical properties are
consistent with time-variable or episodic outflow activity.
If the individual shells correspond to different ejection episodes,
successive episodes may therefore produce shells with different directions
and physical or kinematic properties.
The detached configurations in the outer
shells may arise if shells formed during finite-duration ejection episodes
continue to propagate after the driving ceases, while the transverse
offsets may reflect
subsequent interactions with an inhomogeneous ambient medium or off-axis driving of the protostellar winds.

\end{enumerate}

\section*{Acknowledgments}

This paper makes use of the following ALMA data:
\nolinkurl{ADS/JAO.ALMA\#2018.1.01205.L},
\nolinkurl{ADS/JAO.ALMA\#2019.1.00261.L}, and
\nolinkurl{ADS/JAO.ALMA\#2019.A.00034.S}.
ALMA is a partnership of ESO (representing its member states),
NSF (USA), and NINS (Japan), together with NRC (Canada),
NSTC and ASIAA (Taiwan), and KASI (Republic of Korea),
in cooperation with the Republic of Chile.
The Joint ALMA Observatory is operated by ESO, AUI/NRAO,
and NAOJ. This work is based in part on observations made with the
NASA/ESA/CSA James Webb Space Telescope.
The data were obtained from the Mikulski Archive for Space
Telescopes at the Space Telescope Science Institute, which is
operated by the Association of Universities for Research in
Astronomy, Inc., under NASA contract NAS 5-03127 for JWST.
These observations are associated with program \#2151.
The JWST data used in this paper are available
from MAST \citep{JWST_MAST_2022}.
Y.I. was supported by the ALMA Japan Research Grant of NAOJ ALMA Project, NAOJ-ALMA-409.
S.T. is supported by JSPS KAKENHI
grant Nos. JP21H04495 and JP24K00674 and by NAOJ
ALMA Scientific Research grant No. 2022-20A.
J.S. is supported by JSPS KAKENHI grant Nos. JP26K17198 and JP26K00741. J.S. acknowledges support from the KU-DREAM program of Kagoshima University.

\appendix
\renewcommand{\theHequation}{\Alph{section}.\arabic{equation}}

\section{Testing a Common Inclination for All Shells}
\label{app:inclination_test}

Our Fiducial and Temperature Models adopt smaller inclinations for
the outer shells, B2 and R2 ($i=5^\circ$ and $8^\circ$, respectively),
than for the inner shells, B1 and R1 ($i=19^\circ$ and $15^\circ$,
respectively).
We tested whether the differences of the inclination angles between the inner and outer shells are robust given the possible degeneracy between the inclination and line-of-sight gas motions.

\begin{figure*}[t]
\centering
\hspace*{-2.5em}
\includegraphics[
  height=0.99\textheight,
  keepaspectratio
]{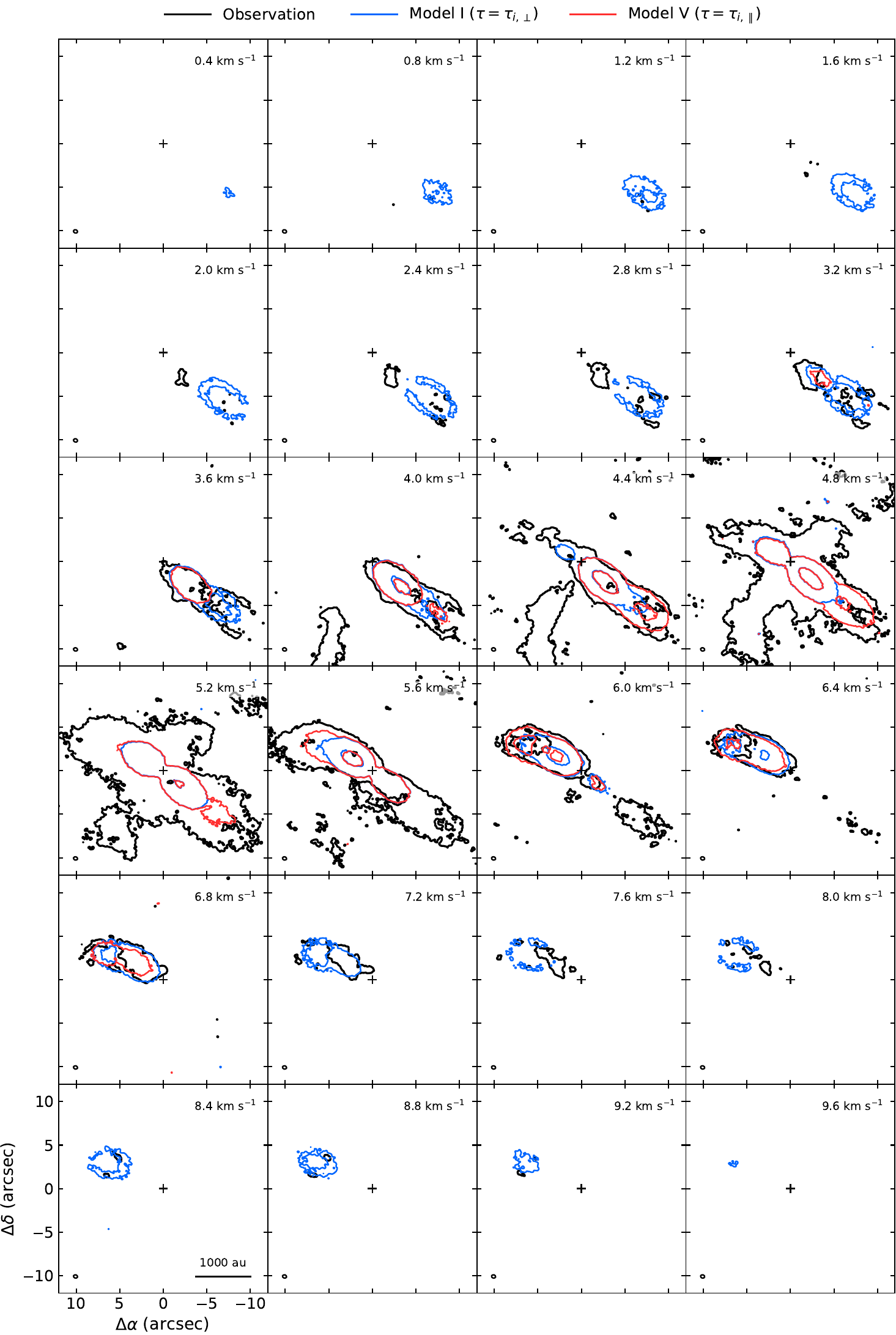}
\caption{
Comparison of the observed H$_2$CO $3_{0,3}$--$2_{0,2}$ channel maps with the
fixed-inclination models at $i=20^\circ$.
Black contours show the observations, while blue and red contours show
Models I and V, respectively.
All contours are drawn at the $5\sigma$ level.
}
\label{fig:obs_model_channelmap_incl20}
\end{figure*}

\begin{figure}
\centering
\includegraphics[width=0.95\textwidth]{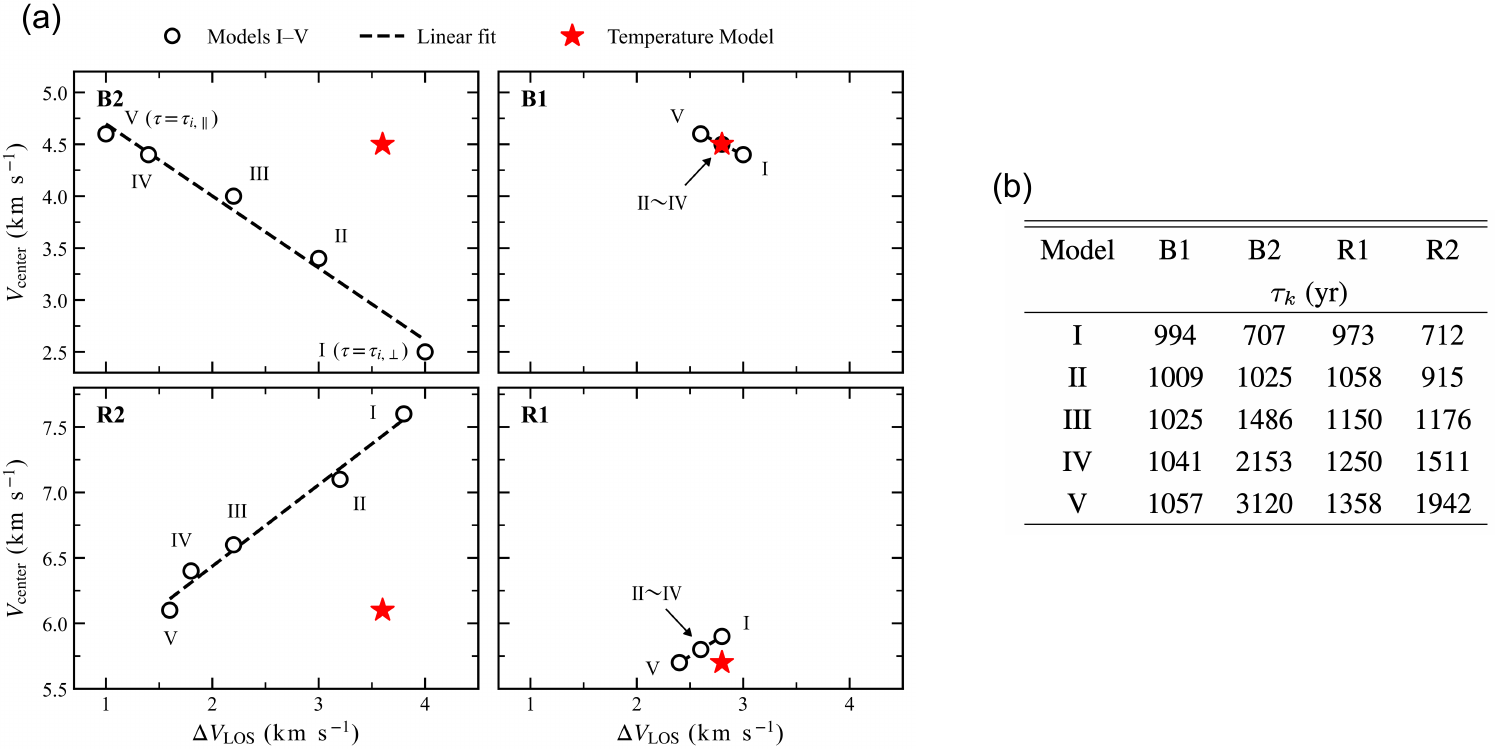}
\caption{
Diagnostic of the fixed-inclination models with $i=20^\circ$, where the inclination angles adopted in both the Fiducial and Temperature Models are
\(i=19^\circ\), \(5^\circ\), \(15^\circ\), and \(8^\circ\)
for B1, B2, R1, and R2, respectively.
(a) Trade-off between the central velocity $V_{\rm cent}$ and the
line-of-sight velocity width $\Delta V_{\rm LOS}$ of the shell-like
structures.
The star indicates the Temperature Model, and the dashed line shows
a linear fit to Models I--V.
(b) Kinematic timescales $\tau_k$ adopted for Models I--V.
Models I--V have $\tau_k$ logarithmically spaced between
$\tau_{k,\perp}$ and $\tau_{k,\parallel}$, with Models I and V
corresponding to these two limits, respectively.
}
\label{fig:inclination_diagnostic}
\end{figure}

For this test, we constructed an alternative set of models in which the
inclinations of all the four shells were fixed to $i=20^\circ$,
the global outflow inclination inferred by
\citet{2014Oya} and \citet{2017Yen}.
For each shell, $a_k$ and $Z_{0,k}$ were rescaled so that the projected
quantities $a_k\cos i$ and $Z_{0,k}\cos i$ were preserved relative to those of both models.
The transverse size $b_k$ is unchanged because it is measured
perpendicular to the projected symmetry axis and is independent of
the inclination.
The shell thickness $d_k$, transverse offset $s_k$, position angle,
density normalization $n_{0,k}$, molecular abundance $X_k(\mathrm{H_2CO})$, and $f_{T,k}$ were also fixed
to their Temperature Model values.
The remaining kinematic parameter considered here is therefore
$\tau_k$.
Inclination and $\tau_k$ affect the projected velocity structure
differently.
For a thin shell with the homologous velocity field adopted in this
study, the characteristic line-of-sight velocity width produced by the
transverse expansion scales as
\[
\Delta V_{\rm LOS}
\propto
\frac{b_k}{\tau_k}\cos i ,
\]
whereas the line-of-sight velocity gradient along the projected outflow
axis scales as
\[
\frac{d v_{\rm los}}{d z_{\rm proj}}
\propto
\frac{\tan i}{\tau_k}.
\]
If $(i_0,\tau_{k,0})$ denotes the inclination and kinematic timescale
common to the Fiducial
and Temperature Models,
the values of $\tau_k$ required to preserve
the transverse velocity extent and the axial velocity gradient after
changing the inclination to $i$ are, respectively,
\begin{equation}
\tau_{k,\perp}
=
\tau_{k,0}\frac{\cos i}{\cos i_0},
\qquad
\tau_{k,\parallel}
=
\tau_{k,0}\frac{\tan i}{\tan i_0}.
\label{eq:inclination_tau_constraints}
\end{equation}
At small inclinations, the fractional change in $\tan i$ is much
larger than that in $\cos i$ for a given change in $i$.
Consequently, changing the inclination of the outer shells produces
a large separation between $\tau_{k,\perp}$ and
$\tau_{k,\parallel}$, making their projected kinematics particularly
sensitive to inclination.
To test this effect with synthetic molecular-line observations, we
constructed five fixed-inclination models (Models I--V) for each shell,
with $\tau_k$ logarithmically spaced between
$\tau_{k,\perp}$ and $\tau_{k,\parallel}$.
Representative channel maps for the fixed-inclination models are
presented in Fig.~\ref{fig:obs_model_channelmap_incl20}, while the
adopted values of $\tau_k$ are summarized in
Fig.~\ref{fig:inclination_diagnostic}(b).

Figure~\ref{fig:obs_model_channelmap_incl20} compares the observed H$_2$CO
channel maps with Models I and V, which represent the two ends of the
tested $\tau_k$ range at fixed $i=20^\circ$.
For the outer shells, smaller $\tau_k$ values produce a
$\Delta V_{\rm LOS}$ comparable to that seen in the observations, but
shift the center of the elliptical emission substantially away from the
systemic velocity (blue contours in Fig.~\ref{fig:obs_model_channelmap_incl20}).
Increasing $\tau_k$ shifts the central velocity toward the observed
value, while simultaneously narrowing $\Delta V_{\rm LOS}$ (red contours).
Thus, at fixed $i=20^\circ$, changing $\tau_k$ improves one kinematic
property at the expense of the other.
Figure~\ref{fig:inclination_diagnostic}(a) quantifies this trade-off.
For Models I--V, we measured the central velocity $V_{\rm cent}$ and the line-of-sight velocity width $\Delta V_{\rm LOS}$
from the minimum and maximum velocities where each shell appears
at the $5\sigma$ level in the channel maps as
$V_{\rm cent}=(V_{\rm max}+V_{\rm min})/2$ and
$\Delta V_{\rm LOS}=V_{\rm max}-V_{\rm min}$.
The same quantities were measured from the Temperature Model for
comparison.
At fixed $i=20^\circ$, varying $\tau_k$ moves Models I--V along a
one-dimensional locus in the
$V_{\rm cent}$--$\Delta V_{\rm LOS}$ plane.
For B2 and R2, the loci of the fixed-inclination models do not pass
through the Temperature Model values.
Smaller $\tau_k$ values approach the Temperature Model
$\Delta V_{\rm LOS}$ but produce a larger discrepancy in $V_{\rm cent}$,
and vice versa.

Thus, within our model framework a unified value of the inclination angle
\(i=20^\circ\) cannot reproduce the observed velocity extents and
the central velocities of the outer shells simultaneously.
For the inner shells, the inclination angles adopted in both models
are close
to $20^\circ$ ($i=19^\circ$ for B1 and $15^\circ$ for R1).
The fixed-inclination locus passes through the Temperature Model value
for B1, while a modest offset remains for R1.
These results support the interpretation that a single global inclination
is insufficient to describe the three-dimensional kinematics of all the four
shells.

\newpage
\section{Additional Results of the Fiducial Model}
\label{app:fiducial_model}

\begin{figure*}[t]
\centering
\hspace*{-2.5em}
\includegraphics[
  height=0.99\textheight,
  keepaspectratio
]{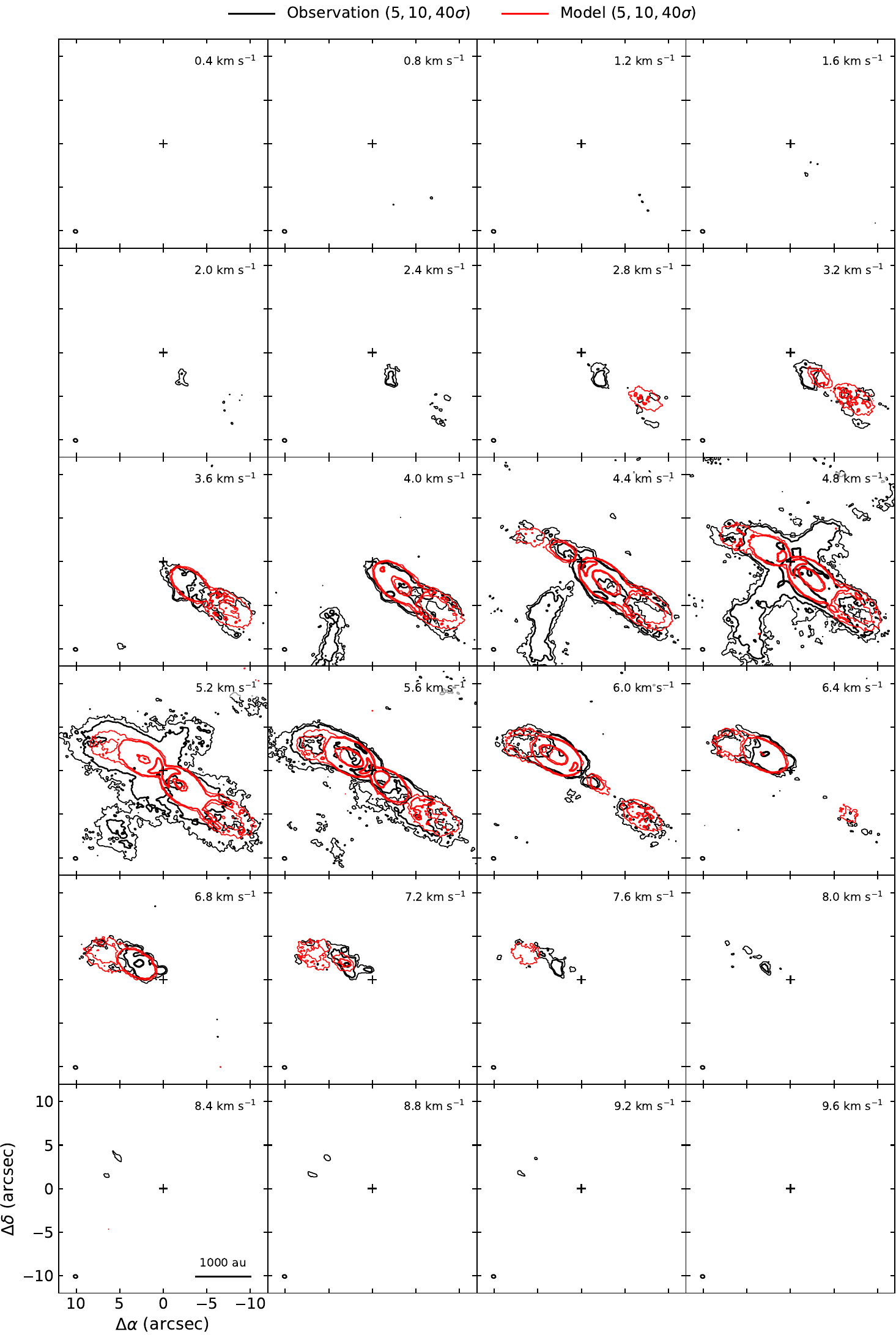}
\caption{
Same as Figure~\ref{fig:obs_model_channelmap_temp},
but for the Fiducial Model.
}
\label{fig:obs_model_channelmap}
\end{figure*}

\begin{figure*}[pt]
\centering
\hspace*{-0.6em}
\includegraphics[width=0.95\textwidth]{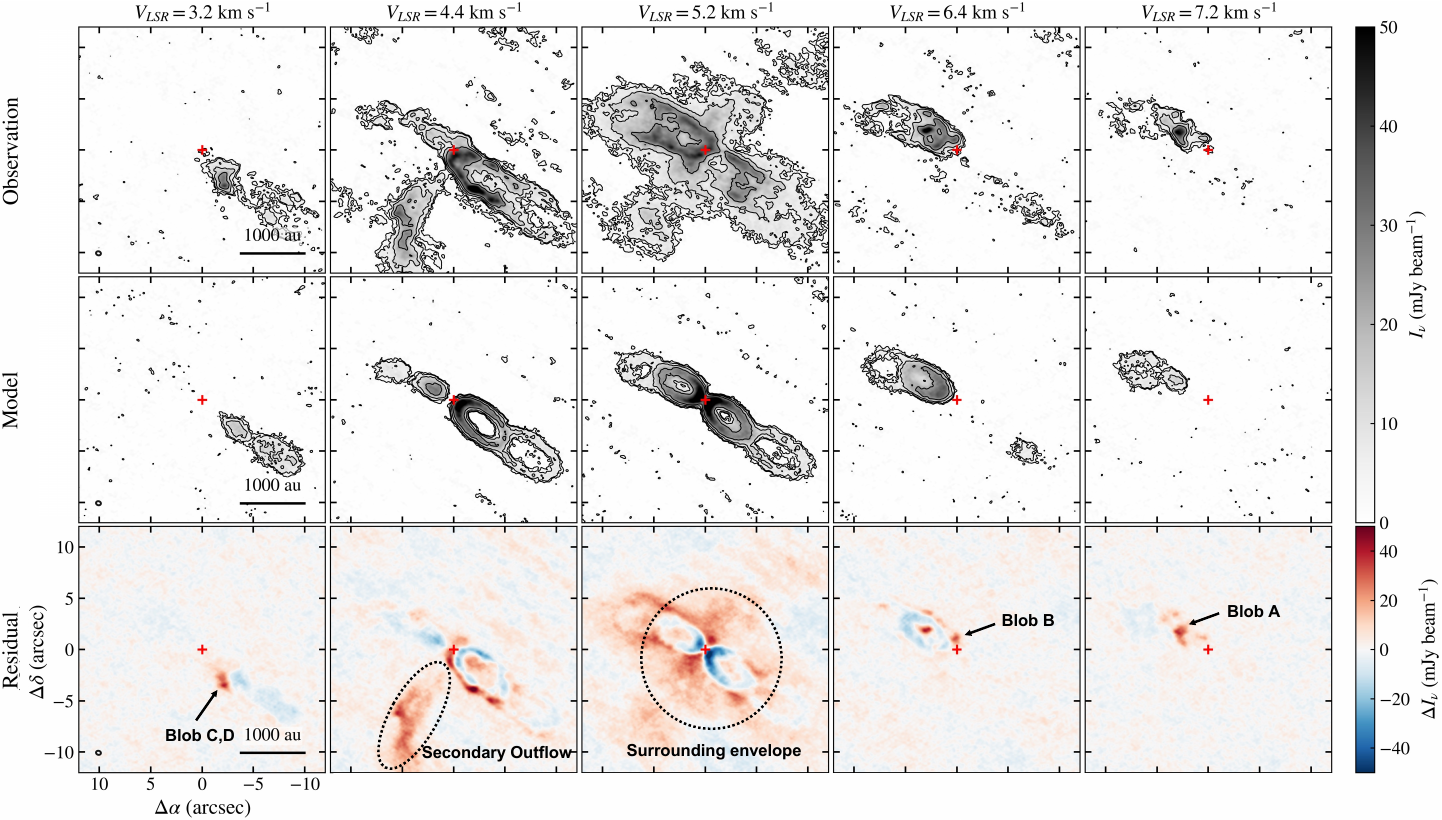}
\caption
{
Same as Figure~\ref{fig:residual_temp},
but for the Fiducial Model.
}
\label{fig:residual}
\end{figure*}

\begin{figure*}[pb]
\centering
\hspace*{-1.1em}
\includegraphics[width=0.93\textwidth]{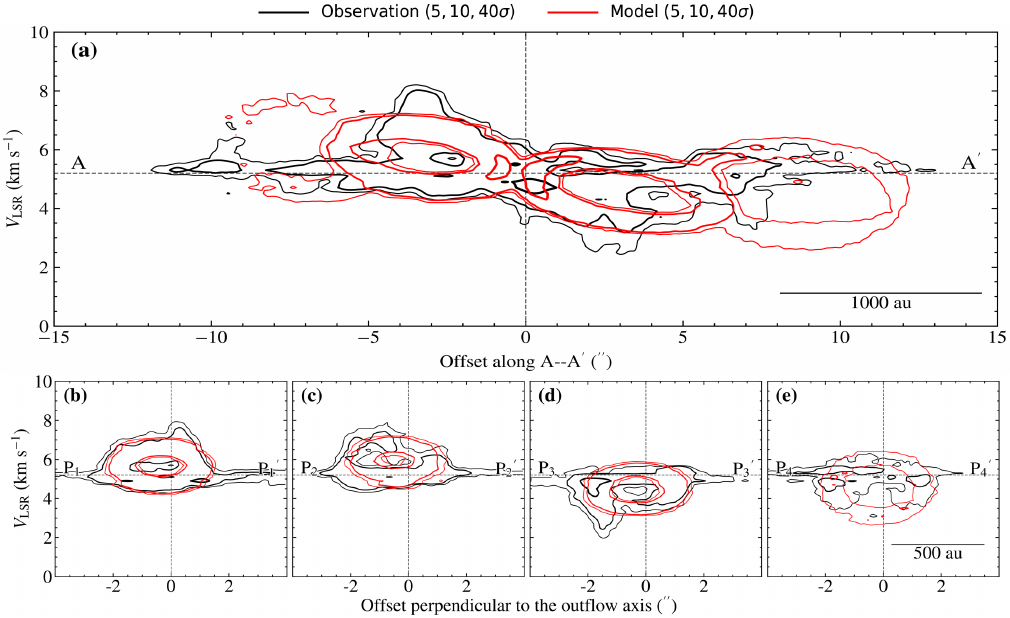}
\caption{
Same as Figure~\ref{fig:obs_model_PV_temp},
but for the Fiducial Model.
}
\label{fig:obs_model_PV}
\end{figure*}

Figures~\ref{fig:obs_model_channelmap},
\ref{fig:residual}, and
\ref{fig:obs_model_PV}
show the velocity channel maps, residuals,
and position--velocity diagrams of the Fiducial Model,
respectively.

\section{Additional Channel Maps}
\label{app:additional_channel_maps}

\begin{figure*}[p]
\centering
\includegraphics[height=0.98\textheight, keepaspectratio]{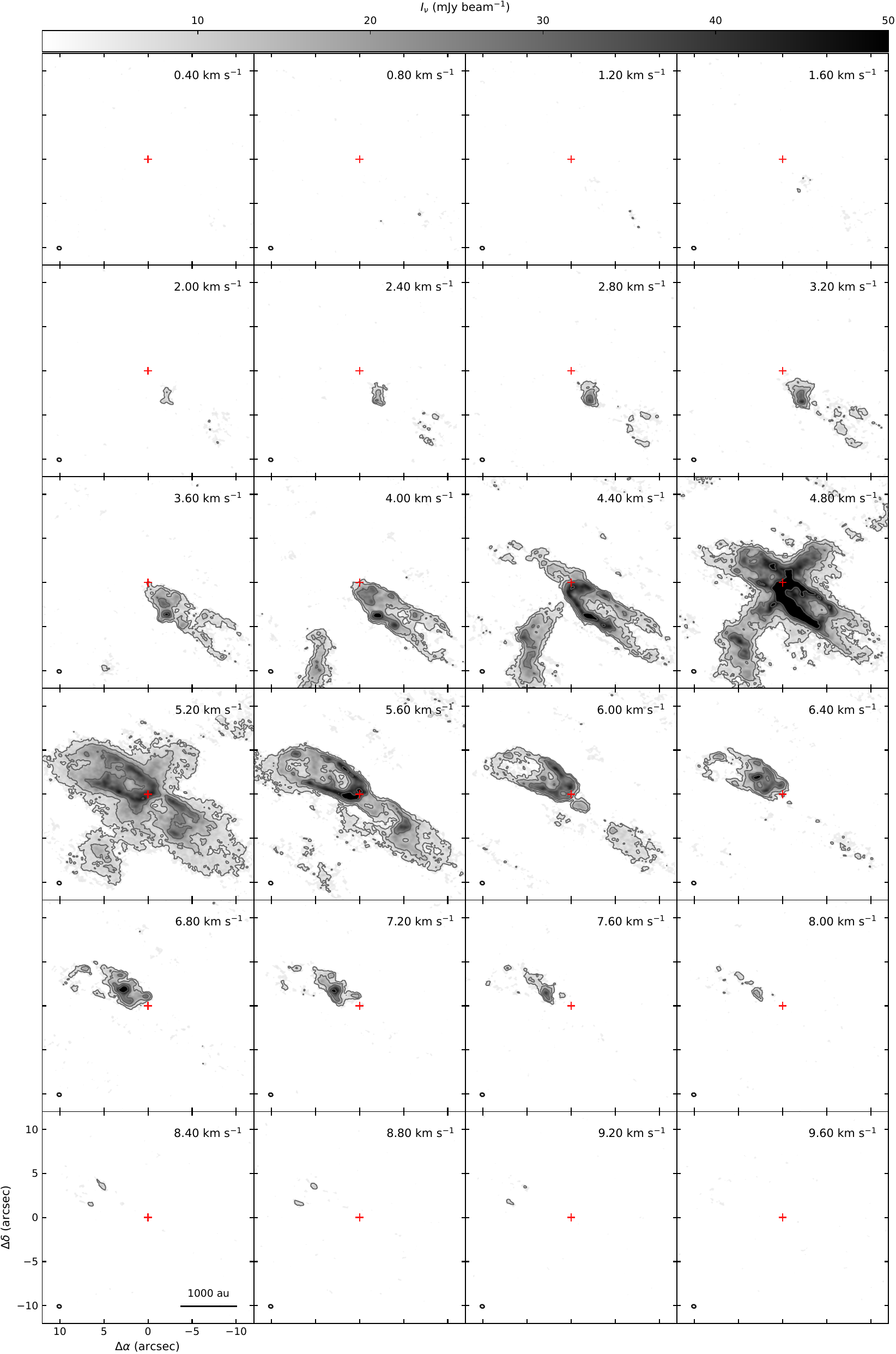}
\caption{
Velocity channel maps of the observed H$_2$CO $3_{0,3}$--$2_{0,2}$ emission.
The grayscale starts from the $3\sigma$ level,
while the contours are at 5, 10, 20, and $40\sigma$.
}
\label{fig:h2co_channel_all_no_fit}
\end{figure*}

\begin{figure*}[p]
\centering
\includegraphics[height=0.98\textheight, keepaspectratio]{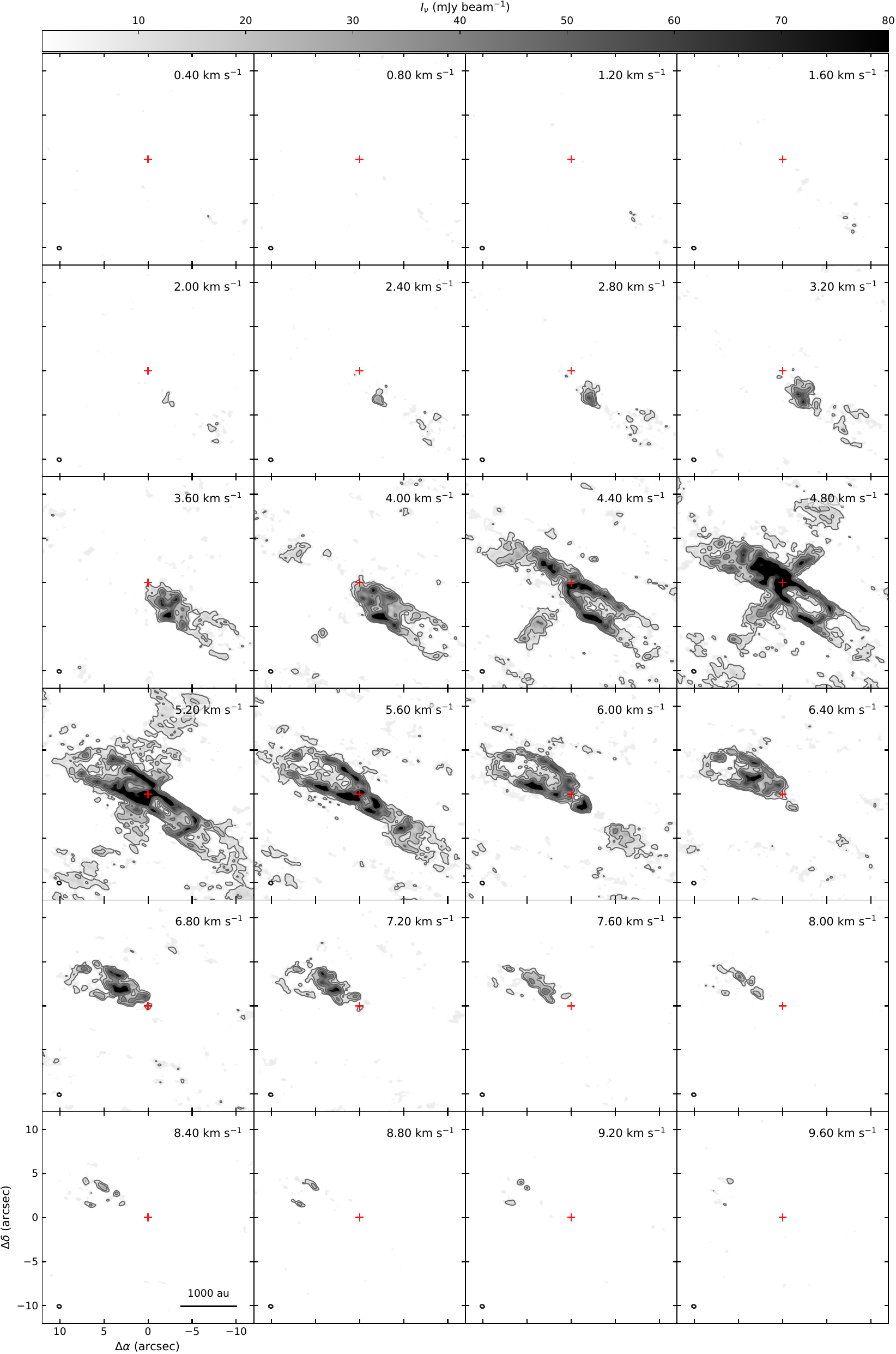}
\caption{
Same as Figure~\ref{fig:h2co_channel_all_no_fit}, but for the observed CS $J=5$--4 emission.
}
\label{fig:cs_channel_all_nofit}
\end{figure*}

\begin{figure*}[t]
\centering
\includegraphics[height=0.37\textheight, keepaspectratio]{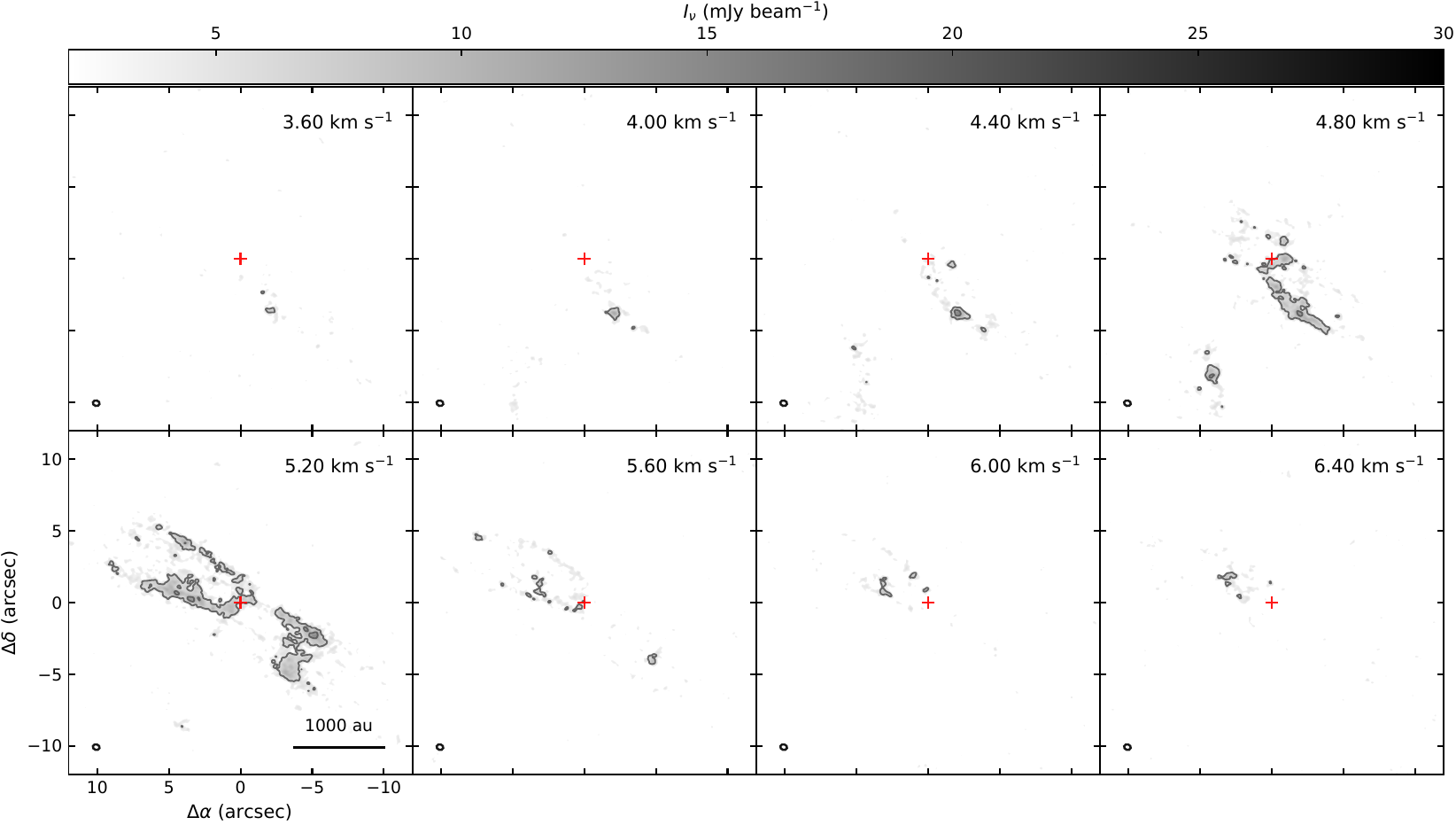}
\caption{
Same as Figure~\ref{fig:h2co_channel_all_no_fit}, but for the observed CH$_3$OH $4_{2,2}$--$3_{1,2}$ emission.
}
\label{fig:ch3oh_channel_all_nofit}
\end{figure*}

\begin{figure*}[b]
\centering
\includegraphics[height=0.53\textheight, keepaspectratio]{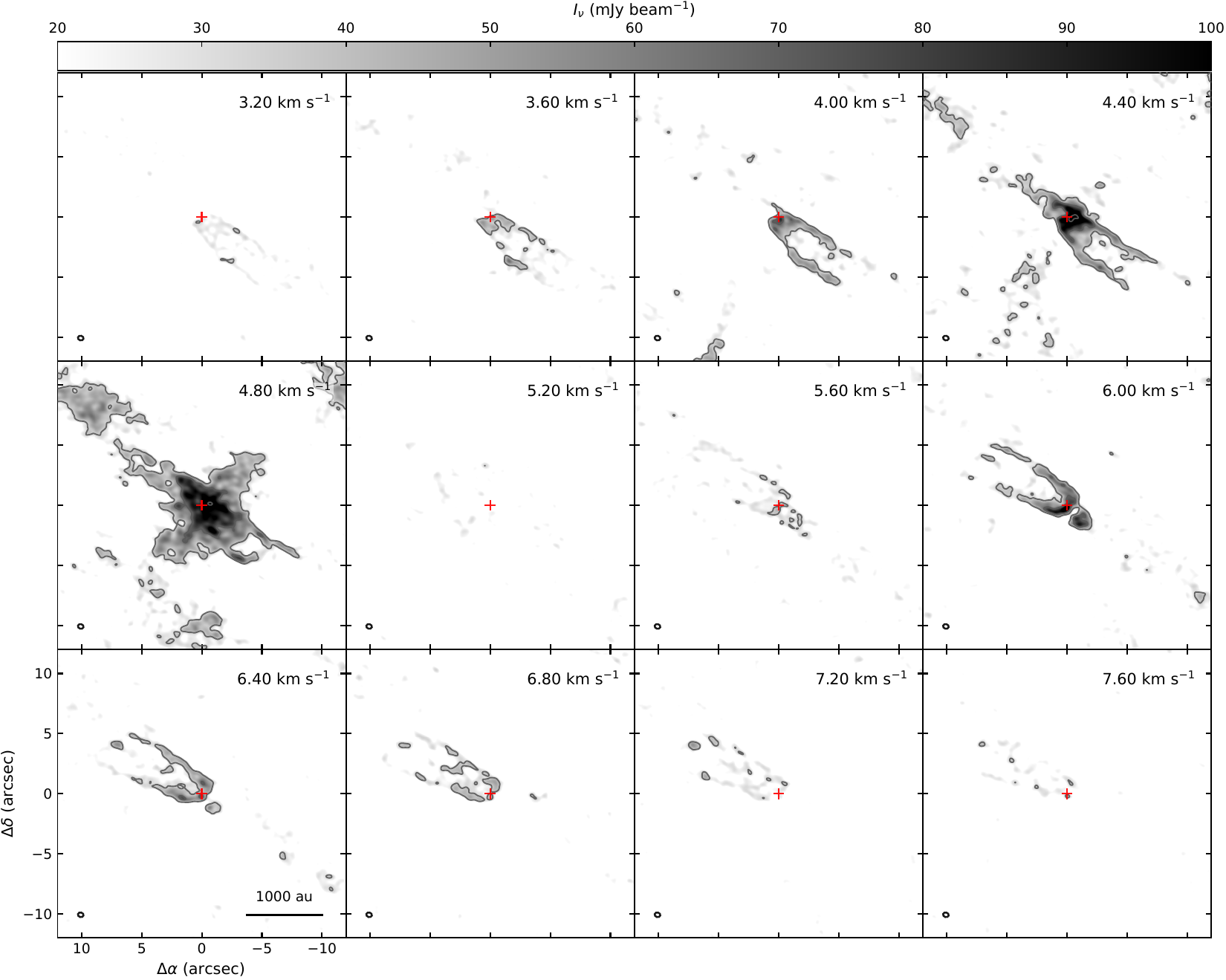}
\caption{
Same as Figure~\ref{fig:h2co_channel_all_no_fit}, but for the observed $^{13}$CO $J=2$--1 emission.
}
\label{fig:13co_channel_all_nofit}
\end{figure*}

\begin{figure*}[p]
\centering
\includegraphics[height=0.98\textheight, keepaspectratio]{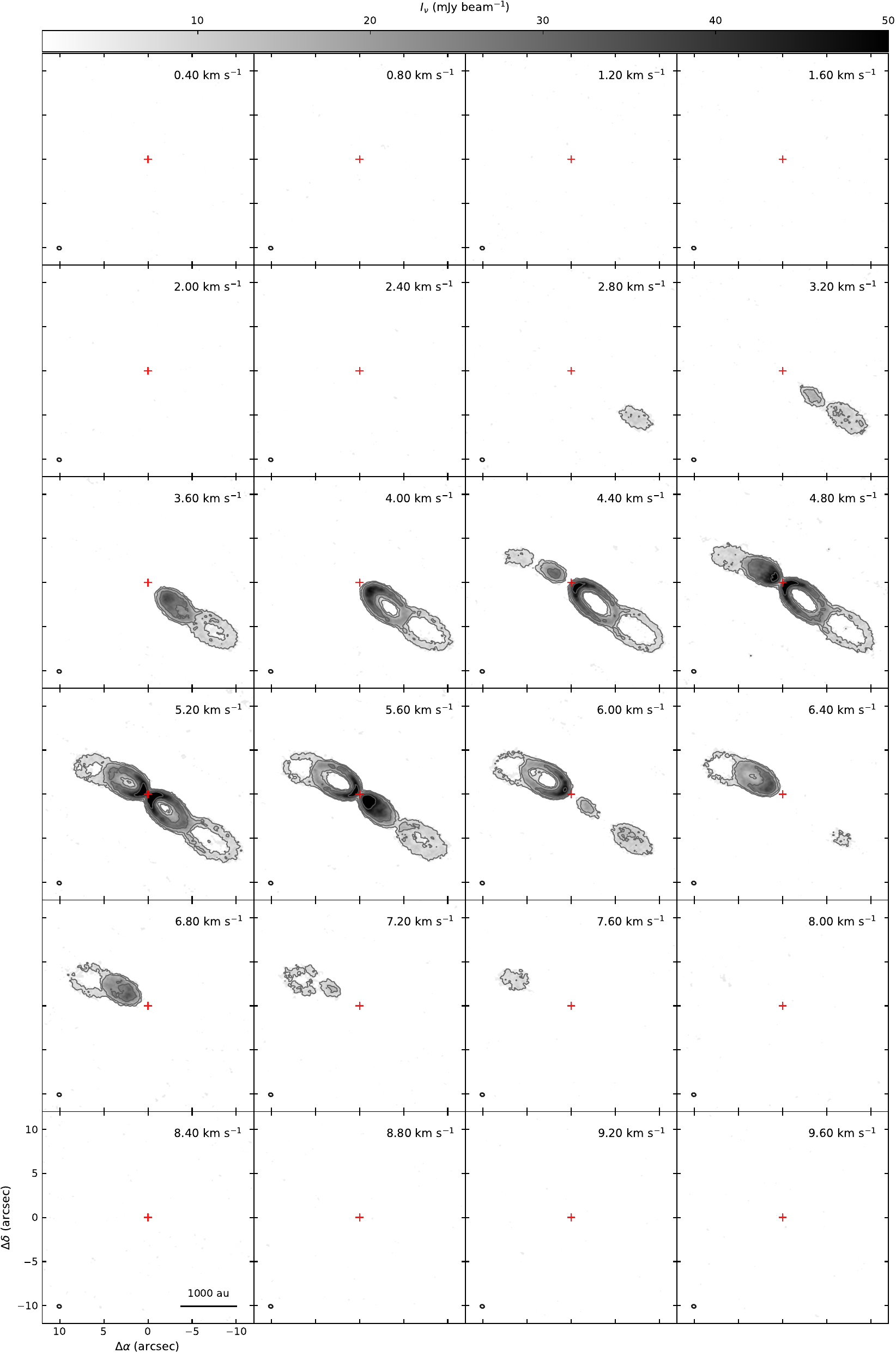}
\caption{
Same as Figure~\ref{fig:h2co_channel_all_no_fit},
but for the model H$_2$CO $3_{0,3}$--$2_{0,2}$ emission
from the Fiducial Model.
}
\label{fig:model_channel_all_no_fit}
\end{figure*}

\begin{figure*}[p]
\centering
\includegraphics[height=0.98\textheight, keepaspectratio]{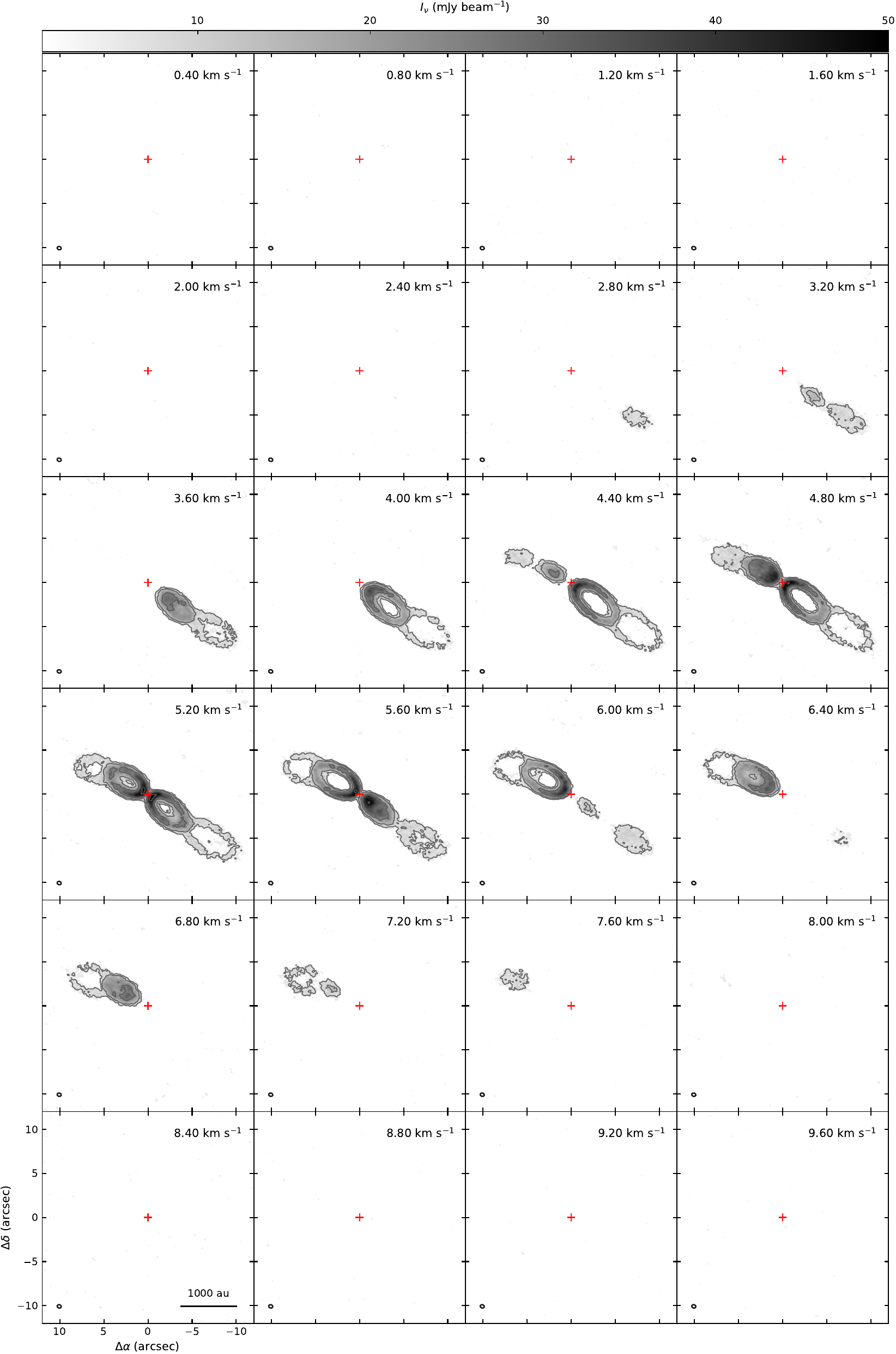}
\caption{
Same as Figure~\ref{fig:h2co_channel_all_no_fit},
but for the model H$_2$CO $3_{0,3}$--$2_{0,2}$ emission
from the Temperature Model.
}
\label{fig:model_channel_all_no_fit_temp}
\end{figure*}

\begin{figure*}[t]
\centering
\includegraphics[height=0.53\textheight, keepaspectratio]{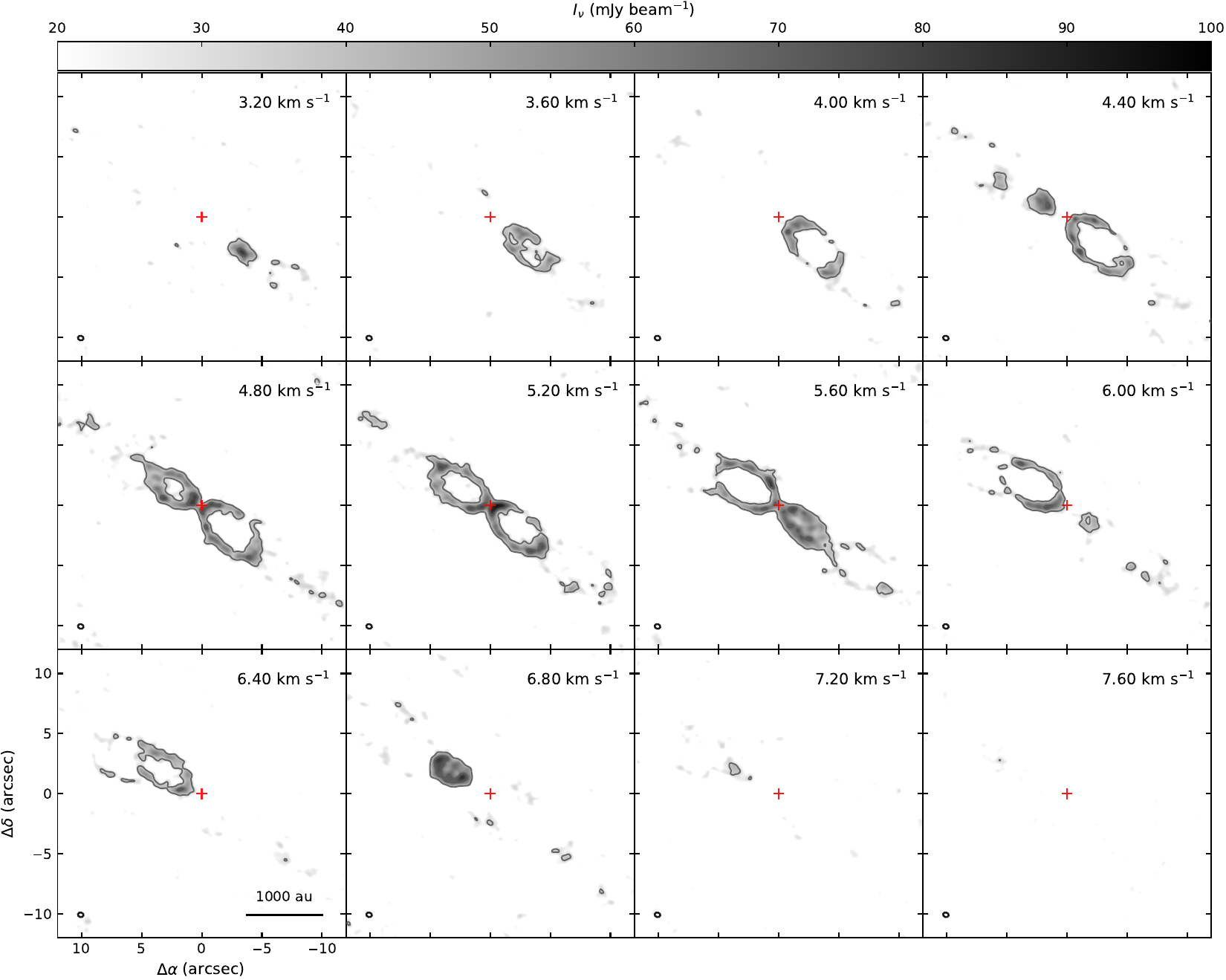}
\caption{
Same as Figure~\ref{fig:h2co_channel_all_no_fit}, but for the model $^{13}$CO $J=2$--1 emission.
}
\label{fig:model_13co_channel_all_nofit}
\end{figure*}

Figures~\ref{fig:h2co_channel_all_no_fit}--\ref{fig:model_13co_channel_all_nofit}
present the full channel maps of the observed H$_2$CO, CS,
CH$_3$OH, and $^{13}$CO emission, together with the modeled
H$_2$CO and $^{13}$CO emission.

\clearpage

\software{
CASA \citep{2007McMullin},
RADMC-3D \citep{2012Dullemond},
NumPy \citep{2020Harris},
SciPy \citep{2020Virtanen},
matplotlib \citep{2007Hunter},
Astropy \citep{2013astropy,2018astropy,2022astropy},
spectral-cube \citep{2019Ginsburg},
PVextractor \citep{2016Ginsburg},
Pandas \citep{2010McKinney}
}
\facility{ALMA, JWST}

\bibliographystyle{aasjournal}
\bibliography{IRAS15398-3359_paper}

@ARTICLE{1990Beckwith,
       author = {{Beckwith}, Steven V.~W. and {Sargent}, Anneila I. and {Chini}, Rolf S. and {Guesten}, Rolf},
        title = {A Survey for Circumstellar Disks around Young Stellar Objects},
      journal = {\aj},
         year = 1990,
        month = mar,
       volume = {99},
        pages = {924},
          doi = {10.1086/115385},
       adsurl = {https://ui.adsabs.harvard.edu/abs/1990AJ.....99..924B}
}

@ARTICLE{2024Basu,
       author = {{Basu}, Shantanu and {Sharkawi}, Mahmoud and {Machida}, Masahiro N.},
        title = "{Outflows Driven from a Magnetic Pseudodisk}",
      journal = {\apj},
         year = 2024,
        month = apr,
       volume = {964},
       number = {2},
          eid = {116},
        pages = {116},
          doi = {10.3847/1538-4357/ad1bf3},
archivePrefix = {arXiv},
       eprint = {2401.04260},
 primaryClass = {astro-ph.SR},
       adsurl = {https://ui.adsabs.harvard.edu/abs/2024ApJ...964..116B}
}

@ARTICLE{2016Bjerkeli_a,
       author = {{Bjerkeli}, P. and {J{\o}rgensen}, J.~K. and {Brinch}, C.},
        title = {A young bipolar outflow from IRAS 15398-3359},
      journal = {\aap},
         year = 2016,
        month = mar,
       volume = {587},
          eid = {A145},
        pages = {A145},
          doi = {10.1051/0004-6361/201527310},
archivePrefix = {arXiv},
       eprint = {1601.06947},
 primaryClass = {astro-ph.GA},
       adsurl = {https://ui.adsabs.harvard.edu/abs/2016A&A...587A.145B}
}

@ARTICLE{2015Bouchet,
       author = {{Bouchet}, Patrice and {Garc{\'\i}a-Mar{\'\i}n}, Macarena and {Lagage}, P.-O. and {Amiaux}, J{\'e}rome and {Augu{\'e}res}, J.-L. and {Bauwens}, Eva and {Blommaert}, J.~A.~D.~L. and {Chen}, C.~H. and {Detre}, {\"O}. H. and {Dicken}, Dan and {Dubreuil}, D. and {Galdemard}, Ph. and {Gastaud}, R. and {Glasse}, A. and {Gordon}, K.~D. and {Gougnaud}, F. and {Guillard}, Phillippe and {Justtanont}, K. and {Krause}, Oliver and {Leboeuf}, Didier and {Longval}, Yuying and {Martin}, Laurant and {Mazy}, Emmanuel and {Moreau}, Vincent and {Olofsson}, G{\"o}ran and {Ray}, T.~P. and {Rees}, J.-M. and {Renotte}, Etienne and {Ressler}, M.~E. and {Ronayette}, Samuel and {Salasca}, Sophie and {Scheithauer}, Silvia and {Sykes}, Jon and {Thelen}, M.~P. and {Wells}, Martyn and {Wright}, David and {Wright}, G.~S.},
        title = {The Mid-Infrared Instrument for the James Webb Space Telescope, III: MIRIM, The MIRI Imager},
      journal = {\pasp},
         year = 2015,
        month = jul,
       volume = {127},
       number = {953},
        pages = {612},
          doi = {10.1086/682254},
archivePrefix = {arXiv},
       eprint = {1508.02488},
 primaryClass = {astro-ph.IM},
       adsurl = {https://ui.adsabs.harvard.edu/abs/2015PASP..127..612B}
}

@ARTICLE{2007Chapman,
       author = {{Chapman}, Nicholas L. and {Lai}, Shih-Ping and {Mundy}, Lee G. and {Evans}, II, Neal J. and {Brooke}, Timothy Y. and {Cieza}, Lucas A. and {Spiesman}, William J. and {Rebull}, Luisa M. and {Stapelfeldt}, Karl R. and {Noriega-Crespo}, Alberto and {Lanz}, Lauranne and {Allen}, Lori E. and {Blake}, Geoffrey A. and {Bourke}, Tyler L. and {Harvey}, Paul M. and {Huard}, Tracy L. and {J{\o}rgensen}, Jes K. and {Koerner}, David W. and {Myers}, Philip C. and {Padgett}, Deborah L. and {Sargent}, Annelia I. and {Teuben}, Peter and {van Dishoeck}, Ewine F. and {Wahhaj}, Zahed and {Young}, Kaisa E.},
        title = {The Spitzer c2d Survey of Large, Nearby, Interstellar Clouds. IV. Lupus Observed with MIPS},
      journal = {\apj},
         year = 2007,
        month = sep,
       volume = {667},
       number = {1},
        pages = {288-302},
          doi = {10.1086/520790},
archivePrefix = {arXiv},
       eprint = {0706.1809},
 primaryClass = {astro-ph},
       adsurl = {https://ui.adsabs.harvard.edu/abs/2007ApJ...667..288C}
}

@ARTICLE{2021Codella,
       author = {{Codella}, C. and {Ceccarelli}, C. and {Chandler}, C. and {Sakai}, N. and {Yamamoto}, S. and {FAUST Team}},
        title = "{Enlightening the chemistry of infalling envelopes and accretion disks around Sun-like protostars: the ALMA FAUST project}",
      journal = {Frontiers in Astronomy and Space Sciences},
         year = 2021,
        month = dec,
       volume = {8},
          eid = {227},
        pages = {227},
          doi = {10.3389/fspas.2021.782006},
archivePrefix = {arXiv},
       eprint = {2111.14121},
 primaryClass = {astro-ph.SR},
       adsurl = {https://ui.adsabs.harvard.edu/abs/2021FrASS...8..227C}
}

@ARTICLE{2007DownesCabrit,
       author = {{Downes}, T.~P. and {Cabrit}, S.},
        title = {Jet-driven molecular outflows from class 0 sources: younger and stronger than they seem?},
      journal = {\aap},
         year = 2007,
        month = sep,
       volume = {471},
       number = {3},
        pages = {873-884},
          doi = {10.1051/0004-6361:20066921},
       adsurl = {https://ui.adsabs.harvard.edu/abs/2007A&A...471..873D}
}

@ARTICLE{1982Frerking,
       author = {{Frerking}, M.~A. and {Langer}, W.~D. and {Wilson}, R.~W.},
        title = {The relationship between carbon monoxide abundance and visual extinction in interstellar clouds.},
      journal = {\apj},
         year = 1982,
        month = nov,
       volume = {262},
        pages = {590-605},
          doi = {10.1086/160451},
       adsurl = {https://ui.adsabs.harvard.edu/abs/1982ApJ...262..590F}
}

@ARTICLE{2026FeeneyJohansson,
       author = {{Feeney-Johansson}, Anton and {Aikawa}, Yuri and {Takakuwa}, Shigehisa and {Ohashi}, Nagayoshi and {Plunkett}, Adele and {J{\o}rgensen}, Jes K. and {Shang}, Hsien and {Li}, Zhi-Yun and {Sharma}, Rajeeb and {Kwon}, Woojin and {Lee}, Jeong-Eun and {Looney}, Leslie W. and {Yang}, Yao-Lun and {Narang}, Mayank and {de Gregorio-Monsalvo}, Itziar and {eDisk Team}},
        title = {Early Planet Formation in Embedded Disks (eDisk). XIX. Structures of Molecular Outflows},
      journal = {\aj},
         year = 2026,
        month = mar,
       volume = {171},
       number = {3},
          eid = {172},
        pages = {172},
          doi = {10.3847/1538-3881/ae3135},
archivePrefix = {arXiv},
       eprint = {2512.21454},
 primaryClass = {astro-ph.SR},
       adsurl = {https://ui.adsabs.harvard.edu/abs/2026AJ....171..172F}
}

@ARTICLE{1984Goldsmith,
       author = {{Goldsmith}, P.~F. and {Snell}, R.~L. and {Hemeon-Heyer}, M. and {Langer}, W.~D.},
        title = {Bipolar outflows in dark clouds.},
      journal = {\apj},
         year = 1984,
        month = nov,
       volume = {286},
        pages = {599-608},
          doi = {10.1086/162635},
       adsurl = {https://ui.adsabs.harvard.edu/abs/1984ApJ...286..599G}
}

@ARTICLE{2024GuzmanCcolque,
       author = {{Guzm{\'a}n Ccolque}, E. and {Fern{\'a}ndez L{\'o}pez}, M. and {Vazzano}, M.~M. and {de Gregorio}, I. and {Plunkett}, A. and {Santamar{\'\i}a-Miranda}, A.},
        title = {Episodicity in accretion-ejection processes associated with IRAS 15398-3359},
      journal = {\aap},
         year = 2024,
        month = jun,
       volume = {686},
          eid = {A143},
        pages = {A143},
          doi = {10.1051/0004-6361/202348816},
archivePrefix = {arXiv},
       eprint = {2403.12841},
 primaryClass = {astro-ph.SR},
       adsurl = {https://ui.adsabs.harvard.edu/abs/2024A&A...686A.143G}
}

@ARTICLE{2025Hanawa,
       author = {{Hanawa}, Tomoyuki and {Okoda}, Yuki and {Yang}, Yao-Lun and {Sakai}, Nami},
        title = "{Origin of the Shell Structure in the Primary Outflow from IRAS 15398{\ensuremath{-}}3359}",
      journal = {\apj},
         year = 2025,
        month = jun,
       volume = {986},
       number = {2},
          eid = {150},
        pages = {150},
          doi = {10.3847/1538-4357/addac0},
archivePrefix = {arXiv},
       eprint = {2505.08470},
 primaryClass = {astro-ph.GA},
       adsurl = {https://ui.adsabs.harvard.edu/abs/2025ApJ...986..150H}
}

@ARTICLE{1989HeyerGraham,
       author = {{Heyer}, Mark H. and {Graham}, J.~A.},
        title = "{Newborn Stars and Stellar Objects in Barnard 228}",
      journal = {\pasp},
         year = 1989,
        month = sep,
       volume = {101},
        pages = {816},
          doi = {10.1086/132502},
       adsurl = {https://ui.adsabs.harvard.edu/abs/1989PASP..101..816H}
}

@ARTICLE{2002Jorgensen,
       author = {{J{\o}rgensen}, J.~K. and {Sch{\"o}ier}, F.~L. and {van Dishoeck}, E.~F.},
        title = {Physical structure and CO abundance of low-mass protostellar envelopes},
      journal = {\aap},
         year = 2002,
        month = jul,
       volume = {389},
        pages = {908-930},
          doi = {10.1051/0004-6361:20020681},
archivePrefix = {arXiv},
       eprint = {astro-ph/0205068},
 primaryClass = {astro-ph},
       adsurl = {https://ui.adsabs.harvard.edu/abs/2002A&A...389..908J}
}

@ARTICLE{2013Jorgensen,
       author = {{J{\o}rgensen}, Jes K. and {Visser}, Ruud and {Sakai}, Nami and {Bergin}, Edwin A. and {Brinch}, Christian and {Harsono}, Daniel and {Lindberg}, Johan E. and {van Dishoeck}, Ewine F. and {Yamamoto}, Satoshi and {Bisschop}, Suzanne E. and {Persson}, Magnus V.},
        title = {A Recent Accretion Burst in the Low-mass Protostar IRAS 15398-3359: ALMA Imaging of Its Related Chemistry},
      journal = {\apjl},
         year = 2013,
        month = dec,
       volume = {779},
       number = {2},
          eid = {L22},
        pages = {L22},
          doi = {10.1088/2041-8205/779/2/L22},
archivePrefix = {arXiv},
       eprint = {1312.0724},
 primaryClass = {astro-ph.SR},
       adsurl = {https://ui.adsabs.harvard.edu/abs/2013ApJ...779L..22J}
}

@ARTICLE{2005Schoier,
       author = {{Sch{\"o}ier}, F.~L. and {van der Tak}, F.~F.~S. and {van Dishoeck}, E.~F. and {Black}, J.~H.},
        title = {An atomic and molecular database for analysis of submillimetre line observations},
      journal = {\aap},
         year = 2005,
        month = mar,
       volume = {432},
       number = {1},
        pages = {369-379},
          doi = {10.1051/0004-6361:20041729},
archivePrefix = {arXiv},
       eprint = {astro-ph/0411110},
 primaryClass = {astro-ph},
       adsurl = {https://ui.adsabs.harvard.edu/abs/2005A&A...432..369S}
}

@ARTICLE{2008Kauffmann,
       author = {{Kauffmann}, J. and {Bertoldi}, F. and {Bourke}, T.~L. and {Evans}, II, N.~J. and {Lee}, C.~W.},
        title = "{MAMBO mapping of Spitzer c2d small clouds and cores}",
      journal = {\aap},
         year = 2008,
        month = sep,
       volume = {487},
       number = {3},
        pages = {993-1017},
          doi = {10.1051/0004-6361:200809481},
archivePrefix = {arXiv},
       eprint = {0805.4205},
 primaryClass = {astro-ph},
       adsurl = {https://ui.adsabs.harvard.edu/abs/2008A&A...487..993K}
}

@ARTICLE{2000Lee,
       author = {{Lee}, Chin-Fei and {Mundy}, Lee G. and {Reipurth}, Bo and {Ostriker}, Eve C. and {Stone}, James M.},
        title = {CO Outflows from Young Stars: Confronting the Jet and Wind Models},
      journal = {\apj},
         year = 2000,
        month = oct,
       volume = {542},
       number = {2},
        pages = {925-945},
          doi = {10.1086/317056},
       adsurl = {https://ui.adsabs.harvard.edu/abs/2000ApJ...542..925L}
}

@ARTICLE{2001Lee,
       author = {{Lee}, Chin-Fei and {Stone}, James M. and {Ostriker}, Eve C. and {Mundy}, Lee G.},
        title = {Hydrodynamic Simulations of Jet- and Wind-driven Protostellar Outflows},
      journal = {\apj},
         year = 2001,
        month = aug,
       volume = {557},
       number = {1},
        pages = {429-442},
          doi = {10.1086/321648},
archivePrefix = {arXiv},
       eprint = {astro-ph/0104373},
 primaryClass = {astro-ph},
       adsurl = {https://ui.adsabs.harvard.edu/abs/2001ApJ...557..429L}
}

@ARTICLE{1999Matzner,
       author = {{Matzner}, Christopher D. and {McKee}, Christopher F.},
        title = {Bipolar Molecular Outflows Driven by Hydromagnetic Protostellar Winds},
      journal = {\apjl},
         year = 1999,
        month = dec,
       volume = {526},
       number = {2},
        pages = {L109-L112},
          doi = {10.1086/312376},
archivePrefix = {arXiv},
       eprint = {astro-ph/9909479},
 primaryClass = {astro-ph},
       adsurl = {https://ui.adsabs.harvard.edu/abs/1999ApJ...526L.109M}
}

@ARTICLE{2005Milam,
       author = {{Milam}, S.~N. and {Savage}, C. and {Brewster}, M.~A. and {Ziurys}, L.~M. and {Wyckoff}, S.},
        title = {The $^{12}$C/$^{13}$C Isotope Gradient Derived from Millimeter Transitions of CN: The Case for Galactic Chemical Evolution},
      journal = {\apj},
         year = 2005,
        month = dec,
       volume = {634},
       number = {2},
        pages = {1126-1132},
          doi = {10.1086/497123},
       adsurl = {https://ui.adsabs.harvard.edu/abs/2005ApJ...634.1126M}
}

@ARTICLE{2020Okoda,
       author = {{Okoda}, Yuki and {Oya}, Yoko and {Sakai}, Nami and {Watanabe}, Yoshimasa and {Yamamoto}, Satoshi},
        title = "{Molecular Distributions of the Protostellar Envelope and the Outflow of IRAS 15398-3359: Principal Component Analysis}",
      journal = {\apj},
         year = 2020,
        month = sep,
       volume = {900},
       number = {1},
          eid = {40},
        pages = {40},
          doi = {10.3847/1538-4357/aba51e},
archivePrefix = {arXiv},
       eprint = {2007.11106},
 primaryClass = {astro-ph.GA},
       adsurl = {https://ui.adsabs.harvard.edu/abs/2020ApJ...900...40O}
}

@ARTICLE{2023Ohashi,
       author = {{Ohashi}, Nagayoshi and {Tobin}, John J. and {J{\o}rgensen}, Jes K. and {Takakuwa}, Shigehisa and {Sheehan}, Patrick and {Aikawa}, Yuri and {Li}, Zhi-Yun and {Looney}, Leslie W. and {Williams}, Jonathan P. and {Aso}, Yusuke and {Sharma}, Rajeeb and {Sai}, Jinshi (Insa Choi) and {Yamato}, Yoshihide and {Lee}, Jeong-Eun and {Tomida}, Kengo and {Yen}, Hsi-Wei and {Encalada}, Frankie J. and {Flores}, Christian and {Gavino}, Sacha and {Kido}, Miyu and {Han}, Ilseung and {Lin}, Zhe-Yu Daniel and {Narayanan}, Suchitra and {Phuong}, Nguyen Thi and {Santamar{\'\i}a-Miranda}, Alejandro and {Thieme}, Travis J. and {van't Hoff}, Merel L.~R. and {de Gregorio-Monsalvo}, Itziar and {Koch}, Patrick M. and {Kwon}, Woojin and {Lai}, Shih-Ping and {Lee}, Chang Won and {Plunkett}, Adele and {Saigo}, Kazuya and {Hirano}, Shingo and {Lam}, Ka Ho and {Mori}, Shoji},
        title = {Early Planet Formation in Embedded Disks (eDisk). I. Overview of the Program and First Results},
      journal = {\apj},
         year = 2023,
        month = jul,
       volume = {951},
       number = {1},
          eid = {8},
        pages = {8},
          doi = {10.3847/1538-4357/acd384},
archivePrefix = {arXiv},
       eprint = {2306.15406},
 primaryClass = {astro-ph.EP},
       adsurl = {https://ui.adsabs.harvard.edu/abs/2023ApJ...951....8O}
}

@ARTICLE{2021Okoda,
       author = {{Okoda}, Yuki and {Oya}, Yoko and {Francis}, Logan and {Johnstone}, Doug and {Inutsuka}, Shu-ichiro and {Ceccarelli}, Cecilia and {Codella}, Claudio and {Chandler}, Claire and {Sakai}, Nami and {Aikawa}, Yuri and {Alves}, Felipe O. and {Balucani}, Nadia and {Bianchi}, Eleonora and {Bouvier}, Mathilde and {Caselli}, Paola and {Caux}, Emmanuel and {Charnley}, Steven and {Choudhury}, Spandan and {De Simone}, Marta and {Dulieu}, Francois and {Dur{\'a}n}, Aurora and {Evans}, Lucy and {Favre}, C{\'e}cile and {Fedele}, Davide and {Feng}, Siyi and {Fontani}, Francesco and {Hama}, Tetsuya and {Hanawa}, Tomoyuki and {Herbst}, Eric and {Hirota}, Tomoya and {Imai}, Muneaki and {Isella}, Andrea and {J{\'\i}menez-Serra}, Izaskun and {Kahane}, Claudine and {Lefloch}, Bertrand and {Loinard}, Laurent and {L{\'o}pez-Sepulcre}, Ana and {Maud}, Luke T. and {Maureira}, Mar{\'\i}a Jos{\'e} and {Menard}, Francois and {Mercimek}, Seyma and {Miotello}, Anna and {Moellenbrock}, George and {Mori}, Shoji and {Murillo}, Nadia M. and {Nakatani}, Riouhei and {Nomura}, Hideko and {Oba}, Yasuhiro and {O'Donoghue}, Ross and {Ohashi}, Satoshi and {Ospina-Zamudio}, Juan and {Pineda}, Jaime E. and {Podio}, Linda and {Rimola}, Albert and {Sakai}, Takeshi and {Segura-Cox}, Dominique and {Shirley}, Yancy and {Svoboda}, Brian and {Taquet}, Vianney and {Testi}, Leonardo and {Vastel}, Charlotte and {Viti}, Serena and {Watanabe}, Naoki and {Watanabe}, Yoshimasa and {Witzel}, Arezu and {Xue}, Ci and {Zhang}, Yichen and {Zhao}, Bo and {Yamamoto}, Satoshi},
        title = {FAUST. II. Discovery of a Secondary Outflow in IRAS 15398-3359: Variability in Outflow Direction during the Earliest Stage of Star Formation?},
      journal = {\apj},
         year = 2021,
        month = mar,
       volume = {910},
       number = {1},
          eid = {11},
        pages = {11},
          doi = {10.3847/1538-4357/abddb1},
archivePrefix = {arXiv},
       eprint = {2101.07404},
 primaryClass = {astro-ph.SR},
       adsurl = {https://ui.adsabs.harvard.edu/abs/2021ApJ...910...11O}
}

@ARTICLE{2025Okoda,
       author = {{Okoda}, Yuki and {Yang}, Yao-Lun and {Evans}, II, Neal J. and {Kim}, Jaeyeong and {Jin}, Mihwa and {Garrod}, Robin T. and {Francis}, Logan and {Johnstone}, Doug and {Ceccarelli}, Cecilia and {Codella}, Claudio and {Chandler}, Claire J. and {Yamamoto}, Satoshi and {Sakai}, Nami},
        title = "{CORINOS. III. Outflow Shocked Regions of the Low-mass Protostellar Source IRAS 15398─3359 with JWST and ALMA}",
      journal = {\apj},
         year = 2025,
        month = apr,
       volume = {982},
       number = {2},
          eid = {149},
        pages = {149},
          doi = {10.3847/1538-4357/adb83f},
archivePrefix = {arXiv},
       eprint = {2503.03050},
 primaryClass = {astro-ph.SR},
       adsurl = {https://ui.adsabs.harvard.edu/abs/2025ApJ...982..149O}
}

@ARTICLE{2014Oya,
       author = {{Oya}, Yoko and {Sakai}, Nami and {Sakai}, Takeshi and {Watanabe}, Yoshimasa and {Hirota}, Tomoya and {Lindberg}, Johan E. and {Bisschop}, Suzanne E. and {J{\o}rgensen}, Jes K. and {van Dishoeck}, Ewine F. and {Yamamoto}, Satoshi},
        title = "{A Substellar-mass Protostar and its Outflow of IRAS 15398-3359 Revealed by Subarcsecond-resolution Observations of H$_{2}$CO and CCH}",
      journal = {\apj},
         year = 2014,
        month = nov,
       volume = {795},
       number = {2},
          eid = {152},
        pages = {152},
          doi = {10.1088/0004-637X/795/2/152},
archivePrefix = {arXiv},
       eprint = {1410.5945},
 primaryClass = {astro-ph.SR},
       adsurl = {https://ui.adsabs.harvard.edu/abs/2014ApJ...795..152O}
}

@ARTICLE{2015Rieke,
       author = {{Rieke}, G.~H. and {Wright}, G.~S. and {B{\"o}ker}, T. and {Bouwman}, J. and {Colina}, L. and {Glasse}, Alistair and {Gordon}, K.~D. and {Greene}, T.~P. and {G{\"u}del}, Manuel and {Henning}, Th. and {Justtanont}, K. and {Lagage}, P.-O. and {Meixner}, M.~E. and {N{\o}rgaard-Nielsen}, H.-U. and {Ray}, T.~P. and {Ressler}, M.~E. and {van Dishoeck}, E.~F. and {Waelkens}, C.},
        title = {The Mid-Infrared Instrument for the James Webb Space Telescope, I: Introduction},
      journal = {\pasp},
         year = 2015,
        month = jul,
       volume = {127},
       number = {953},
        pages = {584},
          doi = {10.1086/682252},
archivePrefix = {arXiv},
       eprint = {1508.02294},
 primaryClass = {astro-ph.IM},
       adsurl = {https://ui.adsabs.harvard.edu/abs/2015PASP..127..584R}
}

@ARTICLE{2024Sai,
       author = {{Sai}, Jinshi and {Yen}, Hsi-Wei and {Machida}, Masahiro N. and {Ohashi}, Nagayoshi and {Aso}, Yusuke and {Maury}, Ana{\"e}lle J. and {Maret}, S{\'e}bastien},
        title = "{Multiple Outflows around a Single Protostar IRAS 15398{\ensuremath{-}}3359}",
      journal = {\apj},
         year = 2024,
        month = may,
       volume = {966},
       number = {2},
          eid = {192},
        pages = {192},
          doi = {10.3847/1538-4357/ad34b7},
archivePrefix = {arXiv},
       eprint = {2403.10857},
 primaryClass = {astro-ph.SR},
       adsurl = {https://ui.adsabs.harvard.edu/abs/2024ApJ...966..192S}
}

@ARTICLE{1977Shu,
       author = {{Shu}, F.~H.},
        title = {Self-similar collapse of isothermal spheres and star formation.},
      journal = {\apj},
         year = 1977,
        month = jun,
       volume = {214},
        pages = {488-497},
          doi = {10.1086/155274},
       adsurl = {https://ui.adsabs.harvard.edu/abs/1977ApJ...214..488S}
}

@ARTICLE{1991Shu,
       author = {{Shu}, Frank H. and {Ruden}, Steven P. and {Lada}, Charles J. and {Lizano}, Susana},
        title = {Star Formation and the Nature of Bipolar Outflows},
      journal = {\apjl},
         year = 1991,
        month = mar,
       volume = {370},
        pages = {L31},
          doi = {10.1086/185970},
       adsurl = {https://ui.adsabs.harvard.edu/abs/1991ApJ...370L..31S}
}

@ARTICLE{2023Thieme,
       author = {{Thieme}, Travis J. and {Lai}, Shih-Ping and {Ohashi}, Nagayoshi and {Tobin}, John J. and {J{\o}rgensen}, Jes K. and {Sai}, Jinshi (Insa Choi) and {Aso}, Yusuke and {Williams}, Jonathan P. and {Yamato}, Yoshihide and {Aikawa}, Yuri and {de Gregorio-Monsalvo}, Itziar and {Han}, Ilseung and {Kwon}, Woojin and {Lee}, Chang Won and {Lee}, Jeong-Eun and {Li}, Zhi-Yun and {Lin}, Zhe-Yu Daniel and {Looney}, Leslie W. and {Narayanan}, Suchitra and {Phuong}, Nguyen Thi and {Plunkett}, Adele L. and {Santamar{\'\i}a-Miranda}, Alejandro and {Sharma}, Rajeeb and {Takakuwa}, Shigehisa and {Yen}, Hsi-Wei},
        title = "{Early Planet Formation in Embedded Disks (eDisk). VIII. A Small Protostellar Disk around the Extremely Low Mass and Young Class 0 Protostar IRAS 15398-3359}",
      journal = {\apj},
         year = 2023,
        month = nov,
       volume = {958},
       number = {1},
          eid = {60},
        pages = {60},
          doi = {10.3847/1538-4357/ad003a},
archivePrefix = {arXiv},
       eprint = {2310.12453},
 primaryClass = {astro-ph.EP},
       adsurl = {https://ui.adsabs.harvard.edu/abs/2023ApJ...958...60T}
}

@ARTICLE{2021Vazzano,
       author = {{Vazzano}, M.~M. and {Fern{\'a}ndez-L{\'o}pez}, M. and {Plunkett}, A. and {de Gregorio-Monsalvo}, I. and {Santamar{\'\i}a-Miranda}, A. and {Takahashi}, S. and {Lopez}, C.},
        title = {Outflows, envelopes, and disks as evolutionary indicators in Lupus young stellar objects},
      journal = {\aap},
         year = 2021,
        month = apr,
       volume = {648},
          eid = {A41},
        pages = {A41},
          doi = {10.1051/0004-6361/202039228},
archivePrefix = {arXiv},
       eprint = {2101.05330},
 primaryClass = {astro-ph.SR},
       adsurl = {https://ui.adsabs.harvard.edu/abs/2021A&A...648A..41V}
}

@ARTICLE{2023Wright,
       author = {{Wright}, Gillian S. and {Rieke}, George H. and {Glasse}, Alistair and {Ressler}, Michael and {Garc{\'\i}a Mar{\'\i}n}, Macarena and {Aguilar}, Jonathan and {Alberts}, Stacey and {{\'A}lvarez-M{\'a}rquez}, Javier and {Argyriou}, Ioannis and {Banks}, Kimberly and {Baudoz}, Pierre and {Boccaletti}, Anthony and {Bouchet}, Patrice and {Bouwman}, Jeroen and {Brandl}, Bernard R. and {Breda}, David and {Bright}, Stacey and {Cale}, Steven and {Colina}, Luis and {Cossou}, Christophe and {Coulais}, Alain and {Cracraft}, Misty and {De Meester}, Wim and {Dicken}, Daniel and {Engesser}, Michael and {Etxaluze}, Mireya and {Fox}, Ori D. and {Friedman}, Scott and {Fu}, Henry and {Gasman}, Danny and {G{\'a}sp{\'a}r}, Andr{\'a}s and {Gastaud}, Ren{\'e} and {Geers}, Vincent and {Glauser}, Adrian Michael and {Gordon}, Karl D. and {Greene}, Thomas and {Greve}, Thomas R. and {Grundy}, Timothy and {G{\"u}del}, Manuel and {Guillard}, Pierre and {Haderlein}, Peter and {Hashimoto}, Ryan and {Henning}, Thomas and {Hines}, Dean and {Holler}, Bryan and {Detre}, {\"O}rs Hunor and {Jahromi}, Amir and {James}, Bryan and {Jones}, Olivia C. and {Justtanont}, Kay and {Kavanagh}, Patrick and {Kendrew}, Sarah and {Klaassen}, Pamela and {Krause}, Oliver and {Labiano}, Alvaro and {Lagage}, Pierre-Olivier and {Lambros}, Scott and {Larson}, Kirsten and {Law}, David and {Lee}, David and {Libralato}, Mattia and {Lorenzo Alverez}, Jose and {Meixner}, Margaret and {Morrison}, Jane and {Mueller}, Migo and {Murray}, Katherine and {Mycroft}, Matthew and {Myers}, Richard and {Nayak}, Omnarayani and {Naylor}, Bret and {Nickson}, Bryony and {Noriega-Crespo}, Alberto and {{\"O}stlin}, G{\"o}ran and {O'Sullivan}, Brian and {Ottens}, Richard and {Patapis}, Polychronis and {Penanen}, Konstantin and {Pietraszkiewicz}, Martin and {Ray}, Tom and {Regan}, Michael and {Roteliuk}, Anthony and {Royer}, Pierre and {Samara-Ratna}, Piyal and {Samuelson}, Bridget and {Sargent}, Beth A. and {Scheithauer}, Silvia and {Schneider}, Analyn and {Schreiber}, J{\"u}rgen and {Shaughnessy}, Bryan and {Sheehan}, Evan and {Shivaei}, Irene and {Sloan}, G.~C. and {Tamas}, Laszlo and {Teague}, Kelly and {Temim}, Tea and {Tikkanen}, Tuomo and {Tustain}, Samuel and {van Dishoeck}, Ewine F. and {Vandenbussche}, Bart and {Weilert}, Mark and {Whitehouse}, Paul and {Wolff}, Schuyler},
        title = {The Mid-infrared Instrument for JWST and Its In-flight Performance},
      journal = {\pasp},
         year = 2023,
        month = apr,
       volume = {135},
       number = {1046},
          eid = {048003},
        pages = {048003},
          doi = {10.1088/1538-3873/acbe66},
       adsurl = {https://ui.adsabs.harvard.edu/abs/2023PASP..135d8003W}
}

@ARTICLE{2022Yang,
       author = {{Yang}, Yao-Lun and {Green}, Joel D. and {Pontoppidan}, Klaus M. and {Bergner}, Jennifer B. and {Cleeves}, L. Ilsedore and {Evans}, II, Neal J. and {Garrod}, Robin T. and {Jin}, Miwha and {Kim}, Chul Hwan and {Kim}, Jaeyeong and {Lee}, Jeong-Eun and {Sakai}, Nami and {Shingledecker}, Christopher N. and {Shope}, Brielle and {Tobin}, John J. and {van Dishoeck}, Ewine F.},
        title = "{CORINOS. I. JWST/MIRI Spectroscopy and Imaging of a Class 0 Protostar IRAS 15398─3359}",
      journal = {\apjl},
         year = 2022,
        month = dec,
       volume = {941},
       number = {1},
          eid = {L13},
        pages = {L13},
          doi = {10.3847/2041-8213/aca289},
archivePrefix = {arXiv},
       eprint = {2208.10673},
 primaryClass = {astro-ph.SR},
       adsurl = {https://ui.adsabs.harvard.edu/abs/2022ApJ...941L..13Y}
}

@ARTICLE{2017Yen,
       author = {{Yen}, Hsi-Wei and {Koch}, Patrick M. and {Takakuwa}, Shigehisa and {Krasnopolsky}, Ruben and {Ohashi}, Nagayoshi and {Aso}, Yusuke},
        title = "{Signs of Early-stage Disk Growth Revealed with ALMA}",
      journal = {\apj},
         year = 2017,
        month = jan,
       volume = {834},
       number = {2},
          eid = {178},
        pages = {178},
          doi = {10.3847/1538-4357/834/2/178},
archivePrefix = {arXiv},
       eprint = {1611.08416},
 primaryClass = {astro-ph.SR},
       adsurl = {https://ui.adsabs.harvard.edu/abs/2017ApJ...834..178Y}
}

@ARTICLE{2020FernandezLopez,
       author = {{Fern{\'a}ndez-L{\'o}pez}, Manuel and {Zapata}, Luis A. and {Rodr{\'\i}guez}, Luis F. and {Vazzano}, Mar{\'\i}a M. and {Guzm{\'a}n}, Andr{\'e}s E. and {L{\'o}pez}, Rosario},
        title = {A Ringed Pole-on Outflow from DO Tauri Revealed by ALMA},
      journal = {\aj},
         year = 2020,
        month = apr,
       volume = {159},
       number = {4},
          eid = {171},
        pages = {171},
          doi = {10.3847/1538-3881/ab7a10},
archivePrefix = {arXiv},
       eprint = {2003.01148},
 primaryClass = {astro-ph.SR},
       adsurl = {https://ui.adsabs.harvard.edu/abs/2020AJ....159..171F}
}

@ARTICLE{1993Raga,
       author = {{Raga}, A. and {Cabrit}, S.},
        title = {Molecular outflows entrained by jet bowshocks.},
      journal = {\aap},
         year = 1993,
        month = oct,
       volume = {278},
        pages = {267-278},
       adsurl = {https://ui.adsabs.harvard.edu/abs/1993A&A...278..267R}
}

@ARTICLE{2019Zhang,
       author = {{Zhang}, Yichen and {Arce}, H{\'e}ctor G. and {Mardones}, Diego and {Cabrit}, Sylvie and {Dunham}, Michael M. and {Garay}, Guido and {Noriega-Crespo}, Alberto and {Offner}, Stella S.~R. and {Raga}, Alejandro C. and {Corder}, Stuartt A.},
        title = {An Episodic Wide-angle Outflow in HH 46/47},
      journal = {\apj},
         year = 2019,
        month = sep,
       volume = {883},
       number = {1},
          eid = {1},
        pages = {1},
          doi = {10.3847/1538-4357/ab3850},
archivePrefix = {arXiv},
       eprint = {1908.00689},
 primaryClass = {astro-ph.GA},
       adsurl = {https://ui.adsabs.harvard.edu/abs/2019ApJ...883....1Z}
}

@ARTICLE{2008Arce,
       author = {{Arce}, H{\'e}ctor G. and {Santiago-Garc{\'\i}a}, Joaqu{\'\i}n and {J{\o}rgensen}, Jes K. and {Tafalla}, Mario and {Bachiller}, Rafael},
        title = {Complex Molecules in the L1157 Molecular Outflow},
      journal = {\apjl},
         year = 2008,
        month = jul,
       volume = {681},
       number = {1},
        pages = {L21},
          doi = {10.1086/590110},
archivePrefix = {arXiv},
       eprint = {0805.2550},
 primaryClass = {astro-ph},
       adsurl = {https://ui.adsabs.harvard.edu/abs/2008ApJ...681L..21A}
}

@ARTICLE{2016Burkhardt,
       author = {{Burkhardt}, Andrew M. and {Dollhopf}, Niklaus M. and {Corby}, Joanna F. and {Carroll}, P. Brandon and {Shingledecker}, Christopher N. and {Loomis}, Ryan A. and {Booth}, Shawn Thomas and {Blake}, Geoffrey A. and {Herbst}, Eric and {Remijan}, Anthony J. and {McGuire}, Brett A.},
        title = {CSO and CARMA Observations of L1157. II. Chemical Complexity in the Shocked Outflow},
      journal = {\apj},
         year = 2016,
        month = aug,
       volume = {827},
       number = {1},
          eid = {21},
        pages = {21},
          doi = {10.3847/0004-637X/827/1/21},
archivePrefix = {arXiv},
       eprint = {1605.09707},
 primaryClass = {astro-ph.GA},
       adsurl = {https://ui.adsabs.harvard.edu/abs/2016ApJ...827...21B}
}

@ARTICLE{2025Codella,
       author = {{Codella}, C. and {Bianchi}, E. and {Podio}, L. and {De Simone}, M. and {L{\'o}pez-Sepulcre}, A. and {Ceccarelli}, C. and {Caselli}, P.},
        title = "{SOLIS: XIX. The chemically rich SVS13-B protostellar jet}",
      journal = {\aap},
         year = 2025,
        month = apr,
       volume = {696},
          eid = {A219},
        pages = {A219},
          doi = {10.1051/0004-6361/202453621},
archivePrefix = {arXiv},
       eprint = {2504.18255},
 primaryClass = {astro-ph.SR},
       adsurl = {https://ui.adsabs.harvard.edu/abs/2025A&A...696A.219C}
}

@ARTICLE{2005Jimenez-Serra,
       author = {{Jim{\'e}nez-Serra}, I. and {Mart{\'\i}n-Pintado}, J. and {Rodr{\'\i}guez-Franco}, A. and {Mart{\'\i}n}, S.},
        title = {Grain Evolution across the Shocks in the L1448-mm Outflow},
      journal = {\apjl},
         year = 2005,
        month = jul,
       volume = {627},
       number = {2},
        pages = {L121-L124},
          doi = {10.1086/432467},
archivePrefix = {arXiv},
       eprint = {astro-ph/0506408},
 primaryClass = {astro-ph},
       adsurl = {https://ui.adsabs.harvard.edu/abs/2005ApJ...627L.121J}
}

@ARTICLE{2026Kim,
       author = {{Kim}, Chul-Hwan and {Lee}, Jeong-Eun and {Johnstone}, Doug and {Herczeg}, Gregory J. and {Lee}, Chin-Fei and {Francis}, Logan and {Sheehan}, Patrick D.},
        title = {Direct evidence for magnetohydrodynamic disk winds driving rotating outflows in protostar HOPS 358},
      journal = {Nature Communications},
         year = 2026,
        month = apr,
       volume = {17},
       number = {1},
          eid = {2957},
        pages = {2957},
          doi = {10.1038/s41467-026-71142-3},
       adsurl = {https://ui.adsabs.harvard.edu/abs/2026NatCo..17.2957K}
}

@ARTICLE{1993PineaudesForets,
       author = {{Pineau des Forets}, G. and {Roueff}, E. and {Schilke}, P. and {Flower}, D.~R.},
        title = {Sulphur-bearing molecules as tracers of shocks in interstellar clouds},
      journal = {\mnras},
         year = 1993,
        month = jun,
       volume = {262},
       number = {4},
        pages = {915-928},
          doi = {10.1093/mnras/262.4.915},
       adsurl = {https://ui.adsabs.harvard.edu/abs/1993MNRAS.262..915P}
}

@ARTICLE{2014Podio,
       author = {{Podio}, L. and {Lefloch}, B. and {Ceccarelli}, C. and {Codella}, C. and {Bachiller}, R.},
        title = {Molecular ions in the protostellar shock L1157-B1},
      journal = {\aap},
         year = 2014,
        month = may,
       volume = {565},
          eid = {A64},
        pages = {A64},
          doi = {10.1051/0004-6361/201322928},
archivePrefix = {arXiv},
       eprint = {1402.2329},
 primaryClass = {astro-ph.GA},
       adsurl = {https://ui.adsabs.harvard.edu/abs/2014A&A...565A..64P}
}

@ARTICLE{2010Tafalla,
       author = {{Tafalla}, M. and {Santiago-Garc{\'\i}a}, J. and {Hacar}, A. and {Bachiller}, R.},
        title = {A molecular survey of outflow gas: velocity-dependent shock chemistry and the peculiar composition of the EHV gas},
      journal = {\aap},
         year = 2010,
        month = nov,
       volume = {522},
          eid = {A91},
        pages = {A91},
          doi = {10.1051/0004-6361/201015158},
archivePrefix = {arXiv},
       eprint = {1007.4549},
 primaryClass = {astro-ph.GA},
       adsurl = {https://ui.adsabs.harvard.edu/abs/2010A&A...522A..91T}
}

@ARTICLE{2026Tafalla,
       author = {{Tafalla}, M. and {Johnstone}, D. and {Santiago-Garc{\'\i}a}, J. and {Zhang}, Q. and {Shang}, H. and {Lee}, C.-F.},
        title = {Interaction of the central jet with the surrounding gas in the protostellar outflow from IRAS 04166+2706},
      journal = {\aap},
         year = 2026,
        month = feb,
       volume = {706},
          eid = {A248},
        pages = {A248},
          doi = {10.1051/0004-6361/202555924},
archivePrefix = {arXiv},
       eprint = {2601.05310},
 primaryClass = {astro-ph.GA},
       adsurl = {https://ui.adsabs.harvard.edu/abs/2026A&A...706A.248T}
}

@ARTICLE{1994Tielens,
       author = {{Tielens}, A.~G.~G.~M. and {McKee}, C.~F. and {Seab}, C.~G. and {Hollenbach}, D.~J.},
        title = {The Physics of Grain-Grain Collisions and Gas-Grain Sputtering in Interstellar Shocks},
      journal = {\apj},
         year = 1994,
        month = aug,
       volume = {431},
        pages = {321},
          doi = {10.1086/174488},
       adsurl = {https://ui.adsabs.harvard.edu/abs/1994ApJ...431..321T}
}

@ARTICLE{2004Watanabe,
       author = {{Watanabe}, Naoki and {Nagaoka}, Akihiro and {Shiraki}, Takahiro and {Kouchi}, Akira},
        title = {Hydrogenation of CO on Pure Solid CO and CO-H$_{2}$O Mixed Ice},
      journal = {\apj},
         year = 2004,
        month = nov,
       volume = {616},
       number = {1},
        pages = {638-642},
          doi = {10.1086/424815},
       adsurl = {https://ui.adsabs.harvard.edu/abs/2004ApJ...616..638W}
}

@ARTICLE{2017Hirota,
       author = {{Hirota}, Tomoya and {Machida}, Masahiro N. and {Matsushita}, Yuko and {Motogi}, Kazuhito and {Matsumoto}, Naoko and {Kim}, Mi Kyoung and {Burns}, Ross A. and {Honma}, Mareki},
        title = {Disk-driven rotating bipolar outflow in Orion Source I},
      journal = {Nature Astronomy},
         year = 2017,
        month = jul,
       volume = {1},
          eid = {0146},
        pages = {0146},
          doi = {10.1038/s41550-017-0146},
archivePrefix = {arXiv},
       eprint = {1712.04606},
 primaryClass = {astro-ph.SR},
       adsurl = {https://ui.adsabs.harvard.edu/abs/2017NatAs...1E.146H}
}

@ARTICLE{2019Machida,
       author = {{Machida}, Masahiro N. and {Basu}, Shantanu},
        title = {The First Two Thousand Years of Star Formation},
      journal = {\apj},
         year = 2019,
        month = may,
       volume = {876},
       number = {2},
          eid = {149},
        pages = {149},
          doi = {10.3847/1538-4357/ab18a7},
archivePrefix = {arXiv},
       eprint = {1904.04424},
 primaryClass = {astro-ph.SR},
       adsurl = {https://ui.adsabs.harvard.edu/abs/2019ApJ...876..149M}
}

@ARTICLE{2020Machida,
       author = {{Machida}, Masahiro N. and {Hirano}, Shingo and {Kitta}, Hideyuki},
        title = {Misalignment of magnetic fields, outflows, and discs in star-forming clouds},
      journal = {\mnras},
         year = 2020,
        month = jan,
       volume = {491},
       number = {2},
        pages = {2180-2197},
          doi = {10.1093/mnras/stz3159},
archivePrefix = {arXiv},
       eprint = {1911.03851},
 primaryClass = {astro-ph.SR},
       adsurl = {https://ui.adsabs.harvard.edu/abs/2020MNRAS.491.2180M}
}

@ARTICLE{2017Tabone,
       author = {{Tabone}, B. and {Cabrit}, S. and {Bianchi}, E. and {Ferreira}, J. and {Pineau des For{\^e}ts}, G. and {Codella}, C. and {Gusdorf}, A. and {Gueth}, F. and {Podio}, L. and {Chapillon}, E.},
        title = {ALMA discovery of a rotating SO/SO$_{2}$ flow in HH212. A possible MHD disk wind?},
      journal = {\aap},
         year = 2017,
        month = nov,
       volume = {607},
          eid = {L6},
        pages = {L6},
          doi = {10.1051/0004-6361/201731691},
archivePrefix = {arXiv},
       eprint = {1710.01401},
 primaryClass = {astro-ph.SR},
       adsurl = {https://ui.adsabs.harvard.edu/abs/2017A&A...607L...6T}
}

@INPROCEEDINGS{2023Tsukamoto,
       author = {{Tsukamoto}, Y. and {Maury}, A. and {Commercon}, B. and {Alves}, F.~O. and {Cox}, E.~G. and {Sakai}, N. and {Ray}, T. and {Zhao}, B. and {Machida}, M.~N.},
        title = {The Role of Magnetic Fields in the Formation of Protostars, Disks, and Outflows},
    booktitle = {Protostars and Planets VII},
         year = 2023,
       editor = {{Inutsuka}, S. and {Aikawa}, Y. and {Muto}, T. and {Tomida}, K. and {Tamura}, M.},
       series = {Astronomical Society of the Pacific Conference Series},
       volume = {534},
        month = jul,
        pages = {317},
          doi = {10.48550/arXiv.2209.13765},
archivePrefix = {arXiv},
       eprint = {2209.13765},
 primaryClass = {astro-ph.SR},
       adsurl = {https://ui.adsabs.harvard.edu/abs/2023ASPC..534..317T}
}

@ARTICLE{2018Birnstiel,
       author = {{Birnstiel}, Tilman and {Dullemond}, Cornelis P. and {Zhu}, Zhaohuan and {Andrews}, Sean M. and {Bai}, Xue-Ning and {Wilner}, David J. and {Carpenter}, John M. and {Huang}, Jane and {Isella}, Andrea and {Benisty}, Myriam and {P{\'e}rez}, Laura M. and {Zhang}, Shangjia},
        title = {The Disk Substructures at High Angular Resolution Project (DSHARP). V. Interpreting ALMA Maps of Protoplanetary Disks in Terms of a Dust Model},
      journal = {\apjl},
         year = 2018,
        month = dec,
       volume = {869},
       number = {2},
          eid = {L45},
        pages = {L45},
          doi = {10.3847/2041-8213/aaf743},
archivePrefix = {arXiv},
       eprint = {1812.04043},
 primaryClass = {astro-ph.SR},
       adsurl = {https://ui.adsabs.harvard.edu/abs/2018ApJ...869L..45B}
}

@MISC{2012Dullemond,
       author = {{Dullemond}, C.~P. and {Juhasz}, A. and {Pohl}, A. and {Sereshti}, F. and {Shetty}, R. and {Peters}, T. and {Commercon}, B. and {Flock}, M.},
        title = {RADMC-3D: A multi-purpose radiative transfer tool},
 howpublished = {Astrophysics Source Code Library, record ascl:1202.015},
         year = 2012,
        month = feb,
          eid = {ascl:1202.015},
archivePrefix = {ascl},
       eprint = {1202.015},
       adsurl = {https://ui.adsabs.harvard.edu/abs/2012ascl.soft02015D}
}

@MISC{2016Ginsburg,
       author = {{Ginsburg}, Adam and {Robitaille}, Thomas and {Beaumont}, Chris},
        title = {pvextractor: Position-Velocity Diagram Extractor},
 howpublished = {Astrophysics Source Code Library, record ascl:1608.010},
         year = 2016,
        month = aug,
          eid = {ascl:1608.010},
archivePrefix = {ascl},
       eprint = {1608.010},
       adsurl = {https://ui.adsabs.harvard.edu/abs/2016ascl.soft08010G}
}

@MISC{2019Ginsburg,
       author = {{Ginsburg}, Adam and {Koch}, Eric and {Robitaille}, Thomas and {Beaumont}, Chris and {Adamginsburg} and {Sip{\H{o}}cz}, Brigitta and {ZuHone}, John and {Patra}, Sushobhana and {Jones}, Craig and {Lim}, P.~L. and {Stern}, Kris and {Rosolowsky}, Erik and {Earl}, Nicholas and {De Val-Borro}, Miguel and {Jrobbfed} and {Shuokong} and {Kepley}, Amanda and {Sokolov}, Vlas and {Badger}, The Gitter and {Maret}, S{\'e}bastien and {Garrido}, Juli{\'a}n and {Booker}, Joseph and {Tollerud}, Erik},
        title = {radio-astro-tools/spectral-cube: Release v0.4.5},
         year = 2019,
        month = nov,
          eid = {10.5281/zenodo.3558614},
          doi = {10.5281/zenodo.3558614},
      version = {v0.4.5},
    publisher = {Zenodo},
       adsurl = {https://ui.adsabs.harvard.edu/abs/2019zndo...3558614G}
}

@ARTICLE{2020Harris,
       author = {{Harris}, Charles R. and {Millman}, K. Jarrod and {van der Walt}, St{\'e}fan J. and {Gommers}, Ralf and {Virtanen}, Pauli and {Cournapeau}, David and {Wieser}, Eric and {Taylor}, Julian and {Berg}, Sebastian and {Smith}, Nathaniel J. and {Kern}, Robert and {Picus}, Matti and {Hoyer}, Stephan and {van Kerkwijk}, Marten H. and {Brett}, Matthew and {Haldane}, Allan and {del R{\'\i}o}, Jaime Fern{\'a}ndez and {Wiebe}, Mark and {Peterson}, Pearu and {G{\'e}rard-Marchant}, Pierre and {Sheppard}, Kevin and {Reddy}, Tyler and {Weckesser}, Warren and {Abbasi}, Hameer and {Gohlke}, Christoph and {Oliphant}, Travis E.},
        title = {Array programming with NumPy},
      journal = {\nat},
         year = 2020,
        month = sep,
       volume = {585},
       number = {7825},
        pages = {357-362},
          doi = {10.1038/s41586-020-2649-2},
archivePrefix = {arXiv},
       eprint = {2006.10256},
 primaryClass = {cs.MS},
       adsurl = {https://ui.adsabs.harvard.edu/abs/2020Natur.585..357H}
}

@ARTICLE{2007Hunter,
       author = {{Hunter}, John D.},
        title = {Matplotlib: A 2D Graphics Environment},
      journal = {Computing in Science and Engineering},
         year = 2007,
        month = jan,
       volume = {9},
       number = {3},
        pages = {90-95},
          doi = {10.1109/MCSE.2007.55},
       adsurl = {https://ui.adsabs.harvard.edu/abs/2007CSE.....9...90H}
}

@software{2010McKinney,
       author = {{McKinney}, Wes},
        title = {Data Structures for Statistical Computing in Python},
 howpublished = {Proceedings of the 9th Python in Science Conference},
         year = 2010,
        month = jan,
          doi = {10.25080/Majora-92bf1922-00a},
       adsurl = {https://ui.adsabs.harvard.edu/abs/2010scpy.soft.....M}
}

@INPROCEEDINGS{2007McMullin,
       author = {{McMullin}, J.~P. and {Waters}, B. and {Schiebel}, D. and {Young}, W. and {Golap}, K.},
        title = {CASA Architecture and Applications},
    booktitle = {Astronomical Data Analysis Software and Systems XVI},
         year = 2007,
       editor = {{Shaw}, R.~A. and {Hill}, F. and {Bell}, D.~J.},
       series = {Astronomical Society of the Pacific Conference Series},
       volume = {376},
        month = oct,
        pages = {127},
       adsurl = {https://ui.adsabs.harvard.edu/abs/2007ASPC..376..127M}
}

@ARTICLE{2020Virtanen,
       author = {{Virtanen}, Pauli and {Gommers}, Ralf and {Oliphant}, Travis E. and {Haberland}, Matt and {Reddy}, Tyler and {Cournapeau}, David and {Burovski}, Evgeni and {Peterson}, Pearu and {Weckesser}, Warren and {Bright}, Jonathan and {van der Walt}, St{\'e}fan J. and {Brett}, Matthew and {Wilson}, Joshua and {Millman}, K. Jarrod and {Mayorov}, Nikolay and {Nelson}, Andrew R.~J. and {Jones}, Eric and {Kern}, Robert and {Larson}, Eric and {Carey}, C.~J. and {Polat}, {\.I}lhan and {Feng}, Yu and {Moore}, Eric W. and {VanderPlas}, Jake and {Laxalde}, Denis and {Perktold}, Josef and {Cimrman}, Robert and {Henriksen}, Ian and {Quintero}, E.~A. and {Harris}, Charles R. and {Archibald}, Anne M. and {Ribeiro}, Ant{\^o}nio H. and {Pedregosa}, Fabian and {van Mulbregt}, Paul and {SciPy 1.  0 Contributors}},
        title = {SciPy 1.0: fundamental algorithms for scientific computing in Python},
      journal = {Nature Methods},
         year = 2020,
        month = feb,
       volume = {17},
        pages = {261-272},
          doi = {10.1038/s41592-019-0686-2},
archivePrefix = {arXiv},
       eprint = {1907.10121},
 primaryClass = {cs.MS},
       adsurl = {https://ui.adsabs.harvard.edu/abs/2020NaMet..17..261V}
}

@ARTICLE{2017Alves,
       author = {{Alves}, F.~O. and {Girart}, J.~M. and {Caselli}, P. and {Franco}, G.~A.~P. and {Zhao}, B. and {Vlemmings}, W.~H.~T. and {Evans}, M.~G. and {Ricci}, L.},
        title = {Molecular outflow launched beyond the disk edge},
      journal = {\aap},
         year = 2017,
        month = jul,
       volume = {603},
          eid = {L3},
        pages = {L3},
          doi = {10.1051/0004-6361/201731077},
archivePrefix = {arXiv},
       eprint = {1707.01501},
 primaryClass = {astro-ph.SR},
       adsurl = {https://ui.adsabs.harvard.edu/abs/2017A&A...603L...3A}
}

@ARTICLE{2022deValon,
       author = {{de Valon}, A. and {Dougados}, C. and {Cabrit}, S. and {Louvet}, F. and {Zapata}, L.~A. and {Mardones}, D.},
        title = {Modeling the CO outflow in DG Tauri B: Swept-up shells versus perturbed MHD disk wind},
      journal = {\aap},
         year = 2022,
        month = dec,
       volume = {668},
          eid = {A78},
        pages = {A78},
          doi = {10.1051/0004-6361/202141316},
archivePrefix = {arXiv},
       eprint = {2212.00514},
 primaryClass = {astro-ph.SR},
       adsurl = {https://ui.adsabs.harvard.edu/abs/2022A&A...668A..78D}
}

@ARTICLE{2012Dunham,
       author = {{Dunham}, Michael M. and {Vorobyov}, Eduard I.},
        title = {Resolving the Luminosity Problem in Low-mass Star Formation},
      journal = {\apj},
         year = 2012,
        month = mar,
       volume = {747},
       number = {1},
          eid = {52},
        pages = {52},
          doi = {10.1088/0004-637X/747/1/52},
archivePrefix = {arXiv},
       eprint = {1112.4789},
 primaryClass = {astro-ph.GA},
       adsurl = {https://ui.adsabs.harvard.edu/abs/2012ApJ...747...52D}
}

@ARTICLE{2025Sharma,
       author = {{Sharma}, Rajeeb and {J{\o}rgensen}, Jes K. and {van't Hoff}, Merel L.~R. and {Lee}, Jeong-Eun and {Aikawa}, Yuri and {Gavino}, Sacha and {Yang}, Yao-Lun and {Ohashi}, Nagayoshi and {Tobin}, John J. and {Koch}, Patrick M. and {Li}, Zhi-Yun and {Looney}, Leslie W. and {Narang}, Mayank and {Narayanan}, Suchitra and {Thieme}, Travis J.},
        title = {Early Planet Formation in Embedded Disks (eDisk): XX. Constraining the chemical tracers of young protostellar sources},
      journal = {\aap},
         year = 2025,
        month = dec,
       volume = {704},
          eid = {A133},
        pages = {A133},
          doi = {10.1051/0004-6361/202553721},
archivePrefix = {arXiv},
       eprint = {2509.19158},
 primaryClass = {astro-ph.EP},
       adsurl = {https://ui.adsabs.harvard.edu/abs/2025A&A...704A.133S}
}

@ARTICLE{2024Takakuwa,
       author = {{Takakuwa}, Shigehisa and {Saigo}, Kazuya and {Kido}, Miyu and {Ohashi}, Nagayoshi and {Tobin}, John J. and {J{\o}rgensen}, Jes K. and {Aikawa}, Yuri and {Aso}, Yusuke and {Gavino}, Sacha and {Han}, Ilseung and {Koch}, Patrick M. and {Kwon}, Woojin and {Lee}, Chang Won and {Lee}, Jeong-Eun and {Li}, Zhi-Yun and {Lin}, Zhe-Yu Daniel and {Looney}, Leslie W. and {Mori}, Shoji and {Sai}, Jinshi, Jinshi (Insa Choi) and {Sharma}, Rajeeb and {Sheehan}, Patrick D. and {Tomida}, Kengo and {Williams}, Jonathan P. and {Yamato}, Yoshihide and {Yen}, Hsi-Wei},
        title = {Early Planet Formation in Embedded Disks (eDisk). XIV. Flared Dust Distribution and Viscous Accretion Heating of the Disk around R CrA IRS 7B-a},
      journal = {\apj},
         year = 2024,
        month = mar,
       volume = {964},
       number = {1},
          eid = {24},
        pages = {24},
          doi = {10.3847/1538-4357/ad1f57},
archivePrefix = {arXiv},
       eprint = {2401.08722},
 primaryClass = {astro-ph.EP},
       adsurl = {https://ui.adsabs.harvard.edu/abs/2024ApJ...964...24T}
}

@ARTICLE{2012Lefloch,
       author = {{Lefloch}, B. and {Cabrit}, S. and {Busquet}, G. and {Codella}, C. and {Ceccarelli}, C. and {Cernicharo}, J. and {Pardo}, J.~R. and {Benedettini}, M. and {Lis}, D.~C. and {Nisini}, B.},
        title = {The CHESS Survey of the L1157-B1 Shock Region: CO Spectral Signatures of Jet-driven Bow Shocks},
      journal = {\apjl},
         year = 2012,
        month = oct,
       volume = {757},
       number = {2},
          eid = {L25},
        pages = {L25},
          doi = {10.1088/2041-8205/757/2/L25},
archivePrefix = {arXiv},
       eprint = {1208.4140},
 primaryClass = {astro-ph.SR},
       adsurl = {https://ui.adsabs.harvard.edu/abs/2012ApJ...757L..25L}
}

@ARTICLE{2025Liu,
       author = {{Liu}, Chun-Fan and {Shang}, Hsien and {Johnstone}, Doug and {Ai}, Tsung-Han and {Lee}, Tsz Ming and {Krasnopolsky}, Ruben and {Hirano}, Naomi and {Dutta}, Somnath and {Hsu}, Shih-Ying and {L{\'o}pez-V{\'a}zquez}, Jes{\'u}s Alejandro and {Liu}, Sheng-Yuan and {Liu}, Tie and {Tatematsu}, Ken'ichi and {Zhang}, Qizhou and {Rawlings}, Mark G. and {Eden}, David and {Ren}, Zhiyuan and {Sanhueza}, Patricio and {Kwon}, Woojin and {Lee}, Chang Won and {Kuan}, Yi-Jehng and {Bandopadhyay}, Somdeb and {V{\"a}is{\"a}l{\"a}}, Miikka S. and {Lee}, Chin-Fei and {Das}, Indrani},
        title = {ALMA Survey of Orion Planck Galactic Cold Clumps (ALMASOP): Nested Morphological and Kinematic Structures of Outflows Revealed in SiO and CO Emission},
      journal = {\apj},
         year = 2025,
        month = jan,
       volume = {979},
       number = {1},
          eid = {17},
        pages = {17},
          doi = {10.3847/1538-4357/ad9275},
archivePrefix = {arXiv},
       eprint = {2411.08827},
 primaryClass = {astro-ph.SR},
       adsurl = {https://ui.adsabs.harvard.edu/abs/2025ApJ...979...17L}
}

@ARTICLE{2018Louvet,
       author = {{Louvet}, F. and {Dougados}, C. and {Cabrit}, S. and {Mardones}, D. and {M{\'e}nard}, F. and {Tabone}, B. and {Pinte}, C. and {Dent}, W.~R.~F.},
        title = {The HH30 edge-on T Tauri star. A rotating and precessing monopolar outflow scrutinized by ALMA},
      journal = {\aap},
         year = 2018,
        month = oct,
       volume = {618},
          eid = {A120},
        pages = {A120},
          doi = {10.1051/0004-6361/201731733},
archivePrefix = {arXiv},
       eprint = {1808.03285},
 primaryClass = {astro-ph.GA},
       adsurl = {https://ui.adsabs.harvard.edu/abs/2018A&A...618A.120L}
}

@ARTICLE{2024LopezVazquez,
       author = {{L{\'o}pez-V{\'a}zquez}, J.~A. and {Lee}, Chin-Fei and {Fern{\'a}ndez-L{\'o}pez}, M. and {Louvet}, Fabien and {Guerra-Alvarado}, O. and {Zapata}, Luis A.},
        title = {Multiple Shells Driven by Disk Winds: ALMA Observations in the HH 30 Outflow},
      journal = {\apj},
         year = 2024,
        month = feb,
       volume = {962},
       number = {1},
          eid = {28},
        pages = {28},
          doi = {10.3847/1538-4357/ad132a},
archivePrefix = {arXiv},
       eprint = {2312.03272},
 primaryClass = {astro-ph.SR},
       adsurl = {https://ui.adsabs.harvard.edu/abs/2024ApJ...962...28L}
}

@ARTICLE{2025Mercimek,
       author = {{Mercimek}, S. and {Codella}, C. and {Podio}, L. and {Caselli}, P. and {Chandler}, C.~J. and {Chahine}, L. and {Ohashi}, S. and {Sabatini}, G. and {Loinard}, L. and {Johnstone}, D. and {Bianchi}, E. and {Zhang}, Y. and {De Simone}, M. and {Ceccarelli}, C. and {Sakai}, N. and {Yamamoto}, S.},
        title = {FAUST-XXII. Deuteration in the VLA1623-2417 protostellar hot-corinos, cavities, and streamers},
      journal = {\mnras},
         year = 2025,
        month = apr,
       volume = {538},
       number = {3},
        pages = {1481-1500},
          doi = {10.1093/mnras/staf367},
archivePrefix = {arXiv},
       eprint = {2502.18675},
 primaryClass = {astro-ph.SR},
       adsurl = {https://ui.adsabs.harvard.edu/abs/2025MNRAS.538.1481M}
}

@ARTICLE{2012Panoglou,
       author = {{Panoglou}, D. and {Cabrit}, S. and {Pineau Des For{\^e}ts}, G. and {Garcia}, P.~J.~V. and {Ferreira}, J. and {Casse}, F.},
        title = "{Molecule survival in magnetized protostellar disk winds. I. Chemical model and first results}",
      journal = {\aap},
         year = 2012,
        month = feb,
       volume = {538},
          eid = {A2},
        pages = {A2},
          doi = {10.1051/0004-6361/200912861},
archivePrefix = {arXiv},
       eprint = {1112.3248},
 primaryClass = {astro-ph.SR},
       adsurl = {https://ui.adsabs.harvard.edu/abs/2012A&A...538A...2P}
}

@ARTICLE{2000Casse,
       author = {{Casse}, Fabien and {Ferreira}, Jonathan},
        title = {Magnetized accretion-ejection structures. IV. Magnetically-driven jets from resistive, viscous, Keplerian discs},
      journal = {\aap},
         year = 2000,
        month = jan,
       volume = {353},
        pages = {1115-1128},
          doi = {10.48550/arXiv.astro-ph/9911471},
archivePrefix = {arXiv},
       eprint = {astro-ph/9911471},
 primaryClass = {astro-ph},
       adsurl = {https://ui.adsabs.harvard.edu/abs/2000A&A...353.1115C}
}

@ARTICLE{2000Shirley,
       author = {{Shirley}, Yancy L. and {Evans}, II, Neal J. and {Rawlings}, Jonathan M.~C. and {Gregersen}, Erik M.},
        title = {Tracing the Mass during Low-Mass Star Formation. I. Submillimeter Continuum Observations},
      journal = {\apjs},
         year = 2000,
        month = nov,
       volume = {131},
       number = {1},
        pages = {249-271},
          doi = {10.1086/317358},
archivePrefix = {arXiv},
       eprint = {astro-ph/0006183},
 primaryClass = {astro-ph},
       adsurl = {https://ui.adsabs.harvard.edu/abs/2000ApJS..131..249S}
}

@ARTICLE{2023Shang,
       author = {{Shang}, Hsien and {Liu}, Chun-Fan and {Krasnopolsky}, Ruben and {Wang}, Liang-Yao},
        title = {A Unified Model for Bipolar Outflows from Young Stars: Kinematic Signatures of Jets, Winds, and Their Magnetic Interplay with the Ambient Toroids},
      journal = {\apj},
         year = 2023,
        month = feb,
       volume = {944},
       number = {2},
          eid = {230},
        pages = {230},
          doi = {10.3847/1538-4357/aca763},
archivePrefix = {arXiv},
       eprint = {2301.07447},
 primaryClass = {astro-ph.SR},
       adsurl = {https://ui.adsabs.harvard.edu/abs/2023ApJ...944..230S}
}

@misc{JWST_MAST_2022,
  doi = {10.17909/WV1N-RF97},
  url = {http://archive.stsci.edu/doi/resolve/resolve.html?doi=10.17909/wv1n-rf97},
  author = {{Yang,  Yao-Lun}},
  title = {IRAS 15398-3359 pipeline products},
  publisher = {STScI/MAST},
  year = {2022}
}

@article{2013astropy,
        Adsurl = {https://adsabs.harvard.edu/abs/2013A%26A...558A..33A},
        Archiveprefix = {arXiv},
        Author = {{Astropy Collaboration} and {Robitaille}, T.~P. and {Tollerud}, E.~J. and {Greenfield}, P. and {Droettboom}, M. and {Bray}, E. and {Aldcroft}, T. and {Davis}, M. and {Ginsburg}, A. and {Price-Whelan}, A.~M. and {Kerzendorf}, W.~E. and {Conley}, A. and {Crighton}, N. and {Barbary}, K. and {Muna}, D. and {Ferguson}, H. and {Grollier}, F. and {Parikh}, M.~M. and {Nair}, P.~H. and {Unther}, H.~M. and {Deil}, C. and {Woillez}, J. and {Conseil}, S. and {Kramer}, R. and {Turner}, J.~E.~H. and {Singer}, L. and {Fox}, R. and {Weaver}, B.~A. and {Zabalza}, V. and {Edwards}, Z.~I. and {Azalee Bostroem}, K. and {Burke}, D.~J. and {Casey}, A.~R. and {Crawford}, S.~M. and {Dencheva}, N. and {Ely}, J. and {Jenness}, T. and {Labrie}, K. and {Lim}, P.~L. and {Pierfederici}, F. and {Pontzen}, A. and {Ptak}, A. and {Refsdal}, B. and {Servillat}, M. and {Streicher}, O.},
        Doi = {10.1051/0004-6361/201322068},
        Eid = {A33},
        Eprint = {1307.6212},
        Journal = {\aap},
        Month = oct,
        Pages = {A33},
        Primaryclass = {astro-ph.IM},
        Title = {{Astropy: A community Python package for astronomy}},
        Volume = 558,
        Year = 2013}

@ARTICLE{2018astropy,
               author = {{Astropy Collaboration} and {Price-Whelan}, A.~M. and
                 {Sip{\H{o}}cz}, B.~M. and {G{\"u}nther}, H.~M. and {Lim}, P.~L. and
                 {Crawford}, S.~M. and {Conseil}, S. and {Shupe}, D.~L. and
                 {Craig}, M.~W. and {Dencheva}, N. and {Ginsburg}, A. and {Vand
                erPlas}, J.~T. and {Bradley}, L.~D. and {P{\'e}rez-Su{\'a}rez}, D. and
                 {de Val-Borro}, M. and {Aldcroft}, T.~L. and {Cruz}, K.~L. and
                 {Robitaille}, T.~P. and {Tollerud}, E.~J. and {Ardelean}, C. and
                 {Babej}, T. and {Bach}, Y.~P. and {Bachetti}, M. and {Bakanov}, A.~V. and
                 {Bamford}, S.~P. and {Barentsen}, G. and {Barmby}, P. and
                 {Baumbach}, A. and {Berry}, K.~L. and {Biscani}, F. and {Boquien}, M. and
                 {Bostroem}, K.~A. and {Bouma}, L.~G. and {Brammer}, G.~B. and
                 {Bray}, E.~M. and {Breytenbach}, H. and {Buddelmeijer}, H. and
                 {Burke}, D.~J. and {Calderone}, G. and {Cano Rodr{\'\i}guez}, J.~L. and
                 {Cara}, M. and {Cardoso}, J.~V.~M. and {Cheedella}, S. and {Copin}, Y. and
                 {Corrales}, L. and {Crichton}, D. and {D'Avella}, D. and {Deil}, C. and
                 {Depagne}, {\'E}. and {Dietrich}, J.~P. and {Donath}, A. and
                 {Droettboom}, M. and {Earl}, N. and {Erben}, T. and {Fabbro}, S. and
                 {Ferreira}, L.~A. and {Finethy}, T. and {Fox}, R.~T. and
                 {Garrison}, L.~H. and {Gibbons}, S.~L.~J. and {Goldstein}, D.~A. and
                 {Gommers}, R. and {Greco}, J.~P. and {Greenfield}, P. and
                 {Groener}, A.~M. and {Grollier}, F. and {Hagen}, A. and {Hirst}, P. and
                 {Homeier}, D. and {Horton}, A.~J. and {Hosseinzadeh}, G. and {Hu}, L. and
                 {Hunkeler}, J.~S. and {Ivezi{\'c}}, {\v{Z}}. and {Jain}, A. and
                 {Jenness}, T. and {Kanarek}, G. and {Kendrew}, S. and {Kern}, N.~S. and
                 {Kerzendorf}, W.~E. and {Khvalko}, A. and {King}, J. and {Kirkby}, D. and
                 {Kulkarni}, A.~M. and {Kumar}, A. and {Lee}, A. and {Lenz}, D. and
                 {Littlefair}, S.~P. and {Ma}, Z. and {Macleod}, D.~M. and
                 {Mastropietro}, M. and {McCully}, C. and {Montagnac}, S. and
                 {Morris}, B.~M. and {Mueller}, M. and {Mumford}, S.~J. and {Muna}, D. and
                 {Murphy}, N.~A. and {Nelson}, S. and {Nguyen}, G.~H. and
                 {Ninan}, J.~P. and {N{\"o}the}, M. and {Ogaz}, S. and {Oh}, S. and
                 {Parejko}, J.~K. and {Parley}, N. and {Pascual}, S. and {Patil}, R. and
                 {Patil}, A.~A. and {Plunkett}, A.~L. and {Prochaska}, J.~X. and
                 {Rastogi}, T. and {Reddy Janga}, V. and {Sabater}, J. and
                 {Sakurikar}, P. and {Seifert}, M. and {Sherbert}, L.~E. and
                 {Sherwood-Taylor}, H. and {Shih}, A.~Y. and {Sick}, J. and
                 {Silbiger}, M.~T. and {Singanamalla}, S. and {Singer}, L.~P. and
                 {Sladen}, P.~H. and {Sooley}, K.~A. and {Sornarajah}, S. and
                 {Streicher}, O. and {Teuben}, P. and {Thomas}, S.~W. and
                 {Tremblay}, G.~R. and {Turner}, J.~E.~H. and {Terr{\'o}n}, V. and
                 {van Kerkwijk}, M.~H. and {de la Vega}, A. and {Watkins}, L.~L. and
                 {Weaver}, B.~A. and {Whitmore}, J.~B. and {Woillez}, J. and
                 {Zabalza}, V. and {Astropy Contributors}},
                title = {The Astropy Project: Building an Open-science Project and Status of the v2.0 Core Package},
              journal = {\aj},
                 year = 2018,
                month = sep,
               volume = {156},
               number = {3},
                  eid = {123},
                pages = {123},
                  doi = {10.3847/1538-3881/aabc4f},
        archivePrefix = {arXiv},
               eprint = {1801.02634},
         primaryClass = {astro-ph.IM},
               adsurl = {https://ui.adsabs.harvard.edu/abs/2018AJ....156..123A}
        }

@ARTICLE{2022astropy,
               author = {{Astropy Collaboration} and {Price-Whelan}, Adrian M. and {Lim}, Pey Lian and {Earl}, Nicholas and {Starkman}, Nathaniel and {Bradley}, Larry and {Shupe}, David L. and {Patil}, Aarya A. and {Corrales}, Lia and {Brasseur}, C.~E. and {N{"o}the}, Maximilian and {Donath}, Axel and {Tollerud}, Erik and {Morris}, Brett M. and {Ginsburg}, Adam and {Vaher}, Eero and {Weaver}, Benjamin A. and {Tocknell}, James and {Jamieson}, William and {van Kerkwijk}, Marten H. and {Robitaille}, Thomas P. and {Merry}, Bruce and {Bachetti}, Matteo and {G{"u}nther}, H. Moritz and {Aldcroft}, Thomas L. and {Alvarado-Montes}, Jaime A. and {Archibald}, Anne M. and {B{'o}di}, Attila and {Bapat}, Shreyas and {Barentsen}, Geert and {Baz{'a}n}, Juanjo and {Biswas}, Manish and {Boquien}, M{'e}d{'e}ric and {Burke}, D.~J. and {Cara}, Daria and {Cara}, Mihai and {Conroy}, Kyle E. and {Conseil}, Simon and {Craig}, Matthew W. and {Cross}, Robert M. and {Cruz}, Kelle L. and {D'Eugenio}, Francesco and {Dencheva}, Nadia and {Devillepoix}, Hadrien A.~R. and {Dietrich}, J{"o}rg P. and {Eigenbrot}, Arthur Davis and {Erben}, Thomas and {Ferreira}, Leonardo and {Foreman-Mackey}, Daniel and {Fox}, Ryan and {Freij}, Nabil and {Garg}, Suyog and {Geda}, Robel and {Glattly}, Lauren and {Gondhalekar}, Yash and {Gordon}, Karl D. and {Grant}, David and {Greenfield}, Perry and {Groener}, Austen M. and {Guest}, Steve and {Gurovich}, Sebastian and {Handberg}, Rasmus and {Hart}, Akeem and {Hatfield-Dodds}, Zac and {Homeier}, Derek and {Hosseinzadeh}, Griffin and {Jenness}, Tim and {Jones}, Craig K. and {Joseph}, Prajwel and {Kalmbach}, J. Bryce and {Karamehmetoglu}, Emir and {Ka{l}uszy{'n}ski}, Miko{l}aj and {Kelley}, Michael S.~P. and {Kern}, Nicholas and {Kerzendorf}, Wolfgang E. and {Koch}, Eric W. and {Kulumani}, Shankar and {Lee}, Antony and {Ly}, Chun and {Ma}, Zhiyuan and {MacBride}, Conor and {Maljaars}, Jakob M. and {Muna}, Demitri and {Murphy}, N.~A. and {Norman}, Henrik and {O'Steen}, Richard and {Oman}, Kyle A. and {Pacifici}, Camilla and {Pascual}, Sergio and {Pascual-Granado}, J. and {Patil}, Rohit R. and {Perren}, Gabriel I. and {Pickering}, Timothy E. and {Rastogi}, Tanuj and {Roulston}, Benjamin R. and {Ryan}, Daniel F. and {Rykoff}, Eli S. and {Sabater}, Jose and {Sakurikar}, Parikshit and {Salgado}, Jes{'u}s and {Sanghi}, Aniket and {Saunders}, Nicholas and {Savchenko}, Volodymyr and {Schwardt}, Ludwig and {Seifert-Eckert}, Michael and {Shih}, Albert Y. and {Jain}, Anany Shrey and {Shukla}, Gyanendra and {Sick}, Jonathan and {Simpson}, Chris and {Singanamalla}, Sudheesh and {Singer}, Leo P. and {Singhal}, Jaladh and {Sinha}, Manodeep and {Sip{H{o}}cz}, Brigitta M. and {Spitler}, Lee R. and {Stansby}, David and {Streicher}, Ole and {{{S}}umak}, Jani and {Swinbank}, John D. and {Taranu}, Dan S. and {Tewary}, Nikita and {Tremblay}, Grant R. and {Val-Borro}, Miguel de and {Van Kooten}, Samuel J. and {Vasovi{'c}}, Zlatan and {Verma}, Shresth and {de Miranda Cardoso}, Jos{'e} Vin{'i}cius and {Williams}, Peter K.~G. and {Wilson}, Tom J. and {Winkel}, Benjamin and {Wood-Vasey}, W.~M. and {Xue}, Rui and {Yoachim}, Peter and {Zhang}, Chen and {Zonca}, Andrea and {Astropy Project Contributors}},
                title = {The Astropy Project: Sustaining and Growing a Community-oriented Open-source Project and the Latest Major Release (v5.0) of the Core Package},
              journal = {\apj},
                 year = 2022,
                month = aug,
               volume = {935},
               number = {2},
                  eid = {167},
                pages = {167},
                  doi = {10.3847/1538-4357/ac7c74},
        archivePrefix = {arXiv},
               eprint = {2206.14220},
         primaryClass = {astro-ph.IM},
               adsurl = {https://ui.adsabs.harvard.edu/abs/2022ApJ...935..167A}
        }

@ARTICLE{2025Sabatini,
       author = {{Sabatini}, G. and {Bianchi}, E. and {Chandler}, C.~J. and {Cacciapuoti}, L. and {Podio}, L. and {Maureira}, M.~J. and {Codella}, C. and {Ceccarelli}, C. and {Sakai}, N. and {Testi}, L. and {Toci}, C. and {Svoboda}, B. and {Sakai}, T. and {Bouvier}, M. and {Caselli}, P. and {Cuello}, N. and {De Simone}, M. and {J{\'\i}menez-Serra}, I. and {Johnstone}, D. and {Loinard}, L. and {Zhang}, Z.~E. and {Yamamoto}, S.},
        title = {FAUST: XXIV. Large dust grains in the protostellar outflow cavity walls of the Class I binary L1551 IRS5},
      journal = {\aap},
         year = 2025,
        month = jun,
       volume = {698},
          eid = {L16},
        pages = {L16},
          doi = {10.1051/0004-6361/202554750},
archivePrefix = {arXiv},
       eprint = {2505.13596},
 primaryClass = {astro-ph.GA},
       adsurl = {https://ui.adsabs.harvard.edu/abs/2025A&A...698L..16S}
}

\end{document}